%% file: main.tex
\documentclass[%
 aps,prb, amsmath,amssymb,
reprint,%
superscriptaddress
]{revtex4-1}

\usepackage{graphicx}
\usepackage{dcolumn}
\usepackage{float}
\usepackage{placeins}
\usepackage{booktabs,xspace,multirow,bm}
\usepackage[dvipsnames]{xcolor}

\usepackage[utf8]{inputenc}
\usepackage[T1]{fontenc}

\DeclareUnicodeCharacter{0301}{\'y}

\newcommand{\rrscan}{r$^2$SCAN\xspace}
\newcommand{\rtt}[1]{\textcolor{black}{#1}}

\begin{document}

\title{\rtt{Workhorse minimally-empirical dispersion-corrected density functional, with tests for weakly-bound systems: \rrscan{}+rVV10}}

\author{Jinliang Ning}
\affiliation{Department of Physics and Engineering Physics, Tulane University, New Orleans, Louisiana 70118, United States}
\author{Manish Kothakonda}
\affiliation{Department of Physics and Engineering Physics, Tulane University, New Orleans, Louisiana 70118, United States}
\author{James W. Furness}
\affiliation{Department of Physics and Engineering Physics, Tulane University, New Orleans, Louisiana 70118, United States}

\author{Aaron D. Kaplan}
\affiliation{Department of Physics, Temple University, Philadelphia, Pennsylvania 19122, United States}
\author{Sebastian Ehlert}
\affiliation{Mulliken Center for Theoretical Chemistry, University of Bonn, Beringstr. 4, 53115 Bonn, Germany}

\author{Jan Gerit Brandenburg}
\affiliation{Chief Science and Technology Office, Merck KGaA, Frankfurter Str. 250, 64293 Darmstadt, Germany}

\author{John P. Perdew}

\affiliation{Department of Physics, Temple University, Philadelphia, Pennsylvania 19122, United States}
\affiliation{Department of Chemistry, Temple University, Philadelphia, Pennsylvania 19122, United States}
\author{Jianwei Sun}
\email{jsun@tulane.edu}
\affiliation{Department of Physics and Engineering Physics, Tulane University, New Orleans, Louisiana 70118, United States}

\date{\today}

\begin{abstract}

SCAN+rVV10 has been demonstrated to be a versatile van der Waals (vdW) density functional that delivers good predictions of both energetic and structural properties for many types of bonding.
Recently, the r$^2$SCAN functional has been devised as a revised form of SCAN with improved numerical stability.
In this work, we refit the rVV10 functional to optimize the r$^2$SCAN+rVV10 vdW density functional, and test its performance for molecular interactions and layered materials.
Our molecular tests demonstrate that r$^2$SCAN+rVV10 outperforms its predecessor SCAN+rVV10 in both efficiency (numerical stability) and accuracy.
This good performance is also found in lattice-constant predictions.
In comparison with benchmark results from higher-level theories or experiments, r$^2$SCAN+rVV10 yields excellent interlayer binding energies and phonon dispersions for layered materials.
\end{abstract}

\maketitle

\section{\label{sec:intro}Introduction}

Quantum fluctuations in the electronic density give rise to instantaneous dipole moments, making van der Waals (vdW) or London dispersion interactions ubiquitous in electronic matter.
Despite its relative small strength, the ubiquitous vdW force plays a fundamental role in diverse fields of both science and industry: from structural biology and polymer science, to nanotechnology and surface science.
It participates in the structural evolution of DNA\cite{DNA_vdw}, proteins\cite{protein_vdw}, and many other complex molecules and their interactions\cite{biology_vdw}, and hence the origination\cite{life_vdw} and physical activities of living beings.
The vdW forces are also crucial for the surface and interfacial reactions controlling artificial and natural catalytic\cite{complex_vdw,Aucat_vdw,enzyme_vdw} and corrosion reactions on alloy surfaces\cite{corrosion_vdW}.
The vdW interactions are even found to be necessary for accurate descriptions of some densely packed systems, suggesting that vdW forces are not as negligible for normal solids as commonly thought \cite{vdwUsyn,vdw_alkali,vdw_solids}.

While vdW interactions are fully captured in the exact density functional theory (DFT)\cite{KohnShamDFT}, their non-local nature means they (or at least their most long-ranged parts) are missed by semi-local exchange-correlation (XC) density functional approximations (DFAs) like the local density approximation (LDA), generalized gradient approximation (GGA), or meta-GGA.
Despite this limitation, semi-local DFAs are the mainstay of modern first-principles electronic structure modelling, achieving useful accuracy at reasonable cost.
While higher-level methods that fully account for vdW forces, such as quantum Monte Carlo (QMC)\cite{QMC}, coupled-cluster singles and doubles with perturbative triples CCSD(T)\cite{CCSDT}, and the adiabatic-connection fluctuation-dissipation theorem within the random-phase approximation (RPA)\cite{RPA}, can provide benchmark references, their poor scaling with system size prohibits large-scale applications.
Instead, a practical choice for improving accuracy is to include vdW interactions in the DFT framework as a modification or correction to a semi-local XC approximation.
Common approaches include the DFT+D series\cite{D2,D3,D4,r2SCAND4}, Tkatchenko-Scheffler (TS) methods\cite{TS1,TS2,TS3}, the Rutgers-Chalmers vdW-DF family\cite{vdw-df}, Vydrov-van Voorhis (VV10)\cite{vv10}, rVV10\cite{rvv10} density functionals, and the Becke--Johnson exchange hole model \cite{Becke2005, Becke2007}.
We should also mention the damped-Zaremba-Kohn (dZK) \cite{tang2020,chowdhury2021} correction, which requires many material-dependent input parameters.

The performance of the vdW-corrected DFA depends upon both the semi-local XC and vdW functionals.
A good example for this case is the SCAN+rVV10 vdW functional.
The strongly constrained and appropriately normed (SCAN) meta-GGA\cite{SCAN}, satisfies all known 17 exact constraints applicable to a meta-GGA, and has shown good accuracy for diverse bonding environments \cite{SCAN_NChem}.
It has been demonstrated that SCAN includes a portion of the intermediate range of vdW interactions, which rationalizes its excellent predictions of structural and energetic properties of water \cite{SCAN_NChem}.
The rVV10 non-local vdW density functional \cite{rvv10} requires only the electron density and its first derivatives as inputs, and contains two empirical parameters, $C$ and $b$.
The final SCAN+rVV10 vdW density functional has been demonstrated to work for general geometries, and achieves an accuracy comparable to that of higher-level methods like RPA and CCSD(T) for various vdW benchmark systems, while scaling  more favorably with system size\cite{SCAN+rVV10}.

Despite these successes, SCAN exhibits undesirable numerical problems \cite{beta_indicator,bartok2019} that harm its computational efficiency and can prevent the self-consistent field process from converging.
To achieve high accuracy for diverse systems, SCAN interpolates between single-orbital and slowly-varying energy densities using a variable $\alpha$ (defined in Ref. \citenum{SCAN}) that is sensitive to the local chemical environment.
$\alpha$ partly contributes to the numerical instability of SCAN\cite{beta_indicator}.

Moreover, SCAN's inclusion of intermediate vdW interactions can be a hindrance when combined with non-local dispersion corrections.
SCAN predicts quantitatively correct lattice parameters for the layered solid PPTA, whereas SCAN+rVV10 strongly overbinds within layers, yielding a much too-small $a$-parameter \cite{Yu2020}.
SCAN's tendency to overbind hydrogen-bound molecules is worsened in both SCAN+rVV10 and SCAN+D3 \cite{scand3,Wiktor2017,Hermann2018}.
When evaluated on the Hartree-Fock density (a kind of ``density correction''), SCAN provides a chemically accurate description of liquid water, whereas dispersion-corrected variants of SCAN still overbind \cite{Dasgupta2021}.

The rSCAN meta-GGA \cite{bartok2019} modifies SCAN to successfully improve numerical stability, but at the price of reduced accuracy \cite{Mejia-Rodriguez2019, Bartok2019a,mejia-rodriguez2020a}.
To remove the divergence in the derivatives of $\alpha$ in single orbital regions ($\alpha \to 0$),\cite{beta_indicator} rSCAN uses a regularized $\alpha^\prime$ that breaks exact coordinate scaling conditions \cite{Levy1985a, Gorling1993a, Pollack2000} and the uniform density limit obeyed by SCAN.
To remove oscillations in the exchange-correlation potential of SCAN induced by the function of $\alpha$ that interpolates between energy densities, rSCAN uses a smooth polynomial for the chemically-relevant range $0\leq \alpha \leq 2.5$.
This choice introduces spurious terms in the slowly-varying ($\alpha\approx 1$) density-gradient expansion that deviate from the exact expansion\cite{Svendsen1996, Perdew1992a} recovered by SCAN.

These shortcomings are remedied by the r$^2$SCAN meta-GGA\cite{r2SCAN}, which modifies the rSCAN regularizations to obey almost every exact constraint SCAN does.
(A higher-order density-gradient expansion for exchange is recovered by SCAN \cite{furness2021}.)
The satisfaction of exact constraints and greater smoothness of \rrscan preserves the accuracy of SCAN and numerical efficiency of rSCAN \cite{r2SCAN, r2SCAND4, mejia-rodriguez2020a, Grimme2021}, permitting construction of meta-GGA pseudopotentials \cite{holzwarth2022}.
Therefore, we expect r$^2$SCAN to be a better candidate for the rVV10 correction.

It should be noted that a variant of \rrscan with a long range D4 \cite{D4} dispersion correction was recently published. \cite{r2SCAND4}
\rrscan{}+D4 more realistically describes non-covalent and hydrogen-bound systems than SCAN+D4,\cite{r2SCAND4}, suggesting that \rrscan includes less of the intermediate vdW interaction than SCAN.
Reference \citenum{r2SCAND4} presented a fitted value $b=12.3$ for \rrscan{}+VV10 (\textit{not} rVV10).
rVV10 was designed to perform like VV10, but at a lower computational cost in plane-wave basis set codes.
We now motivate why an \rrscan{}+rVV10 is needed when a highly-accurate \rrscan{}+D4 exists.

The D and VV10 series of vdW corrections are complementary approaches for describing long-range vdW interactions in real systems.
Both corrections have empirical parts, with the VV10 series requiring two material-independent empirical parameters, and D4 requiring three parameters in its damping function.
The D4 dispersion coefficients are computed on-the-fly from tabulated material-dependent data like the atomic polarizabilities and Mulliken partial charges \cite{D4}.
rVV10 is conceptually simpler than D4 and its reliance on fewer empirical parameters makes it an appealing alternative to D4 for solid-state physics, though both methods find common use.
In a comparison \cite{mardirossian2017a} of 243 non-covalent cluster interactions, SCAN-D3 and SCAN+rVV10 had comparable root mean square deviations from reference values.

The original VV10 \cite{vv10} and subsequent rVV10 \cite{rvv10} vdW corrections differ in subtle ways.
The VV10 kernel is a two-point function, and its evaluation requires a double integral over real space.
Such a correction is challenging to implement in plane-wave codes because of the high numeric cost of this double integral.
The rVV10 kernel approximates the VV10 kernel by interpolation over a set of grid points, drastically reducing the computational overhead in plane-wave basis set codes.

When rVV10 is a good approximation to VV10, the $b$ parameters should not differ substantially.
We confirm this interpretation here.
However, a VV10-corrected DFA which tends to overbind molecules is expected \cite{mardirossian2017} to further overbind when combined with rVV10 using the same $b$ parameter.
When using the same $b$ parameter, the most pronounced differences between VV10 and rVV10 occur in low-density regions \cite{terentjev2019}.
However, the dispersion correction to a meta-GGA like SCAN or \rrscan should be most meaningful in these low-density regions.

A limitation of the VV10 and rVV10 long-range dispersion corrections is that they can describe only two-body interactions between volume elements, ignoring the three-body Axilrod-Teller\cite{axilrod1943} effects.
Here we fit the $b$ parameters in those corrections to the binding energy curve of the Ar dimer, in which the conventional many-body expansion stops at the two-body term.

The vdW interactions are crucial in shaping the structure and properties of 2D/layered materials.
Such materials have seen renewed interest since the exfoliation of graphene in 2004 \cite{Graphene}, and have nurtured new applications promising the next generation of information technology devices\cite{2D}.
As such, we test the newly determined $b$ parameter for \rrscan +rVV10 on standard sets, with a focus on layered materials properties.

\section{\label{sec:methods}Methods}

\subsection{\label{subsec:Parameters}Parameters in r$^2$SCAN+rVV10}

The rVV10\cite{vv10,rvv10} non-local correlation functional is similar in construction to the Rutgers--Chalmers vdW-DF family\cite{vdw-df},
\begin{equation}
\begin{split}
  E_{\mathrm{c}}^{\mathrm{nl}} =
  \int {d\mathbf{r}\,n(\mathbf{r})} \left[ \frac{\hbar}{2} \int {d\mathbf{r^{\prime}}\,\phi(\mathbf{r},\mathbf{r^{\prime}}) n(\mathbf{r^{\prime}})} + \beta \right].
\end{split}
\end{equation}
$\beta$ vanishes for the Rutgers--Chalmers vdW-DFs, and the XC functional reads as
\begin{equation}
E_{\mathrm{xc}} = E_{\mathrm{xc}}^{0} + E_{\mathrm{c}}^{\mathrm{nl}}.
\end{equation}

Here, $n(\mathbf{r})$ is the electron density, $\phi(\mathbf{r},\mathbf{r^{\prime}})$ is the density-density interaction kernel, and $E_{\mathrm{xc}}^{0}$ is the semi-local exchange correlation functionals to be corrected.
$\beta = (3/b^2)^{3/4}/32$ in Hartree is required for zero $E_{\mathrm{c}}^{\mathrm{nl}}$ for the uniform electron gas.
Two empirical dimensionless parameters $C$ and $b$ appear in the kernel $\phi(\mathbf{r},\mathbf{r^{\prime}})$: $C$ is adjusted to recover the accurate $-C_{6}/R^{6}$ asymptotic vdW interaction between molecules at large separation $R$, and $b$ controls the damping of $E_{\mathrm{c}}^{\mathrm{nl}}$ at short range.

The original VV10 and rVV10 functionals \cite{vv10,rvv10} were combined with the semi-local XC functional \cite{PBE,xforvdw} $E_{\mathrm{xc}}^{0} = E_{\mathrm{x}}^{\mathrm{rPW86}} + E_{\mathrm{c}}^{\mathrm{PBE}}$, partly due to the near absence of vdW in rPW86 exchange \cite{xforvdw}.
(For a discussion of how intermediate-range vdW can arise from semilocal exchange, see Ref. \citenum{SCAN+rVV10}.)
For a semi-local $E_{\mathrm{xc}}^{0}$, $C = 0.0093$ was recommended \cite{vv10}, and the $b$ parameter was determined as 5.9 and 6.3 by ﬁtting to the interaction energies of the S22 set \cite{S22,S22A} for the original VV10 and rVV10, respectively.
Increasing $C$ or $b$ generally results in a smaller vdW correction.
There is a conventional many-body expansion \cite{hankins1979} of the dispersion interaction within a collection of bodies (atoms or molecules) that includes two-body and higher-order many-body effective interactions.
By construction, the VV10 and rVV10 long-range corrections explicitly account for only pairwise interactions between volume elements.
The Ar dimer has only conventional two-atom interactions, whereas the S22 has many-atom interactions.
Fitting rVV10 to systems with many-atom interactions would average over the two- and many-atom interactions \cite{dobson2014}.

\begin{figure}
\includegraphics[width=0.45\textwidth]{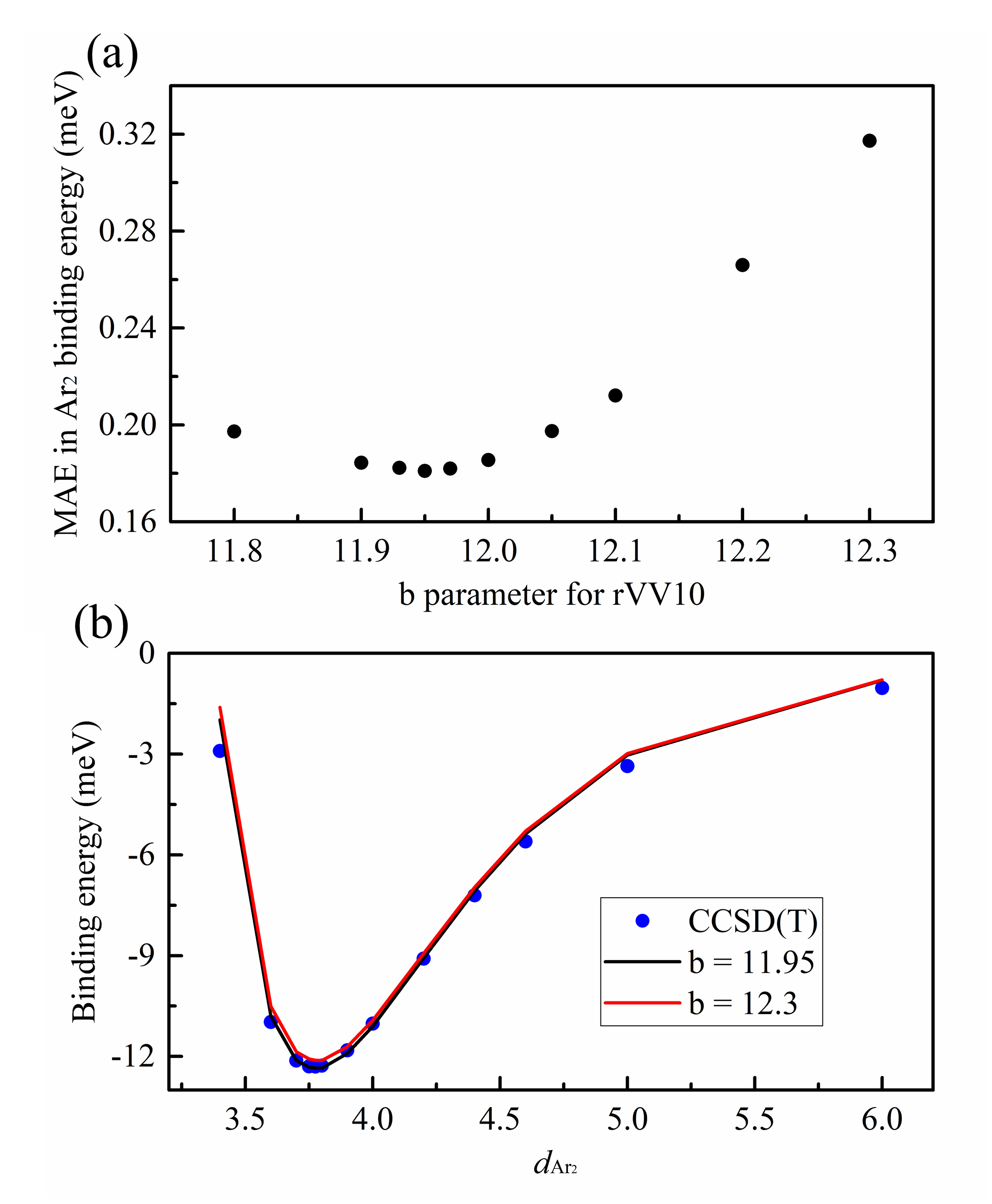}
\vspace{-0.4cm}
\caption{\footnotesize (a) Mean absolute error of the Ar dimer binding energy as a function of the $b$ parameter in r$^2$SCAN+rVV10. (b) The binding curve for Ar dimer from r$^2$SCAN+rVV10 (red solid line) compared to CCSD(T) curve\cite{Ar2CCSD_1, Ar2CCSD_2} as the reference (blue dots) as a function of their separation $d_{\text{Ar}_2}$ in \AA{}.}
\label{fig:bfitAr2}
\end{figure}

Here, we refit $b=11.95$ for \rrscan{}+rVV10 by adjusting it to best recover the binding-energy curve of the argon dimer with bond lengths between 3.5 and 6.0 \AA, as shown in Fig. \ref{fig:bfitAr2}(a).
Using the \rrscan{}+VV10 (MAE 0.32 kcal/mol for S22 \cite{r2scan_d4_repo}) value $b=12.3$ \cite{r2SCAND4}, the mean absolute error (MAE) in the binding energy curve of Ar$_2$ increases \rtt{negligibly by} 0.2 meV (0.0046 kcal/mol).
Note that \rrscan{}-D4 makes a 0.29 kcal/mol MAE on the S22 set \cite{r2scan_d4_repo}, \rtt{virtually identical to} both variants of \rrscan{}+rVV10.

\subsection{\label{subsec:Computational}Computational details}
The DFT calculations in this work were performed using the Vienna Ab initio Simulation Package (VASP) \cite{VASP} \rtt{version 5.4.4, with user corrections for the meta-GGA correlation potential in spin-unrestricted calculations, and to the rVV10 stress tensor \cite{vasp_stress_bug}.}
For computational details of the Ar dimer binding-energy curve, the S22 molecular interaction energies \cite{S22, S22A, S22B}, and the L28 layered material database \cite{L28_jcp}, refer to the the original publication of SCAN+rVV10 \cite{SCAN+rVV10}, with the following adjustments.
We follow the practice \cite{SCAN+rVV10} of using hard pseudopotentials for the S22 set,
due to their better accuracy for molecules with short bonds, and as recommended by the VASP manual\cite{VASP}.
\rtt{All input and output files for the Ar$_2$, S22, and L28 calculations can be found at the public code repository \cite{ning2022}.}

Error statistics for the inter-layer binding energies and lattice parameters of the L28 set are presented in Tables \ref{tab:table2} and \ref{tab:table3}.
Values for individual solids are presented in Tables \ref{tab:table_A2} and \ref{tab:table_A3} of the Appendix.
The same methods used to validate SCAN+rVV10\cite{SCAN+rVV10} are used here for calculation of the L28 binding energies (compared to reference RPA \cite{Eb_RPA} calculations): the intra-layer lattice constants were fixed to their experimental values, and the inter-layer lattice constants were relaxed only for the bulk structures.
Only atomic coordinates were relaxed for the mono-layer model, as in the RPA calculations.
The calculated lattice constants in Tables \ref{tab:table2}, \ref{tab:table3}, \ref{tab:table_A2} and \ref{tab:table_A3} are from full relaxations.

Non-magnetic ground states were used in the current calculations for these compounds, except the three vanadium-based compounds, where a ferromagnetic ordering was used instead.
For the SCAN+rVV10 and r$^2$SCAN+rVV10 results in Table \ref{tab:table4}, intra-layer lattice constants were also relaxed for the bulk and mono-layer models, although the difference in binding energy was negligible.
For the phonon calculations of graphite and MoS$_2$, we used the Phonopy code \cite{phonopy} to obtain the harmonic force constants from VASP atomic force calculations within the finite displacement method (0.015 \AA).

For solid poly(\textit{p}-phenylene terephthalamide) (PPTA), $\bm{k}$-point spacing of 0.15 \AA{}$^{-1}$ (yielding a $6\times 9 \times 4$ $\bm{k}$-grid), and a plane-wave cutoff of 900 eV were used.

\section{\label{sec:Results}Results and discussions}

\subsection{Dispersion interactions in molecules \label{sec:results_mols}}
\FloatBarrier

To evaluate the performance of r$^2$SCAN+rVV10 with the newly fit $b = 11.95$, we tested it on both molecular systems (S22 data set) and layered materials. We are especially interested in the efficiency and accuracy of r$^2$SCAN+rVV10, in comparison with its predecessor SCAN+rVV10.

\input{Tables/table1_simplified}

We assessed the accuracy of SCAN+rVV10 and r$^2$SCAN+rVV10 predicted interaction energies for the S22 molecular complexes data set.
The S22 set includes seven hydrogen-bonded, eight dispersion-bound, and seven mixed-binding complexes. Table \ref{tab:table1} presents the error statistics of SCAN+rVV10 and r$^2$SCAN+rVV10 for the S22 set, relative to CCSD(T) benchmarks \cite{S22B}.
Table \ref{tab:table_A1} of the Appendix complements Table \ref{tab:table1}, presenting values for each molecule in the S22 set, and comparing our SCAN+rVV10 and r$^2$SCAN+rVV10 results with Perdew-Burke-Ernzerhof (PBE) \cite{PBE}, SCAN, and vdW-DF2\cite{vdW_DF2} predictions.

To further demonstrate the improved numeric stability of \rrscan over SCAN, we also present results using smaller grid sizes in these tables.
Two parameters, ENCUT and ENAUG, control the size of the plane-wave basis sets used by VASP.
A plane-wave basis set offers a systematic approach to converged total energies by adding more reciprocal lattice vectors $\bm{G}$ to the set.
ENCUT (in eV) controls how many $\bm{G}$ are used to represent the valence electron density by accepting only those $|\bm G + \bm k|^2 < 2(\text{ENCUT})$ for each $\bm k$-point.
In the pseudopotential approach used by VASP, the potential due to core states is represented by a non-local potential within an ``augmentation'' radius.
ENAUG controls the number of $\bm G$ used to represent the orbitals within the augmentation radius, in the same fashion as ENCUT.
\rtt{We have noticed a strong sensitivity of SCAN-like meta-GGAs to the ENAUG setting, which we have set at an appropriately high value (2000 eV) to ensure well-converged results.
Similar grid sensitivities were noted \cite{gould2018} for SCAN and SCAN + rVV10 applied to different arrangements of the benzene dimer.
Note also that VASP permits compilation with a precompiler flag, DnoAugXCmeta, that does not use the augmented charge density in meta-GGA calculations.
That flag was not used in this work.}

With the refit $b = 11.95$, r$^2$SCAN+rVV10 outperforms SCAN+rVV10 in all three subgroups and overall for the S22 binding energy database, and has an accuracy competitive with the original rVV10 functional.
When compared to its excellent performance for dispersion-bound and mixed complexes, though improvement is noteworthy, r$^2$SCAN+rVV10 still tends to over-bind hydrogen-bonded systems.
This is rationalized as a density-driven error, rather than an error inherent to rVV10.
For example, the hydrogen-bonded water dimer is over-bound by 0.44 kcal/mol or 9\% in SCAN, and this error is reduced to 0.13 kcal/mol when SCAN is applied to the more accurate Hartree-Fock electron density, and not to its own self-consistent density \cite{dasgupta2022}.
That fact speaks for fitting the $b$ parameter of rVV10 to the binding energy curve of the Ar dimer (as done here) or to the eight dispersion-bound complexes in S22, and not to the whole S22 set.
For the eight dispersion-bound complexes of S22, \rrscan{}+rVV10 is quite accurate (see Table \ref{tab:table1}).
The superior numerical performance of r$^2$SCAN over SCAN is consistent with other works studying molecular systems \cite{r2SCAN}, lattice dynamics of solid-state systems \cite{r2SCAN_phonon}, and in combination with the D4 vdW functional\cite{r2SCAND4}.

Column (d) of Table \ref{tab:table1} presents the S22 error scores of \rrscan{}+rVV10 with the VV10 value $b=12.3$.
The 0.0--0.04 kcal/mol differences in the converged S22 mean absolute errors (MAEs) using both $b$ parameters are comparable to the error in the reference CCSD(T) values, which used \cite{S22B} small triple-$\zeta$ grids.
Thus we cannot definitively say that one value of $b$ is better for describing common noncovalent interactions.
The method \cite{D4} that fitted $b=12.3$ used larger sets of dispersion-bound dimers as a function of the inter-monomer separation, yet yields essentially the same average errors as $b=11.95$, fitted to the Ar dimer.
As explicated in Ref. \citenum{mardirossian2017}, we advocate for using different parameters in VV10- and rVV10-corrected DFAs; thus we recommend using $b=12.3$ for \rrscan{}+VV10, and $b=11.95$ for \rrscan{}+rVV10.

\begin{figure}
	\includegraphics[width=0.45\textwidth]{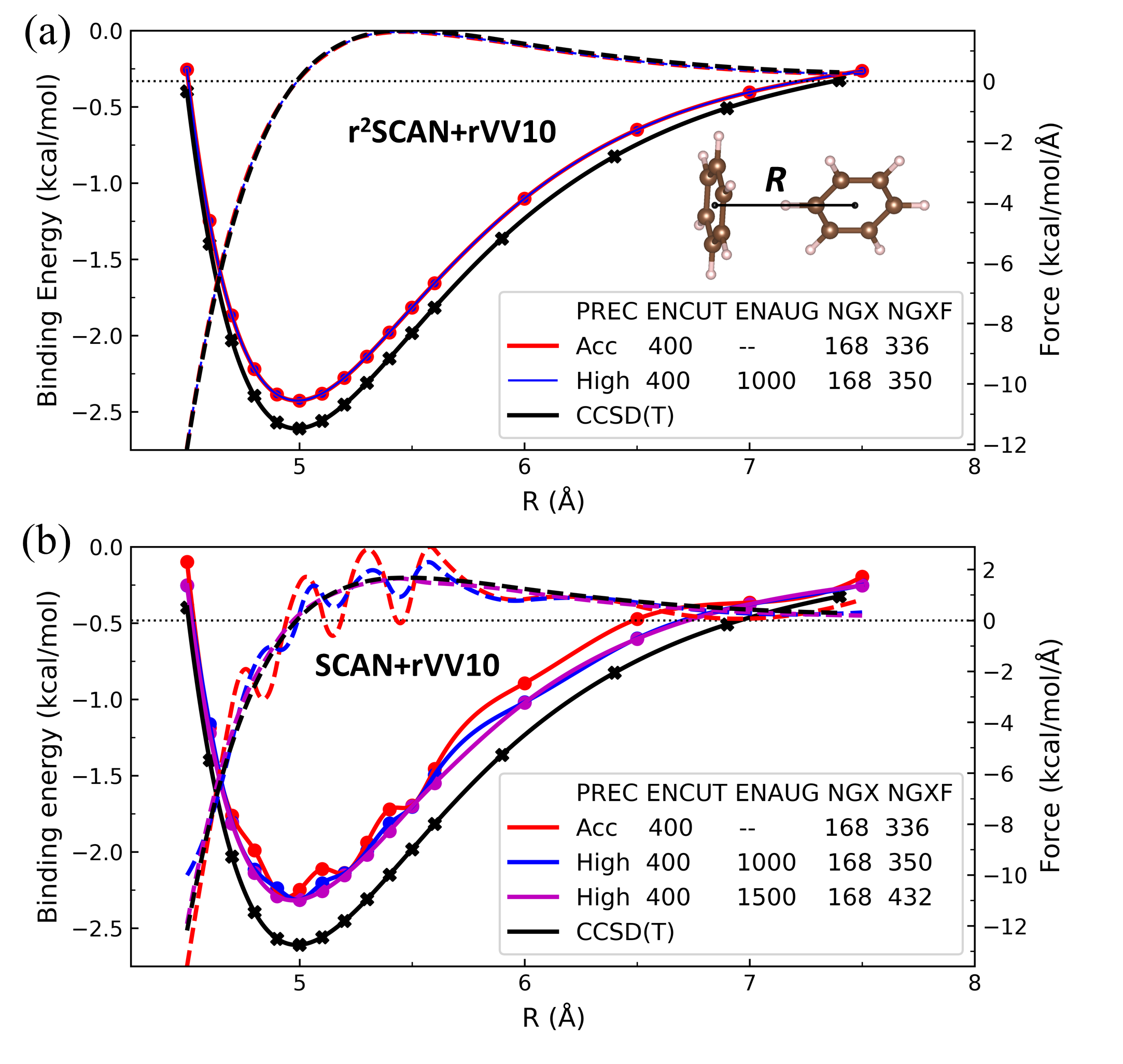}
\caption{\rtt{The binding energy curves (solid lines) and forces (dashed lines) for the T configuration of benzene dimer from (a) r$^2$SCAN+rVV10 and (b) SCAN+rVV10 compared to the CCSD(T) results \cite{Tben_CCSDT} as the reference, as a function of their separation R in \AA{}.
As in Ref. \citenum{gould2018}, forces are computed using a spline interpolation of the binding energy data.
}}
	\label{fig:Tben}
\end{figure}

\rtt{A previous study \cite{gould2018} demonstrated that SCAN+rVV10 produces significant oscillations in the interaction energy and force curves of the benzene dimer, which persist even with a large energy cutoff. In this work, we consider the T benzene dimer and confirm that removing such oscillations requires denser real space grids. Specifically for VASP users, we recommend using a high ENAUG ($\sim$1500) at certain ENCUT with PREC=High, instead of increasing ENCUT with PREC=Accurate. The r$^2$SCAN+rVV10 binding energy and force curves don't show oscillations even with low accuracy settings, as shown in Fig. \ref{fig:Tben}. However, r$^2$SCAN as a meta-GGA is still much more complicated than LDA and PBE, and thus may still need dense real space grids for certain applications \cite{r2SCAN_phonon}. To ensure stable convergence behavior, we recommend using dense real space grids (PREC=High; ENAUG=1500 or 2000) for SCAN-like metaGGA's and functionals based on them.}

\subsection{Layered materials}
\FloatBarrier

\input{Tables/table2}

\input{Tables/table3}

\begin{figure}
	\includegraphics[width=0.45\textwidth]{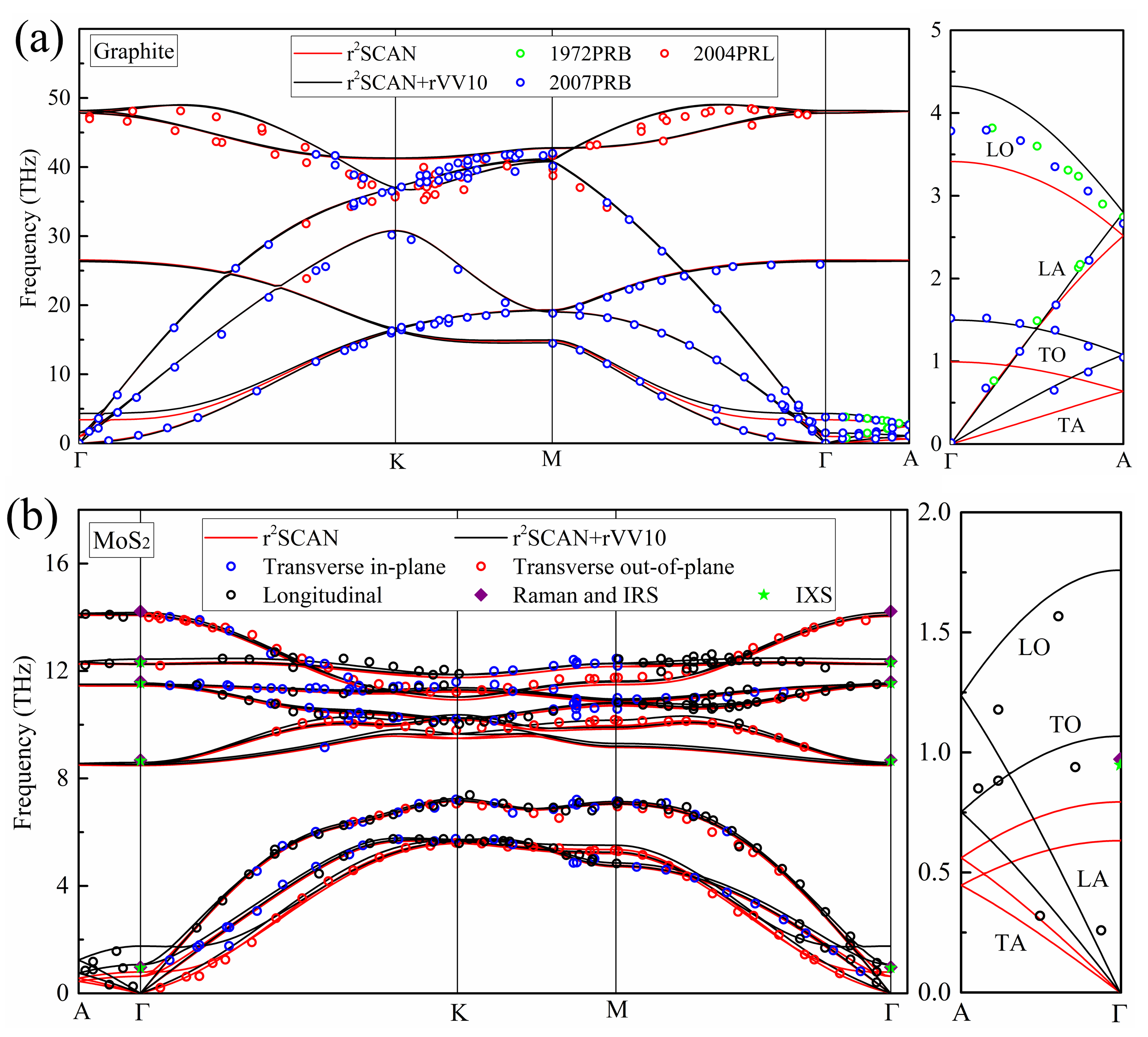}
\caption{Phonon dispersion in (a) graphite, (b) MoS$_2$, highlighting the improvements on the phonon branches along $\Gamma$-A (along the $c$-axis, or interlayer direction) from vdW corrections, compared with available experimental data from Refs. \citenum{G_phonon_PRL,G_phonon_PRB,G_phonon_1972PRB} for graphite and Refs. \citenum{MoS2_2019phonon, MoS2_R2015phonon, MoS2_IR1971phonon, MoS2_R1974phonon} for MoS$_2$.
Calculations were performed at the relaxed lattice parameters.
For an analogous figure using PBE and PBE-D4, see Fig. \ref{fig:pbe_phonon_disp} of the Appendix.
}
	\label{fig:phonon}
\end{figure}

\input{Tables/table4_flipped}

We also tested the predictions of SCAN+rVV10 and r$^2$SCAN+rVV10 for geometry and inter-layer binding properties for 28 layered materials (L28). As shown in Tables \ref{tab:table2} and \ref{tab:table3}, r$^2$SCAN+rVV10 more accurately predicts lattice constants than SCAN+rVV10 for this test set, and converges quicker with respect to plane-wave basis truncation and the size of the real space integration grid, the \texttt{ENCUT} and \texttt{ENAUG} settings in VASP respectively. This is clearly shown by the $c$ lattice constants in Tables \ref{tab:table2} and \ref{tab:table3}.
For values of the individual solids in the L28 set, refer to Tables \ref{tab:table_A2} and \ref{tab:table_A3} of the Appendix.

SCAN+rVV10 and \rrscan{}+rVV10 predict much longer $c$ lattice constants for PtSe$_2$, WSe$_2$, MoTe$_2$, NbS$_2$, NbSe$_2$, and NbTe$_2$ than those found experimentally \cite{L28_jcp}. We expect this may be due to the complicated electronic ground states of these materials, featuring charge density wave or superconductive phases \cite{NbS2_sup, NbSe2_CDW,NbTe2_CDW}, which were not considered in the present calculations. The effect of vdW functional corrections on these properties warrants further examination, but is beyond the scope of the current work.

To assess inter-layer binding energies for the L28 set in Table \ref{tab:table2}, we must use RPA reference values\cite{L28_jcp}, as those from more sophisticated methods [like the CCSD(T) references \cite{S22, S22A, S22B} for S22] are unavailable.
Select exceptions will be discussed further.
While the RPA includes long-range vdW interactions \cite{dobson1999}, it lacks an accurate description of short-range correlation \cite{kurth1999} and tends to underestimate $C_6$ vdW coefficients \cite{gould2012}.
RPA may tend to underbind layered materials.

Table \ref{tab:table4} presents inter-layer binding energies for a few solids where high-level QMC \cite{G_QMC,G2L_QMC,hBN2L_QMC,TiS2_QMC,BlackP_QMC} and silver-standard RPA values are available.
\rtt{No gold-standard correlated wavefunction calculations [such as CCSD(T)] for these solids have been undertaken at the time of writing.}
The QMC and experimental benchmarks show that RPA underbinds bulk graphite, MoS$_2$, and TiS$_2$ by 5-10 meV/\AA$^2$.
SCAN+rVV10 and \rrscan{}+rVV10 are slightly more accurate than RPA for these three bulk materials, but overestimate the bilayer binding energies of graphite and MoS$_2$.
SCAN+rVV10 often predicts larger binding energies than the RPA, and \rrscan{}+rVV10 often predicts larger binding energies than SCAN+rVV10.

With these findings, we may tentatively say that r$^2$SCAN+rVV10 is more accurate than RPA and SCAN+rVV10 for layered materials, though further benchmark studies with expanded comparison to high accuracy QMC calculations would be beneficial.

Alongside accurate static structural properties, dynamical lattice properties are also essential for materials design applications. We have recently shown that while SCAN gives accurate static structural properties, its accuracy for dynamical properties is limited by its numerical sensitivity, while \rrscan maintains good performance for both static and dynamical properties\cite{r2SCAN_phonon}. With this in mind, phonon dispersion in graphite and MoS$_2$ are presented in Fig. \ref{fig:phonon}. For both systems, our r$^2$SCAN+rVV10 results are in excellent agreement with the experimental data, especially for the lowest longitudinal acoustic (LA), longitudinal optical (LO), transverse acoustic (TA), and transverse optical (TO) phonon branches along the $\Gamma-A$ (interlayer or $c$-axis) direction. The calculated strengths of these branches are dominated by the inter-layer binding forces, and are thus sensitive to vdW corrections. Without the rVV10 correction, the uncorrected r$^2$SCAN severely underestimates these phonon branches.

\subsection{Complex materials: PPTA}

\FloatBarrier
\input{Tables/table5}

\begin{figure}
	\includegraphics[width=0.45\textwidth]{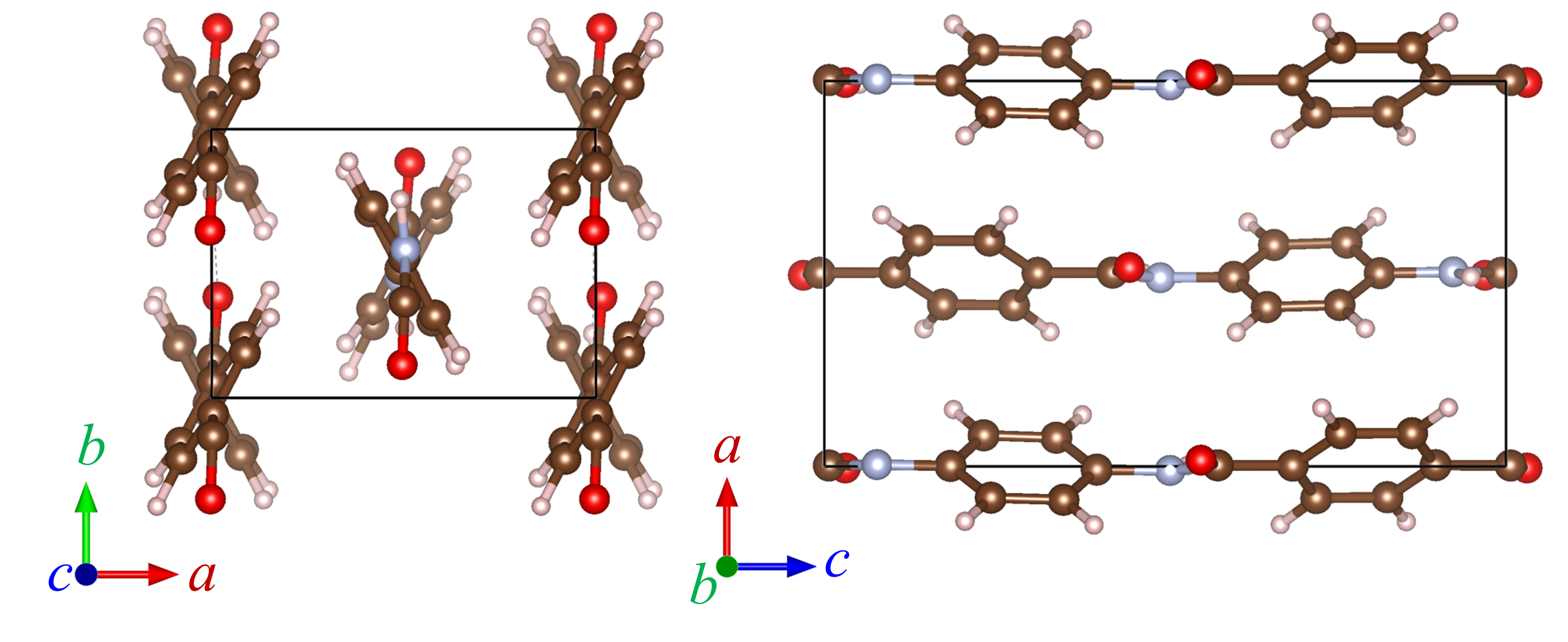}
\caption{\rtt{PPTA crystal structure in view of the ab plane and ac plane. Carbon, nitrogen, oxygen, and hydrogen atoms are rendered in brown, gray, red, and white, respectively.
}}
	\label{fig:PPTA}
\end{figure}

Last, we present calculations for poly(\textit{p}-phenylene terephthalamide) (PPTA), a layered material that is challenging for standard DFAs. PPTA, \rtt{as shown in Fig. \ref{fig:PPTA},} is primarily vdW-bonded along its $a$-axis, hydrogen-bonded along its $b$-axis, and covalently bonded along its $c$-axis \cite{Yu2020} -- a robust test for general-purpose DFAs.
Table \ref{tab:table5} presents the equilibrium structure of PPTA determined by SCAN, \rrscan, and their rVV10 variants.
The SCAN+rVV10 lattice parameters computed in Ref. \citenum{Yu2020} (and included in Table \ref{tab:table5}) used an older version of VASP where the rVV10 stress tensor elements were not correctly computed \cite{vasp_stress_bug}.
The calculations performed here use a corrected version of VASP and different  computational parameters than those of Ref. \citenum{Yu2020}.
We used a $6\times 9 \times 4$ $\bm{k}$-point grid (corresponding to $\bm{k}$-point spacing of 0.15 \AA{}$^{-1}$) and a plane-wave cutoff of 900 eV, whereas Ref. \citenum{Yu2020} used a $6\times 6 \times 6$ $\bm{k}$-point grid and plane-wave cutoff of 520 eV.
The number of grid points along $\bm{c}$ is well-converged at 4 points.

The effects of incorrect stress tensor elements are pronounced: the minima in the energy curves as a function of strained lattice parameters in Fig. 2 of Ref. \citenum{Yu2020} do not coincide with the values in their Table 1.
As their relaxed values of $b$ and $c$ for SCAN and SCAN+rVV10 are similar to ours, we refit their energy data as a function of $a$ at fixed $b=5.10$ \AA{} and $c=12.96$ \AA{} for SCAN, and $b=5.08$ \AA{} and $c=12.95$ \AA{} for SCAN+rVV10.
We find $a=7.92$ \AA{} for SCAN, and $a=7.42$ \AA{} for SCAN+rVV10, more comparable to our values in Table \ref{tab:table5}.

Although the $a$ axis is the vdW-bonded axis in PPTA, the uncorrected SCAN provides the most correct description of inter-layer binding in PPTA.
SCAN+rVV10 and \rrscan{}+rVV10 severely overbind along the $a$ axis, and do not provide substantive corrections to the parent meta-GGA along the $b$ and $c$ axes.

\FloatBarrier

\section{\label{sec:conclusion}Conclusions}

We have optimized the r$^2$SCAN+rVV10 vdW density functional and tested its performance against both molecular (S22) and layered material databases. The global $b$ parameter is adjusted to 11.95 by fitting to the Ar dimer binding energy curve. This is somewhat smaller than the VV10 $b=12.3$ parameter in Ref. \citenum{r2SCAND4}, and considerably smaller than the 15.7 used in SCAN+rVV10, suggesting that \rrscan requires more vdW correction than SCAN. With $b = 11.95$, r$^2$SCAN+rVV10 is more accurate than SCAN+rVV10 for the S22 binding energy database, and is competitive with the original rVV10 functional.

For the L28 layered material data set, \rrscan{}+rVV10 also outperforms SCAN+rVV10 in accuracy and efficiency for lattice constants predictions. For inter-layer binding energies, r$^2$SCAN+rVV10 shows stronger binding than SCAN+rVV10, which suggests over-binding when compared with RPA and available QMC benchmarks. In extended systems like layered bulk materials and bilayers, important many-atom/screening effects may be present in QMC that are missing in \rrscan{}+rVV10. However, \rrscan{}+rVV10 accurately accounts for phonon dispersion in layered bulk materials, improving substantially over \rrscan.
The study of PPTA demonstrates that care must be taken when using vdW-corrected DFAs.
The uncorrected parent DFA may sufficiently describe intermediate vdW interactions, leading to overbinding when the rVV10 correction is included.
We also highlight that r$^2$SCAN+rVV10 inherits the good numerical stability of r$^2$SCAN, and recommend r$^2$SCAN+rVV10 as a versatile vdW XC functional.

\begin{acknowledgments}

J.N. and M.K. acknowledge the support of the U.S. Department of Energy (DOE), Office of Science (OS), Basic Energy Sciences (BES), Grant No. DE-SC0014208.
J.S. acknowledges the support of the U.S. National Science Foundation (NSF) under Grant No. DMR-2042618.
J.W.F. acknowledges support from DOE grant DE-SC0019350.
A.D.K. acknowledges the support of the U.S. DOE, Office of Science, BES, through Grant No. DE-SC0012575 to the Energy Frontier Research Center: Center for Complex Materials from First Principles, and also support from Temple University.
J.P.P. acknowledges the support of the US NSF under Grant No. DMR-1939528.
We thank J. Yu for discussions on PPTA.

\end{acknowledgments}

\bibliographystyle{apsrev4-1}
\bibliography{literature}

\clearpage
\onecolumngrid
\appendix*
\section{Additional data sets and figures}

\input{Tables/table1}

\input{Tables/S22_supp_tab}

\input{Tables/L28_supp_tab}

\input{Tables/L28_grid_conv}

\begin{figure}[!h]
    \centering
    \includegraphics[width=0.8\columnwidth]{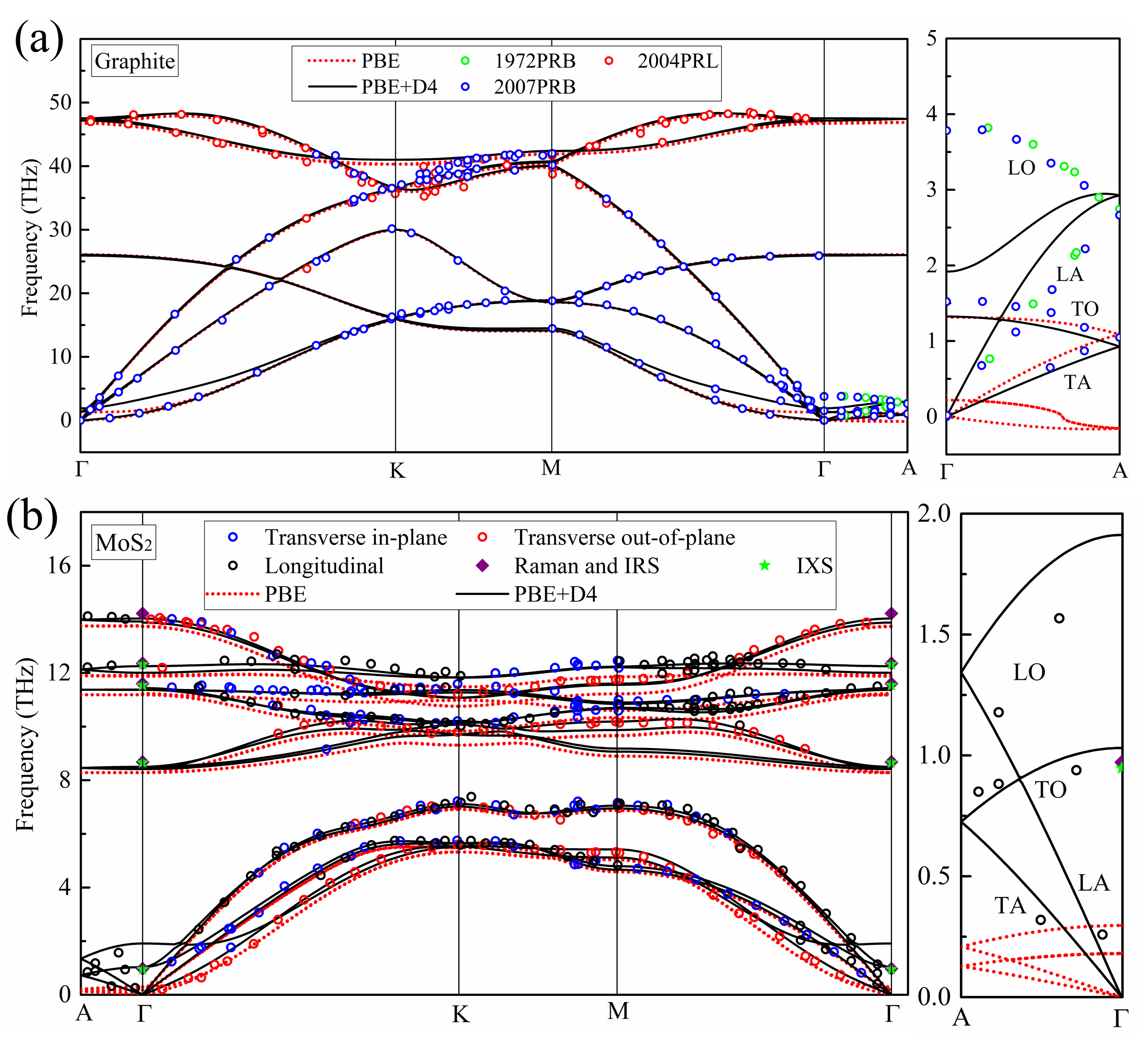}
    \caption{Phonon dispersion in graphite and MoS$_2$ analogous to Fig. \ref{fig:phonon}, but using PBE \cite{PBE} and PBE-D4 \cite{D4} instead of \rrscan.
    Just as for \rrscan, adding a dispersion correction to PBE produces a more realistic phonon dispersion, especially along the inter-layer direction A-$\Gamma$.
    In MoS$_2$, PBE-D4 is in good accord with available experimental phonon dispersion data, however \rrscan{}+rVV10 provides a more realistic description of phonons in graphite along the inter-layer direction.
    }
    \label{fig:pbe_phonon_disp}
\end{figure}

\begin{figure}
    \centering
    \includegraphics[width=0.8\columnwidth]{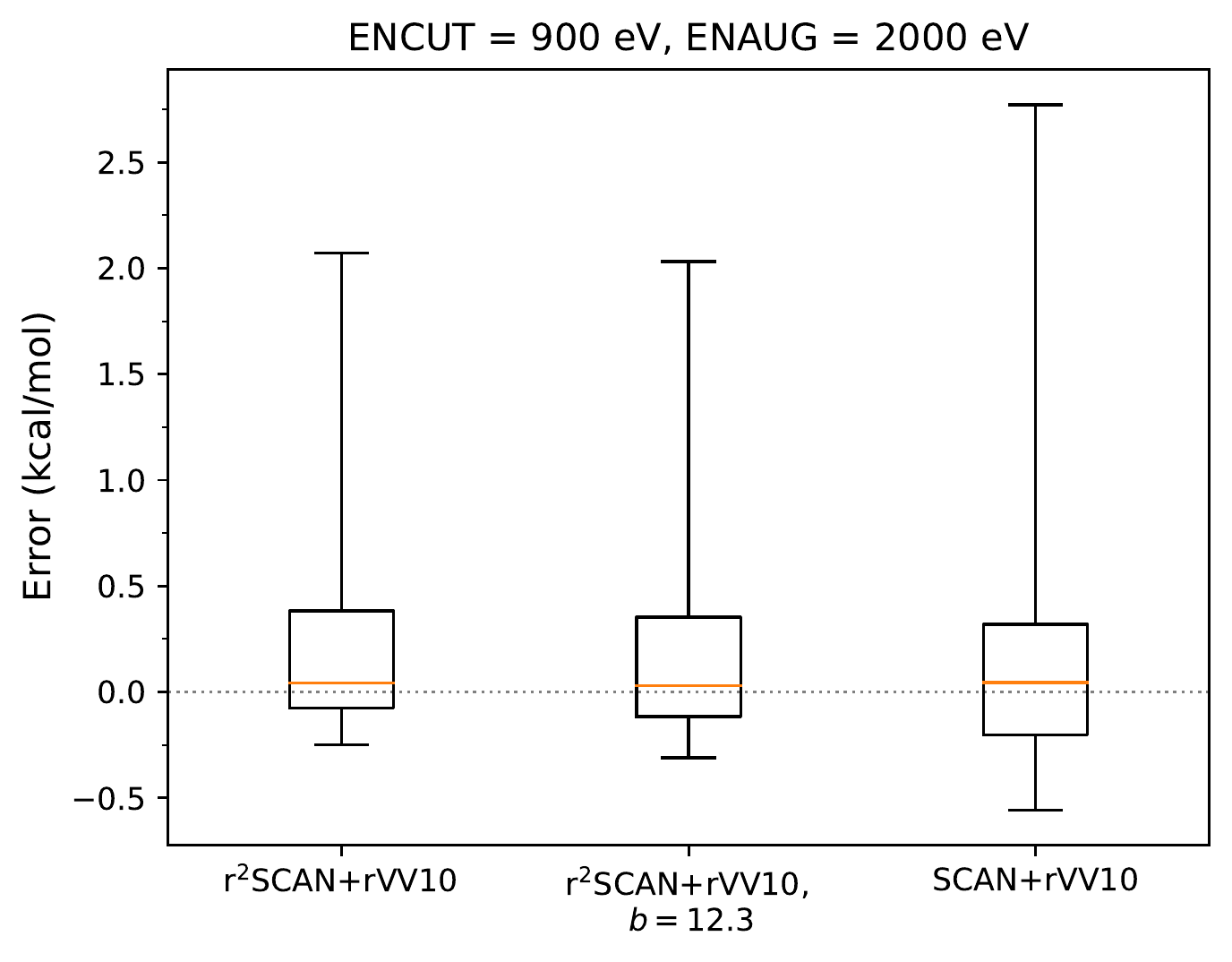}
    \caption{Box and whisker plot of the S22 errors (kcal/mol) for \rrscan{}+rVV10 with the presently-fitted value $b=11.95$, with the VV10-fitted $b=12.3$, and SCAN+rVV10.
    See Tables \ref{tab:table1} and \ref{tab:table_A1} for tabulated errors.} 
    \label{fig:S22_BNW}
\end{figure}

\begin{figure}
    \centering
    \includegraphics[width=0.8\columnwidth]{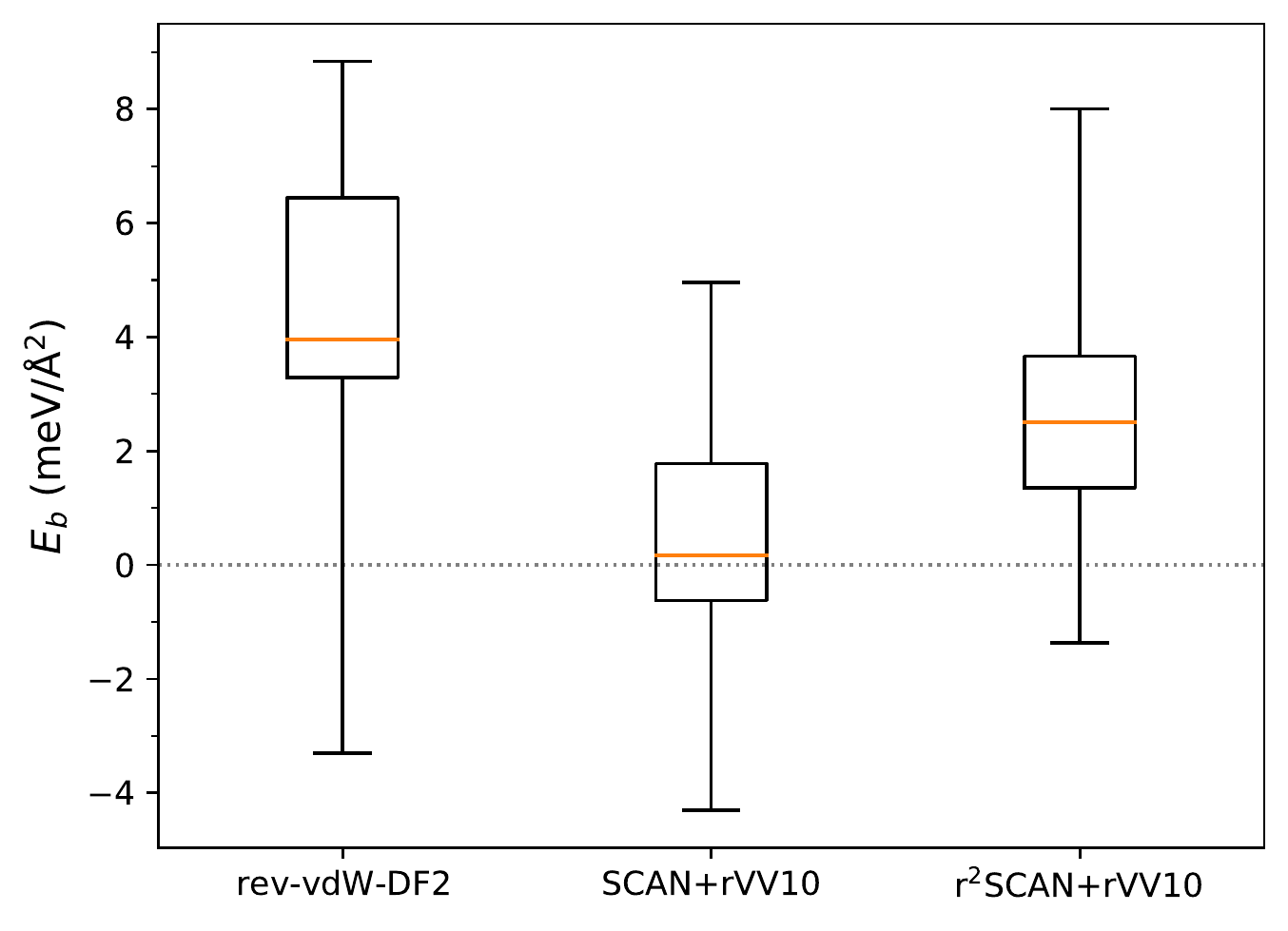}
    \caption{Box and whisker plot of the L28 set binding energy $E_b$ errors (meV/\AA{}$^2$) for rev-vdW-DF2, \rrscan{}+rVV10 with the presently-fitted value $b=11.95$, and SCAN+rVV10.
    See Tables \ref{tab:table2} and \ref{tab:table_A2} for tabulated errors.} 
    \label{fig:L28_Eb_BNW}
\end{figure}

\begin{figure}
    \centering
    \includegraphics[width=0.8\columnwidth]{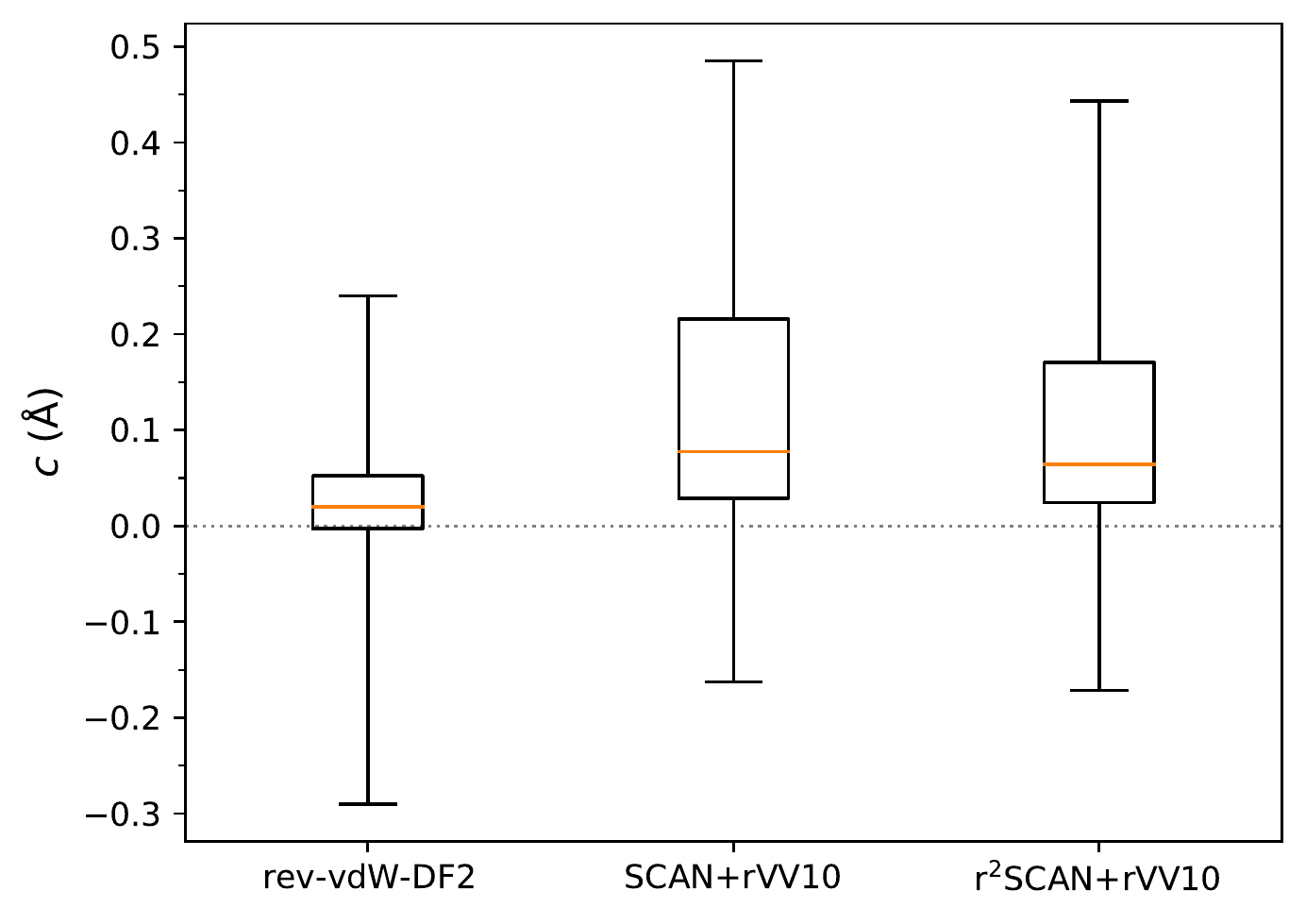}
    \caption{Box and whisker plot of the L28 set out-of-plane lattice constant $c$ errors (\AA{}) for rev-vdW-DF2, \rrscan{}+rVV10 with the presently-fitted value $b=11.95$, and SCAN+rVV10.
    See Tables \ref{tab:table2} and \ref{tab:table_A2} for tabulated errors.} 
    \label{fig:L28_c_BNW}
\end{figure}

\begin{figure}
    \centering
    \includegraphics[width=0.8\columnwidth]{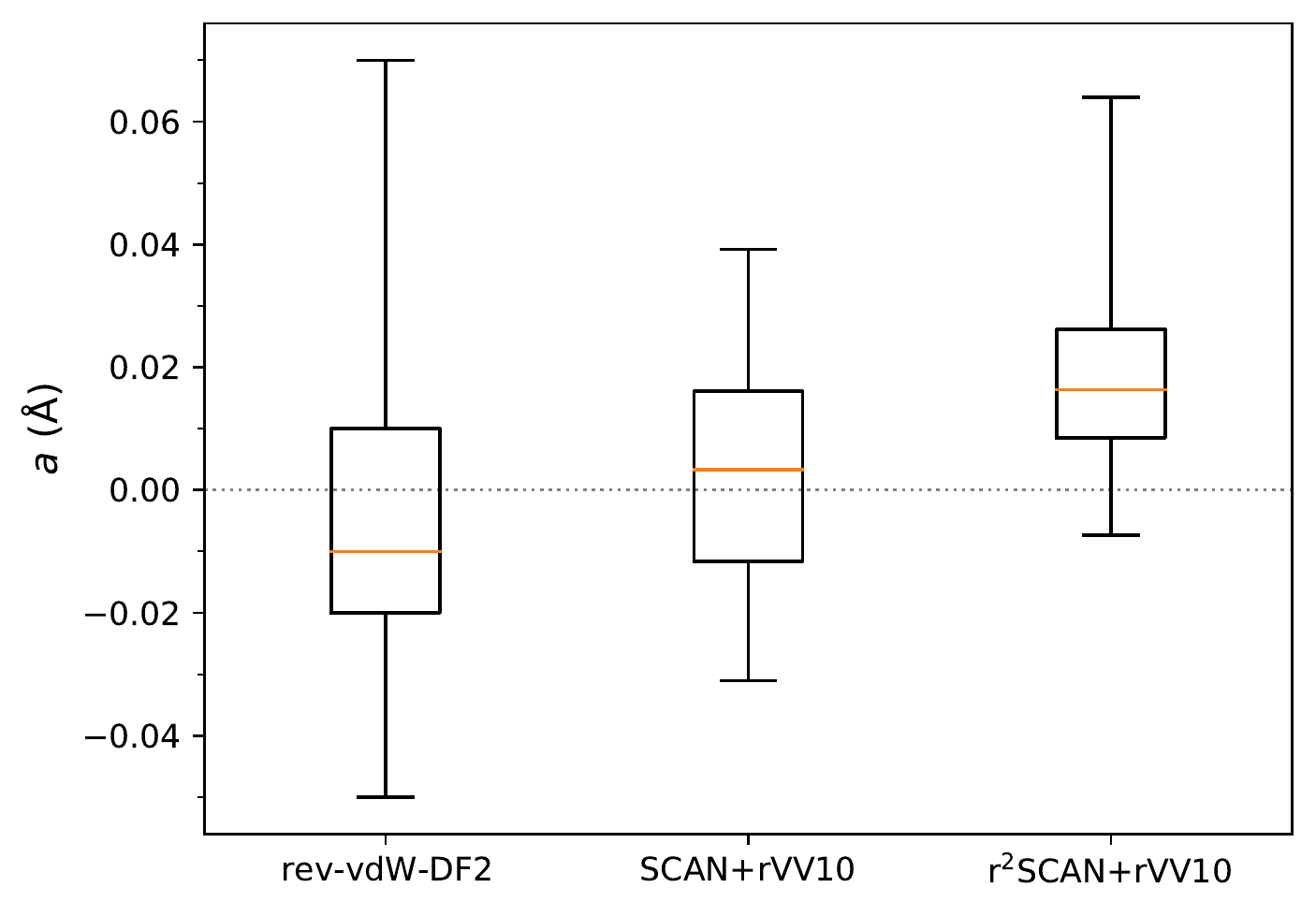}
    \caption{Box and whisker plot of the L28 set in-plane lattice constant $a$ errors (\AA{}) for rev-vdW-DF2, \rrscan{}+rVV10 with the presently-fitted value $b=11.95$, and SCAN+rVV10.
    See Tables \ref{tab:table2} and \ref{tab:table_A2} for tabulated errors.} 
    \label{fig:L28_a_BNW}
\end{figure}

\end{document}

%% file: Tables/table1_simplified.tex
\begin{ruledtabular}
  \begin{table}
    \caption{
    Mean errors (ME, kcal/mol) and mean absolute errors (MAE, kcal/mol) in the unsigned interaction energies of the S22 data set, taken with respect to CCSD(T) results \cite{S22B}.
    Different (\texttt{ENCUT}, \texttt{ENAUG}) settings (described in Section \ref{sec:results_mols}) are tested for r$^2$SCAN+rVV10 and SCAN+rVV10; both values are in eV.
    Users who need less accuracy can use lower settings.
    \rtt{Table \ref{tab:table1_full} in the Appendix presents S22 data for another (\texttt{ENCUT}, \texttt{ENAUG}) setting intermediate to those shown here, as well as percentage errors.}
    Table \ref{tab:table_A1} in the Appendix presents interaction energies for each molecule in the S22 set, the CCSD(T) reference values, as well as values for other density functional approximations.
    \label{tab:table1}}
    \centering
    \begin{tabular}{lrr|rr|r}
      & \multicolumn{2}{c|}{SCAN + rVV10} & \multicolumn{3}{c}{r$^2$SCAN + rVV10} \\
      & & & \multicolumn{2}{c|}{$b=11.95$} & $b=12.3$ \\
       & (600,600) & (900,2k) & (600,600) & (900,2k) & (900,2k) \\ \hline
      \multicolumn{6}{c}{\textit{7 hydrogen-bonded complexes}} \\
     MAE	&	0.99	&	0.89	&	0.54	&	0.62	&	0.58	\\
     ME 	&	0.99	&	0.89	&	0.54	&	0.62	&	0.58	\\ \hline

      \multicolumn{6}{c}{\textit{8 dispersion-bound complexes}} \\
     MAE	&	0.42	&	0.22	&	0.14	&	0.18	&	0.18	\\
     ME 	&	-0.11	&	-0.16	&	0.09	&	0.08	&	0.00	\\ \hline

      \multicolumn{6}{c}{\textit{7 mixed complexes}}\\
     MAE	&	0.36	&	0.20	&	0.23	&	0.18	&	0.20	\\
     ME 	&	-0.02	&	-0.06	&	-0.01	&	-0.02	&	-0.06	\\ \hline

     \multicolumn{6}{c}{\textit{Total}}\\
     MAE	&	0.58	&	0.43	&	0.30	&	0.32	&	0.31	\\
     ME 	&	0.27	&	0.21	&	0.20	&	0.22	&	0.17	\\

    \end{tabular}
  \end{table}
\end{ruledtabular}

%% file: Tables/table2.tex
\begin{ruledtabular}
  \begin{table}
  \caption{ Unsigned layer-layer binding energy $E_b$ in meV/\AA{}$^2$, lattice constants $c$ and $a$ in \AA{}, for 28 layered materials (L28) from SCAN+rVV10 and $r^2$SCAN+rVV10.
    Mean \rtt{deviations (MDs) and mean absolute deviations (MADs)} are taken with respect to the RPA \cite{Eb_RPA} (an uncertain reference; see Table \ref{tab:table4}) for $E_b$, and experiment \cite{L28_jcp} for the lattice constants $c$ and $a$.
    Table \ref{tab:table_A2} in the Appendix presents values for each material in the set, the reference values, and values for other density functional approximations.
    \label{tab:table2}}
    \centering
    \begin{tabular}{lrrr|rrr}
      & \multicolumn{3}{c|}{SCAN+rVV10} & \multicolumn{3}{c}{r$^2$SCAN+rVV10} \\
      &	$E_b$	&	$c$	&	$a$	&	$E_b$	&	$c$	&	$a$ \\ \hline
      MAD	&	1.527	&	0.167	&	0.019	&	2.786	&	0.139	&	0.018 \\
      MD	&	0.476	&	0.132	&	-0.007	&	2.670	&	0.108	&	0.009 \\
    \end{tabular}
  \end{table}
\end{ruledtabular}

%% file: Tables/table3.tex
\begin{ruledtabular}
  \begin{table}
    \caption{
    Convergence of lattice constants $c$ and $a$ in \AA{} for 28 layered materials from SCAN+rVV10 and r$^2$SCAN+rVV10.
    Different (ENCUT, ENAUG) settings are presented; both values are in eV.
    Mean \rtt{deviations (MDs) and mean absolute deviations (MADs)} are taken with respect to the largest ENCUT, 800 eV, and ENAUG, 2000 eV, setting.
    r$^2$SCAN+rVV10 approaches its converged values more rapidly than does SCAN+rVV10.
    For the lattice parameters of each solid in the set, refer to Table \ref{tab:table_A3} in the Appendix.
    \label{tab:table3}}
    \centering
    \begin{tabular}{lrr|rr|rr|rr}
      & \multicolumn{4}{c|}{SCAN+rVV10} & \multicolumn{4}{c}{r$^2$SCAN+rVV10} \\
      & \multicolumn{2}{c|}{(500,600)} &	\multicolumn{2}{c|}{(500,1k)} & \multicolumn{2}{c|}{(500,600)} &	\multicolumn{2}{c}{(500,1k)} \\
      & $c$ & $a$ & $c$ & $a$ & $c$ & $a$ & $c$ & $a$ \\ \hline
      MAD	&	0.024	&	0.002	&	0.010	&	0.001	&	0.009	&	0.000	&	0.008	&	0.000	\\
      MD	&	0.006	&	-0.001	&	-0.008	&	0.000	&	-0.007	&	0.000	&	-0.006	&	0.000 \\
    \end{tabular}
  \end{table}
\end{ruledtabular}

%% file: Tables/table4_flipped.tex
\begin{table*}
\caption{\label{tab:table4} Unsigned layer-layer binding energy $E_b$ in meV/Å$^2$ of graphite, hexagonal boron nitride (h-BN), MoS$_2$, TiS$_2$ and black phosphorous, calculated from SCAN+rVV10 and r$^2$SCAN+rVV10 compared with available data from experiments and other computational methods.
A high-level, finite cluster CCSD(T) calculation \cite{BlackP_LdrCCD} for bulk black phosphorous found its exfoliation energy to be 25.81 meV/\AA$^2$.
We also report values for the rev-vdW-DF2 \cite{hamada2014} vdW-corrected GGA when available.
}
\begin{ruledtabular}
	\begin{tabular}{llllllll}
		& & Expt. & QMC & RPA & rev-vdW-DF2 & SCAN + rVV10 & r$^2$SCAN + rVV10 \\ \hline
	\multirow{2}{*}{Graphite} & Bulk & $23.28\pm 1.91$ \cite{G_expt}
		& $22.91\pm 1.91$\cite{G_QMC} & 18.32\cite{lebegue2010} & 23.45 \cite{SCAN+rVV10} & 20.01 & 22.85 \\
	& Bilayer & & $13.51 \pm 0.69$\cite{G2L_QMC} & &  & 17.64 & 20.13 \\ \hline
	\multirow{2}{*}{h-BN} & Bulk & & & 14.49\cite{Eb_RPA} & 21.15 \cite{SCAN+rVV10} & 20.62 & 22.55 \\
	& Bilayer & & $15.02\pm 0.46$ \cite{hBN2L_QMC} & & & 17.03 & 19.70 \\ \hline
	MoS$_2$ & Bulk & $34.33 \pm 8.11 $\cite{MoS2_expt} & & 20.53\cite{Eb_RPA} & 23.53 \cite{SCAN+rVV10} &
		20.15 & 23.07 \\ \hline
	\multirow{2}{*}{TiS$_2$} & Bulk & & $27.2\pm 0.8$ \cite{TiS2_QMC} & 18.88\cite{Eb_RPA} &
		24.8 \cite{TiS2_QMC} & 18.97 & 21.49 \\
	& Bilayer & & $24.9\pm 1.6$ \cite{TiS2_QMC} & & 23.8 \cite{TiS2_QMC} & 17.71	& 20.06 \\ \hline
	\multirow{2}{*}{Black P} & Bulk & & $22.4\pm 1.6$ \cite{BlackP_QMC} & & & 22.59 & 25.46 \\
	& Bilayer & & $16.6\pm 2.2$ \cite{BlackP_QMC} & & & 21.28 & 23.97 \\
	\end{tabular}
\end{ruledtabular}
\end{table*}


%% file: Tables/Table5.tex
\begin{ruledtabular}
  \begin{table}
    \caption{Equilibrium lattice constants of PPTA, found by stress minimization within the VASP code. Computed and experimental values from Ref. \citenum{Yu2020} are included for comparison.
    Unlike other layered materials, the inter-layer or vdW direction in PPTA is the $a$ axis.
    \label{tab:table5}}
    \centering
    \begin{tabular}{p{1.5cm}lllll}
	& Methods		&	$a$(Å)	&	$b$(Å)	&	$c$(Å)	&	$\alpha(\deg)$	\\ \hline
 \multirow{3}{1.5cm}{Ref. \citenum{Yu2020}}	&	Expt.	&	7.87	&	5.18	&	12.9	&	90	\\
	&	SCAN	&	7.75	&	5.10	&	12.96	&	90.2	\\
	&	SCAN+rVV10	&	7.21	&	5.08	&	12.95	&	90	\\ \hline
\multirow{4}{1.5cm}{This Work}
&	SCAN	&	7.86	&	5.09	&	12.96	&	90.3	\\
&	SCAN+rVV10	&	7.43	&	5.10	&	12.96	&	90.1	\\
&	\rrscan	&	7.99	&	5.14	&	12.96	&	90.2	\\
&	\rrscan{}+rVV10	&	7.35	&	5.15	&	12.99	&	90.1	\\

    \end{tabular}
  \end{table}
\end{ruledtabular}

%% file: Tables/table1.tex
\begin{ruledtabular}
  \begin{table*}[!h]
    \caption{
    Mean errors (ME, kcal/mol), mean absolute errors (MAE, kcal/mol), mean percentage errors (MPE), and mean absolute percentage errors (MAPE) in the unsigned interaction energies of the S22 data set, taken with respect to CCSD(T) results \cite{S22B}.
    Different (\texttt{ENCUT}, \texttt{ENAUG}) settings (described in Section \ref{sec:results_mols}) are tested for r$^2$SCAN+rVV10 and SCAN+rVV10; both values are in eV.
    Users who need less accuracy can use lower settings.
    \rtt{For a concise presentation of this data, refer to Table \ref{tab:table1}.}
    Table \ref{tab:table_A1} in the Appendix presents interaction energies for each molecule in the S22 set, the CCSD(T) reference values, as well as values for other density functional approximations.
    \label{tab:table1_full}}
    \centering
    \begin{tabular}{lrrr|rrr|r}
      & \multicolumn{3}{c|}{SCAN+rVV10} & \multicolumn{4}{c}{r$^2$SCAN+rVV10} \\
      & & & & \multicolumn{3}{c|}{$b=11.95$} & $b=12.3$ \\
      & (600, 600) & (600, 1000) & (900, 2000)	&	(600, 600) & (600, 1000) &	(900, 2000)	&	(900, 2000)  \\ \hline
      \multicolumn{4}{l|}{\textit{7 hydrogen-bonded complexes}} & & & & \\
     MAE	&	0.99	&	0.95	&	0.89	&	0.54	&	0.56	&	0.62	&	0.58	\\
     ME 	&	0.99	&	0.95	&	0.89	&	0.54	&	0.56	&	0.62	&	0.58	\\
     MAPE	&	7.78	&	7.54	&	6.64	&	3.90	&	3.99	&	4.38	&	4.11	\\
     MPE	&	7.78	&	7.54	&	6.64	&	3.90	&	3.99	&	4.38	&	4.11	\\ \hline

      \multicolumn{4}{l|}{\textit{8 dispersion-bound complexes}} & & & &  \\
     MAE	&	0.42	&	0.17	&	0.22	&	0.14	&	0.13	&	0.18	&	0.18	\\
     ME 	&	-0.11	&	-0.11	&	-0.16	&	0.09	&	0.08	&	0.08	&	0.00	\\
     MAPE	&	14.00	&	6.17	&	6.65	&	3.01	&	2.56	&	3.65	&	4.66	\\
     MPE	   &	-3.93	&	-2.11	&	-6.02	&	0.14	&	-0.10	&	-1.06	&	-3.00	\\  \hline

      \multicolumn{4}{l|}{\textit{7 mixed complexes}} & & & &  \\
     MAE	&	0.36	&	0.31	&	0.20	&	0.23	&	0.23	&	0.18	&	0.20	\\
     ME 	&	-0.02	&	-0.04	&	-0.06	&	-0.01	&	-0.03	&	-0.02	&	-0.06	\\
     MAPE	&	10.32	&	8.37	&	5.69	&	6.44	&	6.32	&	5.32	&	5.54	\\
     MPE	&	1.88	&	1.32	&	-0.69	&	1.75	&	1.24	&	0.64	&	-0.33	\\ \hline

      \multicolumn{4}{l|}{\textit{Total}} & & & & \\
     MAE	&	0.58	&	0.46	&	0.43	&	0.30	&	0.30	&	0.32	&	0.31	\\
     ME 	&	0.27	&	0.25	&	0.21	&	0.20	&	0.20	&	0.22	&	0.17	\\
     MAPE	&	10.85	&	7.31	&	6.34	&	4.39	&	4.21	&	4.42	&	4.76	\\
     MPE	&	1.65	&	2.05	&	-0.29	&	1.85	&	1.63	&	1.21	&	0.11	\\

    \end{tabular}
  \end{table*}
\end{ruledtabular}

%% file: Tables/S22_supp_tab.tex

\begin{ruledtabular}
\begin{table*}
\scriptsize
\caption{\label{tab:table_A1} Positive interaction \rtt{energy errors (approximate minus the CCSD(T) reference)}, in kcal/mol, for the molecular dimers in the S22 data set from PBE, rVV10, vdW-DF2 (numerical results from Ref. \citenum{vv10}), SCAN results from Ref.
\citenum{SCAN+rVV10}, SCAN+rVV10 and r$^2$SCAN+rVV10 with respect to the CCSD(T) results \cite{S22B}.
Different (ENCUT, ENAUG) settings are tested for r$^2$SCAN+rVV10 and SCAN+rVV10; both values are in eV.
\rtt{Absolute errors that are greater than twice the corresponding MAD are \textbf{bold-faced}.}
}
\begin{tabular}{lrrrrrrrrrrr}
 & & & & & & \multicolumn{3}{c}{r$^2$SCAN+rVV10} & \multicolumn{3}{c}{SCAN+rVV10} \\ 
 & CCSD(T) & PBE & rVV10 & vdW-DF2 & SCAN & (600,600) & (600,1000) & (900,2000) & (600,600) & (600,1000) & (900,2000) \\ \hline 
7 hydrogen-bound complexes & & & & & & & & & & & \\ 
NH$_3$ dimer (C2h) & 3.13 & -0.32 & 0.28 & -0.16 & -0.01 & 0.04 & 0.05 & 0.03 & 0.28 & 0.29 & 0.15 \\ 
H$_2$O dimer (Cs) & 4.99 & -0.05 & 0.52 & -0.21 & 0.44 & 0.34 & 0.35 & 0.39 & 0.55 & 0.56 & 0.56 \\ 
Formic acid dimer (C2h) & 18.75 & -0.51 & \textbf{1.22} & \textbf{-1.98} & \textbf{2.18} & \textbf{1.84} & \textbf{1.87} & \textbf{2.07} & \textbf{3.00} & \textbf{2.86} & \textbf{2.77} \\ 
Formamide dimer (C2h) & 16.06 & -1.28 & \textbf{0.66} & -1.63 & 0.48 & 0.35 & 0.39 & 0.60 & 0.93 & 0.92 & \textbf{0.99} \\ 
Uracil dimer (C2h) & 20.64 & -2.10 & 0.48 & \textbf{-1.95} & -0.15 & 0.27 & 0.32 & 0.36 & 0.24 & 0.64 & 0.62 \\ 
2-pyridone–2-aminopyridine (C1) & 16.93 & -1.56 & \textbf{1.13} & -1.56 & -0.08 & \textbf{0.66} & \textbf{0.68} & \textbf{0.68} & 0.58 & 0.88 & \textbf{0.88} \\ 
Adenine–thymine WC (C1) & 16.66 & -2.31 & \textbf{0.76} & \textbf{-1.92} & -0.67 & 0.30 & 0.26 & 0.21 & \textbf{1.37} & 0.51 & 0.25 \\ \hline 
MAE [REF CCSD(T)] & & 1.16 & 0.72 & 1.35 & 0.57 & 0.54 & 0.56 & 0.62 & 0.99 & 0.95 & 0.89 \\ 
ME [REF CCSD(T)] & & -1.16 & 0.72 & -1.35 & 0.31 & 0.54 & 0.56 & 0.62 & 0.99 & 0.95 & 0.89 \\ 
STD DEV [REF CCSD(T)] & & 0.82 & 0.32 & 0.75 & 0.84 & 0.56 & 0.56 & 0.63 & 0.90 & 0.80 & 0.82 \\ 
MAE [REF (900,2000)] & & & & & & 0.09 & 0.07 & 0.00 & 0.16 & 0.05 & 0.00 \\ \hline 
8 dispersion-bound complexes & & & & & & & & & & & \\ 
CH$_4$ dimer (D3d) & 0.53 & -0.43 & -0.04 & 0.15 & -0.18 & -0.02 & -0.01 & -0.00 & -0.14 & -0.03 & -0.01 \\ 
C$_2$H$_4$ dimer (D2d) & 1.47 & -1.14 & -0.06 & -0.15 & -0.45 & -0.06 & -0.06 & -0.11 & -0.34 & -0.16 & -0.09 \\ 
Benzene–CH$_4$ (C3) & 1.45 & -1.40 & -0.01 & -0.16 & -0.58 & -0.00 & -0.03 & -0.09 & 0.23 & 0.21 & -0.22 \\ 
Benzene dimer (C2h) & 2.65 & -4.50 & 0.07 & -0.50 & -1.58 & 0.12 & 0.05 & 0.02 & -0.10 & 0.05 & -0.24 \\ 
Pyrazine dimer (Cs) & 4.25 & -4.93 & -0.22 & -0.96 & -1.60 & -0.11 & -0.10 & -0.02 & 1.03 & -0.30 & -0.22 \\ 
Uracil dimer (C2) & 9.80 & \textbf{-7.07} & -0.08 & -1.04 & -1.84 & 0.45 & 0.50 & 0.62 & \textbf{-1.32} & -0.26 & 0.25 \\ 
Indole–benzene (C1) & 4.52 & \textbf{-6.69} & 0.01 & -1.08 & \textbf{-2.40} & 0.05 & 0.00 & -0.18 & -0.22 & -0.29 & -0.56 \\ 
Adenine–thymine (C1) & 11.73 & \textbf{-10.31} & -0.31 & \textbf{-2.15} & \textbf{-3.08} & 0.31 & 0.31 & 0.36 & -0.01 & -0.07 & -0.16 \\ \hline 
MAE [REF CCSD(T)] & & 4.56 & 0.10 & 0.78 & 1.47 & 0.14 & 0.13 & 0.18 & 0.42 & 0.17 & 0.22 \\ 
ME [REF CCSD(T)] & & -4.56 & -0.08 & -0.74 & -1.47 & 0.09 & 0.08 & 0.08 & -0.11 & -0.11 & -0.16 \\ 
STD DEV [REF CCSD(T)] & & 3.22 & 0.12 & 0.69 & 0.94 & 0.18 & 0.20 & 0.26 & 0.61 & 0.17 & 0.21 \\ 
MAE [REF (900,2000)] & & & & & & 0.09 & 0.07 & 0.00 & 0.52 & 0.21 & 0.00 \\ \hline 
7 mixed complexes & & & & & & & & & & & \\ 
C$_2$H$_4$–C$_2$H$_2$ (C2v) & 1.50 & -0.32 & 0.17 & 0.03 & -0.16 & 0.07 & 0.06 & 0.05 & 0.24 & 0.06 & 0.06 \\ 
Benzene–H$_2$O (Cs) & 3.27 & -1.25 & 0.04 & -0.48 & 0.00 & 0.43 & 0.40 & 0.39 & 0.58 & 0.54 & 0.34 \\ 
Benzene–NH$_3$ (Cs) & 2.31 & -1.38 & -0.04 & -0.32 & -0.32 & 0.21 & 0.20 & 0.13 & 0.14 & 0.26 & 0.05 \\ 
Benzene–HCN (Cs) & 4.54 & -1.71 & -0.27 & -0.99 & -0.48 & 0.07 & 0.06 & -0.00 & -0.07 & 0.09 & 0.04 \\ 
Benzene dimer (C2v) & 2.72 & -2.59 & -0.17 & -0.66 & -1.24 & -0.17 & -0.19 & -0.24 & -0.30 & -0.26 & -0.33 \\ 
Indole–benzene (Cs) & 5.63 & -3.57 & -0.35 & -1.43 & -1.56 & -0.19 & -0.19 & -0.25 & -0.94 & -0.46 & -0.47 \\ \hline 
MAE [REF CCSD(T)] & & 1.80 & 0.17 & 0.65 & 0.63 & 0.19 & 0.18 & 0.18 & 0.38 & 0.28 & 0.22 \\ 
ME [REF CCSD(T)] & & -1.80 & -0.10 & -0.64 & -0.62 & 0.07 & 0.06 & 0.01 & -0.06 & 0.04 & -0.05 \\ 
STD DEV [REF CCSD(T)] & & 1.03 & 0.18 & 0.47 & 0.57 & 0.21 & 0.21 & 0.22 & 0.48 & 0.33 & 0.27 \\ 
MAE [REF (900,2000)] & & & & & & 0.05 & 0.03 & 0.00 & 0.11 & 0.09 & 0.00 \\ \hline 
Total & & & & & & & & & & & \\ 
MAE [REF CCSD(T)] & & 2.67 & 0.32 & 0.94 & 0.94 & 0.30 & 0.30 & 0.32 & 0.58 & 0.46 & 0.43 \\ 
ME [REF CCSD(T)] & & -2.67 & 0.17 & -0.92 & -0.66 & 0.20 & 0.20 & 0.22 & 0.27 & 0.25 & 0.21 \\ 
STD DEV [REF CCSD(T)] & & 2.55 & 0.43 & 0.71 & 1.09 & 0.44 & 0.45 & 0.49 & 0.84 & 0.70 & 0.68 \\ 
MAE [REF (900,2000)] & & & & & & 0.10 & 0.08 & 0.00 & 0.36 & 0.15 & 0.00 \\ 
\end{tabular}

\end{table*}

\end{ruledtabular}


%% file: Tables/L28_supp_tab.tex

\begin{ruledtabular}
\begin{table*}
\scriptsize
\caption{\label{tab:table_A2} Positive layer-layer binding energy $E_b$ in meV/\AA{}$^2$, lattice constants $c$ and $a$ in \AA{} for 28 layered materials \rtt{(L28 test set)} from SCAN+rVV10 and $r^2$SCAN+rVV10. The reference values are $E_b$ from RPA calculations \cite{Eb_RPA} and lattice constants $c$ and $a$ from experiment \cite{L28_jcp}.
$\Delta E_b$, $\Delta a$, and $\Delta c$ are the \rtt{deviations} in the interlayer binding energy, $a$ lattice parameter, and $c$ lattice parameter, respectively.
\rtt{The mean deviations (MDs), mean absolute deviations (MADs), and standard deviations (STD DEVs) are also presented.
Absolute errors that are greater than twice the corresponding MAD are \textbf{bold-faced}.}
}
\begin{tabular}{l|rrr|rrr|rrr|rrr|rrr}
  & RPA & \multicolumn{2}{c}{Expt.} & \multicolumn{3}{c}{rev-vdW-DF2} & \multicolumn{3}{c}{SCAN} & \multicolumn{3}{c}{SCAN+rVV10} & \multicolumn{3}{c}{r$^2$SCAN+rVV10} \\ 
 & $E_b$ & $c$ & $a$ & $\Delta E_b$ & $\Delta c$ & $\Delta a$ & $\Delta E_b$ & $\Delta c$ & $\Delta a$ & $\Delta E_b$ & $\Delta c$ & $\Delta a$ & $\Delta E_b$ & $\Delta c$ & $\Delta a$ \\ \hline 
h-BN & 14.49 & 6.54 & 2.51 & 6.66 & 0.00 & 0.00 & -7.20 & 0.30 & -0.01 & \textbf{4.96} & 0.00 & -0.01 & \textbf{8.00} & -0.04 & -0.01 \\ 
Graphite & 18.32 & 6.70 & 2.46 & 5.13 & -0.11 & 0.00 & -10.40 & 0.16 & -0.01 & 1.63 & -0.05 & -0.01 & 4.53 & -0.07 & -0.00 \\ 
HfS$_2$ & 16.13 & 5.84 & 3.63 & 3.77 & -0.01 & -0.02 & -10.94 & 0.20 & 0.00 & -0.09 & 0.04 & -0.02 & 2.09 & 0.04 & -0.01 \\ 
HfSe$_2$ & 17.09 & 6.16 & 3.75 & 3.33 & 0.02 & -0.01 & -11.66 & 0.24 & 0.00 & -0.82 & 0.06 & -0.01 & 1.30 & 0.05 & -0.00 \\ 
HfTe$_2$ & 18.68 & 6.65 & 3.96 & 4.48 & 0.04 & -0.03 & -11.68 & 0.28 & 0.01 & -0.50 & 0.13 & -0.01 & 1.37 & 0.13 & 0.02 \\ 
MoS$_2$ & 20.53 & 12.30 & 3.16 & 3.00 & 0.04 & 0.01 & -14.86 & 0.52 & 0.01 & -0.32 & 0.18 & 0.01 & 2.71 & 0.13 & 0.02 \\ 
MoSe$_2$ & 19.63 & 12.93 & 3.29 & 3.45 & 0.12 & 0.01 & -14.01 & 0.61 & 0.02 & 0.08 & 0.24 & 0.01 & 2.91 & 0.19 & 0.02 \\ 
MoTe$_2$ & 20.80 & 13.97 & 3.52 & 3.30 & 0.11 & 0.01 & -13.95 & 0.66 & 0.00 & -0.18 & 0.30 & -0.01 & 2.22 & 0.25 & 0.03 \\ 
NbS$_2$ & 17.58 & 17.91 & 3.33 & 7.58 & \textbf{0.24} & -0.01 & -10.65 & \textbf{0.93} & 0.01 & 2.94 & \textbf{0.46} & 0.00 & \textbf{5.65} & \textbf{0.40} & 0.01 \\ 
NbSe$_2$ & 19.57 & 12.55 & 3.44 & 7.82 & -0.06 & 0.01 & -11.93 & 0.50 & 0.03 & 0.45 & \textbf{0.49} & 0.02 & 3.04 & \textbf{0.44} & 0.02 \\ 
NbTe$_2$ & 23.03 & 6.61 & 3.68 & 4.14 & \textbf{0.20} & -0.01 & -14.37 & 0.57 & -0.02 & -1.24 & \textbf{0.33} & -0.03 & 1.03 & \textbf{0.27} & 0.02 \\ 
PbO & 20.25 & 5.00 & 3.96 & -3.30 & 0.05 & \textbf{0.07} & -8.43 & 0.10 & 0.03 & \textbf{3.08} & -0.07 & 0.03 & 1.40 & 0.01 & 0.03 \\ 
PdTe$_2$ & 40.17 & 5.11 & 4.02 & 3.44 & 0.05 & \textbf{0.05} & -14.98 & -0.07 & 0.03 & 2.25 & -0.08 & 0.02 & -0.25 & 0.06 & 0.04 \\ 
PtS$_2$ & 20.55 & 5.04 & 3.54 & 2.85 & -0.13 & \textbf{0.05} & -15.14 & 0.50 & -0.01 & -1.39 & 0.17 & -0.01 & 1.46 & 0.09 & 0.01 \\ 
PtSe$_2$ & 19.05 & 5.11 & 3.73 & 5.86 & -0.13 & \textbf{0.06} & -13.14 & 0.62 & -0.04 & 0.34 & 0.29 & -0.03 & 3.06 & 0.20 & 0.00 \\ 
TaS$_2$ & 17.68 & 5.90 & 3.36 & 8.29 & 0.00 & -0.01 & -10.30 & 0.24 & 0.00 & \textbf{3.74} & 0.06 & -0.01 & \textbf{6.43} & 0.05 & 0.01 \\ 
TaSe$_2$ & 19.44 & 6.27 & 3.48 & 6.37 & 0.02 & -0.01 & -12.12 & 0.25 & 0.00 & 2.69 & 0.06 & -0.01 & 5.24 & 0.04 & 0.01 \\ 
TiS$_2$ & 18.88 & 5.90 & 3.41 & 5.47 & \textbf{-0.25} & -0.02 & -11.98 & -0.02 & 0.01 & 0.14 & -0.14 & 0.00 & 2.66 & -0.16 & 0.00 \\ 
TiSe$_2$ & 17.39 & 6.27 & 3.54 & 7.38 & \textbf{-0.29} & -0.02 & -10.50 & 0.01 & 0.01 & 1.42 & -0.16 & 0.00 & 3.86 & -0.17 & 0.01 \\ 
TiTe$_2$ & 19.76 & 6.50 & 3.78 & 7.11 & 0.02 & -0.03 & -12.06 & 0.32 & -0.01 & 0.19 & 0.15 & -0.02 & 2.35 & 0.11 & 0.01 \\ 
VS$_2$ & 25.61 & 5.75 & 3.22 & 1.17 & 0.06 & \textbf{-0.05} & -18.40 & 0.32 & -0.03 & \textbf{-4.30} & 0.01 & \textbf{0.03} & -1.37 & -0.00 & 0.04 \\ 
VSe$_2$ & 22.26 & 6.11 & 3.36 & 3.26 & 0.05 & -0.04 & -15.62 & 0.38 & -0.03 & -2.64 & 0.08 & \textbf{0.04} & 0.14 & 0.03 & \textbf{0.05} \\ 
VTe$_2$ & 20.39 & 6.58 & 3.64 & 6.27 & 0.01 & \textbf{-0.05} & -12.89 & 0.55 & \textbf{-0.09} & -0.56 & 0.10 & 0.02 & 1.35 & 0.01 & \textbf{0.06} \\ 
WS$_2$ & 20.24 & 12.32 & 3.15 & 3.69 & 0.09 & 0.02 & -12.15 & 0.32 & 0.03 & 0.56 & 0.21 & 0.01 & 3.60 & 0.16 & 0.02 \\ 
WSe$_2$ & 19.98 & 12.96 & 3.28 & 3.45 & 0.13 & 0.02 & -13.29 & 0.44 & 0.03 & 0.25 & 0.25 & 0.01 & 3.06 & 0.22 & 0.03 \\ 
ZrS$_2$ & 16.98 & 5.81 & 3.66 & 3.09 & 0.02 & -0.01 & -11.55 & 0.21 & 0.03 & -0.85 & 0.06 & 0.02 & 1.35 & 0.05 & 0.01 \\ 
ZrSe$_2$ & 18.53 & 6.13 & 3.77 & 2.55 & 0.02 & 0.00 & -12.66 & 0.24 & 0.03 & -1.84 & 0.08 & 0.02 & 0.34 & 0.06 & 0.02 \\ 
ZrTe$_2$ & 16.34 & 6.66 & 3.95 & 8.84 & 0.01 & -0.02 & -8.33 & 0.26 & \textbf{0.05} & \textbf{3.33} & 0.08 & 0.03 & 5.23 & 0.07 & \textbf{0.05} \\ \hline 
MD & & & & 4.59 & 0.01 & -0.00 & -12.33 & 0.34 & 0.00 & 0.48 & 0.12 & 0.00 & 2.67 & 0.09 & 0.02 \\ 
MAD & & & & 4.82 & 0.08 & 0.02 & 12.33 & 0.35 & 0.02 & 1.53 & 0.15 & 0.02 & 2.79 & 0.13 & 0.02 \\ 
STD DEV & & & & 2.50 & 0.11 & 0.03 & 2.36 & 0.22 & 0.03 & 2.01 & 0.16 & 0.02 & 2.08 & 0.14 & 0.02 \\ 
\end{tabular}

\end{table*}
\end{ruledtabular}


%% file: Tables/L28_grid_conv.tex
\begin{ruledtabular}
\begin{table*}
\caption{\label{tab:table_A3} Lattice constants $c$ and $a$ in \AA{} for 28 layered materials \rtt{(L28 data set)} from SCAN+rVV10 and r$^2$SCAN+rVV10.
\rtt{Deviations are reported under $\Delta c$ and $\Delta a$ columns.}
Different (ENCUT, ENAUG) settings are presented; both values are in eV.
The experimental values of $c$ and $a$ are included for comparison \cite{L28_jcp}.
Mean \rtt{deviations (MDs) and mean absolute deviations (MADs)} are taken with respect to the largest ENCUT, 800 eV, and ENAUG, 2000 eV, setting.
r$^2$SCAN+rVV10 approaches its converged values more rapidly than does SCAN+rVV10.}
\begin{tabular}{lrr|rrrrrr|rrrrrr}
 & & & \multicolumn{6}{c|}{SCAN+rVV10} & \multicolumn{6}{c}{r$^2$SCAN+rVV10} \\ 
 & \multicolumn{2}{c|}{Expt.} & \multicolumn{2}{c}{(500,600)} & \multicolumn{2}{c}{(500,1000)} & \multicolumn{2}{c|}{(800,2000)} & \multicolumn{2}{c}{(500,600)} & \multicolumn{2}{c}{(500,1000)} & \multicolumn{2}{c}{(800,2000)} \\ 
 & c & a & $\Delta c$ & $\Delta a$ & $\Delta c$ & $\Delta a$ & c & a & $\Delta c$ & $\Delta a$ & $\Delta c$ & $\Delta a$ & c & a \\ \hline 
h-BN & 6.54 & 2.51 & \textbf{-0.08} & -0.00 & \textbf{-0.07} & -0.00 & 6.54 & 2.50 & \textbf{-0.07} & -0.00 & \textbf{-0.06} & -0.00 & 6.50 & 2.50 \\ 
Graphite & 6.70 & 2.46 & 0.04 & 0.00 & 0.02 & 0.00 & 6.65 & 2.45 & \textbf{0.03} & 0.00 & \textbf{0.03} & \textbf{0.00} & 6.63 & 2.46 \\ 
HfS$_2$ & 5.84 & 3.63 & 0.04 & 0.00 & -0.01 & -0.00 & 5.87 & 3.61 & -0.00 & -0.00 & -0.00 & -0.00 & 5.88 & 3.62 \\ 
HfSe$_2$ & 6.16 & 3.75 & -0.01 & -0.00 & 0.00 & 0.00 & 6.22 & 3.74 & 0.00 & 0.00 & -0.00 & -0.00 & 6.21 & 3.75 \\ 
HfTe$_2$ & 6.65 & 3.96 & -0.01 & -0.00 & 0.00 & 0.00 & 6.79 & 3.95 & -0.00 & 0.00 & -0.00 & -0.00 & 6.78 & 3.98 \\ 
MoS$_2$ & 12.30 & 3.16 & 0.02 & -0.00 & -0.00 & -0.00 & 12.47 & 3.17 & -0.01 & -0.01 & -0.01 & -0.00 & 12.43 & 3.18 \\ 
MoSe$_2$ & 12.93 & 3.29 & 0.02 & -0.00 & 0.00 & -0.00 & 13.17 & 3.30 & 0.00 & -0.00 & 0.00 & -0.00 & 13.12 & 3.31 \\ 
MoTe$_2$ & 13.97 & 3.52 & 0.04 & -0.00 & 0.00 & -0.00 & 14.26 & 3.51 & -0.01 & 0.00 & -0.01 & -0.00 & 14.22 & 3.55 \\ 
NbS$_2$ & 17.91 & 3.33 & -0.01 & -0.00 & \textbf{-0.03} & -0.00 & 18.34 & 3.33 & \textbf{-0.02} & 0.00 & \textbf{-0.02} & -0.00 & 18.31 & 3.34 \\ 
NbSe$_2$ & 12.55 & 3.44 & 0.00 & -0.00 & -0.01 & -0.00 & 13.01 & 3.46 & -0.01 & -0.00 & -0.01 & -0.00 & 12.99 & 3.46 \\ 
NbTe$_2$ & 6.61 & 3.68 & 0.00 & -0.00 & 0.00 & -0.00 & 6.95 & 3.65 & -0.00 & 0.00 & -0.00 & -0.00 & 6.88 & 3.70 \\ 
PbO & 5.00 & 3.96 & -0.04 & \textbf{0.01} & -0.02 & -0.00 & 4.94 & 3.99 & \textbf{-0.02} & -0.00 & \textbf{-0.02} & -0.00 & 5.01 & 3.99 \\ 
PdTe$_2$ & 5.11 & 4.02 & 0.00 & -0.00 & 0.00 & -0.00 & 5.03 & 4.04 & -0.00 & 0.00 & -0.00 & -0.00 & 5.17 & 4.06 \\ 
PtS$_2$ & 5.04 & 3.54 & \textbf{0.06} & -0.00 & -0.01 & 0.00 & 5.19 & 3.53 & -0.01 & -0.00 & -0.01 & -0.00 & 5.13 & 3.55 \\ 
PtSe$_2$ & 5.11 & 3.73 & -0.01 & 0.00 & -0.01 & 0.00 & 5.40 & 3.70 & -0.00 & -0.00 & -0.01 & 0.00 & 5.31 & 3.73 \\ 
TaS$_2$ & 5.90 & 3.36 & -0.03 & -0.00 & -0.01 & -0.00 & 5.96 & 3.35 & -0.00 & 0.00 & -0.00 & -0.00 & 5.95 & 3.37 \\ 
TaSe$_2$ & 6.27 & 3.48 & -0.01 & -0.00 & 0.00 & -0.00 & 6.34 & 3.47 & 0.00 & -0.00 & 0.00 & 0.00 & 6.31 & 3.49 \\ 
TiS$_2$ & 5.90 & 3.41 & 0.00 & 0.00 & -0.01 & -0.00 & 5.77 & 3.41 & -0.00 & -0.00 & -0.01 & -0.00 & 5.74 & 3.41 \\ 
TiSe$_2$ & 6.27 & 3.54 & 0.02 & \textbf{-0.01} & -0.01 & 0.00 & 6.12 & 3.55 & -0.01 & 0.00 & -0.01 & -0.00 & 6.10 & 3.55 \\ 
TiTe$_2$ & 6.50 & 3.78 & 0.01 & -0.00 & 0.00 & -0.00 & 6.64 & 3.77 & -0.01 & -0.00 & -0.00 & -0.00 & 6.61 & 3.79 \\ 
VS$_2$ & 5.75 & 3.22 & \textbf{0.05} & -0.00 & -0.01 & -0.00 & 5.76 & 3.25 & 0.00 & 0.00 & -0.01 & -0.00 & 5.75 & 3.26 \\ 
VSe$_2$ & 6.11 & 3.36 & -0.02 & -0.00 & 0.00 & -0.00 & 6.19 & 3.40 & 0.00 & -0.00 & 0.00 & \textbf{-0.00} & 6.14 & 3.41 \\ 
VTe$_2$ & 6.58 & 3.64 & \textbf{0.07} & \textbf{-0.01} & \textbf{-0.05} & \textbf{0.02} & 6.68 & 3.66 & -0.02 & -0.00 & -0.01 & 0.00 & 6.59 & 3.70 \\ 
WS$_2$ & 12.32 & 3.15 & 0.01 & -0.00 & -0.01 & -0.00 & 12.51 & 3.16 & -0.01 & -0.00 & -0.01 & -0.00 & 12.48 & 3.17 \\ 
WSe$_2$ & 12.96 & 3.28 & -0.02 & 0.00 & -0.01 & -0.00 & 13.22 & 3.29 & -0.00 & 0.00 & -0.01 & -0.00 & 13.18 & 3.31 \\ 
ZrS$_2$ & 5.81 & 3.66 & 0.01 & -0.00 & -0.01 & -0.00 & 5.88 & 3.68 & -0.00 & -0.00 & -0.00 & -0.00 & 5.86 & 3.67 \\ 
ZrSe$_2$ & 6.13 & 3.77 & -0.01 & -0.00 & 0.00 & 0.00 & 6.21 & 3.79 & 0.00 & -0.00 & 0.00 & 0.00 & 6.19 & 3.79 \\ 
ZrTe$_2$ & 6.66 & 3.95 & 0.02 & 0.00 & 0.00 & 0.00 & 6.74 & 3.98 & -0.00 & 0.00 & -0.00 & -0.00 & 6.73 & 4.00 \\ \hline 
MAD & & & 0.024 & 0.002 & 0.010 & 0.001 & 0.000 & 0.000 & 0.009 & 0.003 & 0.008 & 0.000 & 0.000 & 0.000 \\ 
MD & & & 0.006 & -0.001 & -0.008 & 0.000 & 0.000 & 0.000 & -0.007 & -0.001 & -0.006 & -0.000 & 0.000 & 0.000 \\ 
STD DEV & & & 0.032 & 0.003 & 0.017 & 0.003 & 0.000 & 0.000 & 0.015 & 0.003 & 0.013 & 0.000 & 0.000 & 0.000 \\ 
\end{tabular}

\end{table*}
\end{ruledtabular}

%% file: main.bbl
\begin{thebibliography}{101}%
\makeatletter
\providecommand \@ifxundefined [1]{%
 \@ifx{#1\undefined}
}%
\providecommand \@ifnum [1]{%
 \ifnum #1\expandafter \@firstoftwo
 \else \expandafter \@secondoftwo
 \fi
}%
\providecommand \@ifx [1]{%
 \ifx #1\expandafter \@firstoftwo
 \else \expandafter \@secondoftwo
 \fi
}%
\providecommand \natexlab [1]{#1}%
\providecommand \enquote  [1]{``#1''}%
\providecommand \bibnamefont  [1]{#1}%
\providecommand \bibfnamefont [1]{#1}%
\providecommand \citenamefont [1]{#1}%
\providecommand \href@noop [0]{\@secondoftwo}%
\providecommand \href [0]{\begingroup \@sanitize@url \@href}%
\providecommand \@href[1]{\@@startlink{#1}\@@href}%
\providecommand \@@href[1]{\endgroup#1\@@endlink}%
\providecommand \@sanitize@url [0]{\catcode `\\12\catcode `\$12\catcode
  `\&12\catcode `\#12\catcode `\^12\catcode `\_12\catcode `\%12\relax}%
\providecommand \@@startlink[1]{}%
\providecommand \@@endlink[0]{}%
\providecommand \url  [0]{\begingroup\@sanitize@url \@url }%
\providecommand \@url [1]{\endgroup\@href {#1}{\urlprefix }}%
\providecommand \urlprefix  [0]{URL }%
\providecommand \Eprint [0]{\href }%
\providecommand \doibase [0]{http://dx.doi.org/}%
\providecommand \selectlanguage [0]{\@gobble}%
\providecommand \bibinfo  [0]{\@secondoftwo}%
\providecommand \bibfield  [0]{\@secondoftwo}%
\providecommand \translation [1]{[#1]}%
\providecommand \BibitemOpen [0]{}%
\providecommand \bibitemStop [0]{}%
\providecommand \bibitemNoStop [0]{.\EOS\space}%
\providecommand \EOS [0]{\spacefactor3000\relax}%
\providecommand \BibitemShut  [1]{\csname bibitem#1\endcsname}%
\let\auto@bib@innerbib\@empty
\bibitem [{\citenamefont {DiStasio}\ \emph {et~al.}(2012)\citenamefont
  {DiStasio}, \citenamefont {von Lilienfeld},\ and\ \citenamefont
  {Tkatchenko}}]{DNA_vdw}%
  \BibitemOpen
  \bibfield  {author} {\bibinfo {author} {\bibfnamefont {R.~A.}\ \bibnamefont
  {DiStasio}}, \bibinfo {author} {\bibfnamefont {O.~A.}\ \bibnamefont {von
  Lilienfeld}}, \ and\ \bibinfo {author} {\bibfnamefont {A.}~\bibnamefont
  {Tkatchenko}},\ }\href@noop {} {\bibfield  {journal} {\bibinfo  {journal}
  {Proc. Natl. Acad. Sci. U.S.A.}\ }\textbf {\bibinfo {volume} {109}},\
  \bibinfo {pages} {14791} (\bibinfo {year} {2012})}\BibitemShut {NoStop}%
\bibitem [{\citenamefont {Roth}\ \emph {et~al.}(1996)\citenamefont {Roth},
  \citenamefont {Neal},\ and\ \citenamefont {Lenhoff}}]{protein_vdw}%
  \BibitemOpen
  \bibfield  {author} {\bibinfo {author} {\bibfnamefont {C.~M.}\ \bibnamefont
  {Roth}}, \bibinfo {author} {\bibfnamefont {B.~L.}\ \bibnamefont {Neal}}, \
  and\ \bibinfo {author} {\bibfnamefont {A.~M.}\ \bibnamefont {Lenhoff}},\
  }\href@noop {} {\bibfield  {journal} {\bibinfo  {journal} {Biophys. J.}\
  }\textbf {\bibinfo {volume} {70}},\ \bibinfo {pages} {977} (\bibinfo {year}
  {1996})}\BibitemShut {NoStop}%
\bibitem [{\citenamefont {Israelachvili}(1973)}]{biology_vdw}%
  \BibitemOpen
  \bibfield  {author} {\bibinfo {author} {\bibfnamefont {J.~N.}\ \bibnamefont
  {Israelachvili}},\ }\href@noop {} {\bibfield  {journal} {\bibinfo  {journal}
  {Q. Rev. Biophys.}\ }\textbf {\bibinfo {volume} {6}},\ \bibinfo {pages} {341}
  (\bibinfo {year} {1973})}\BibitemShut {NoStop}%
\bibitem [{\citenamefont {Sowerby}\ \emph {et~al.}(2001)\citenamefont
  {Sowerby}, \citenamefont {Cohn}, \citenamefont {Heckl},\ and\ \citenamefont
  {Holm}}]{life_vdw}%
  \BibitemOpen
  \bibfield  {author} {\bibinfo {author} {\bibfnamefont {S.~J.}\ \bibnamefont
  {Sowerby}}, \bibinfo {author} {\bibfnamefont {C.~A.}\ \bibnamefont {Cohn}},
  \bibinfo {author} {\bibfnamefont {W.~M.}\ \bibnamefont {Heckl}}, \ and\
  \bibinfo {author} {\bibfnamefont {N.~G.}\ \bibnamefont {Holm}},\ }\href@noop
  {} {\bibfield  {journal} {\bibinfo  {journal} {Proc. Natl. Acad. Sci.
  U.S.A.}\ }\textbf {\bibinfo {volume} {98}},\ \bibinfo {pages} {820} (\bibinfo
  {year} {2001})}\BibitemShut {NoStop}%
\bibitem [{\citenamefont {Siegbahn}\ \emph {et~al.}(2010)\citenamefont
  {Siegbahn}, \citenamefont {Blomberg},\ and\ \citenamefont
  {Chen}}]{complex_vdw}%
  \BibitemOpen
  \bibfield  {author} {\bibinfo {author} {\bibfnamefont {P.~E.}\ \bibnamefont
  {Siegbahn}}, \bibinfo {author} {\bibfnamefont {M.~R.}\ \bibnamefont
  {Blomberg}}, \ and\ \bibinfo {author} {\bibfnamefont {S.-L.}\ \bibnamefont
  {Chen}},\ }\href@noop {} {\bibfield  {journal} {\bibinfo  {journal} {J. Chem.
  Theory Comput.}\ }\textbf {\bibinfo {volume} {6}},\ \bibinfo {pages} {2040}
  (\bibinfo {year} {2010})}\BibitemShut {NoStop}%
\bibitem [{\citenamefont {Rodriguez-Reyes}\ \emph {et~al.}(2014)\citenamefont
  {Rodriguez-Reyes}, \citenamefont {Siler}, \citenamefont {Liu}, \citenamefont
  {Tkatchenko}, \citenamefont {Friend},\ and\ \citenamefont
  {Madix}}]{Aucat_vdw}%
  \BibitemOpen
  \bibfield  {author} {\bibinfo {author} {\bibfnamefont {J.~C.~F.}\
  \bibnamefont {Rodriguez-Reyes}}, \bibinfo {author} {\bibfnamefont {C.~G.}\
  \bibnamefont {Siler}}, \bibinfo {author} {\bibfnamefont {W.}~\bibnamefont
  {Liu}}, \bibinfo {author} {\bibfnamefont {A.}~\bibnamefont {Tkatchenko}},
  \bibinfo {author} {\bibfnamefont {C.~M.}\ \bibnamefont {Friend}}, \ and\
  \bibinfo {author} {\bibfnamefont {R.~J.}\ \bibnamefont {Madix}},\ }\href@noop
  {} {\bibfield  {journal} {\bibinfo  {journal} {J. Am. Chem. Soc.}\ }\textbf
  {\bibinfo {volume} {136}},\ \bibinfo {pages} {13333} (\bibinfo {year}
  {2014})}\BibitemShut {NoStop}%
\bibitem [{\citenamefont {Nilofer}\ \emph {et~al.}(2017)\citenamefont
  {Nilofer}, \citenamefont {Sukhwal}, \citenamefont {Mohanapriya},\ and\
  \citenamefont {Kangueane}}]{enzyme_vdw}%
  \BibitemOpen
  \bibfield  {author} {\bibinfo {author} {\bibfnamefont {C.}~\bibnamefont
  {Nilofer}}, \bibinfo {author} {\bibfnamefont {A.}~\bibnamefont {Sukhwal}},
  \bibinfo {author} {\bibfnamefont {A.}~\bibnamefont {Mohanapriya}}, \ and\
  \bibinfo {author} {\bibfnamefont {P.}~\bibnamefont {Kangueane}},\ }\href@noop
  {} {\bibfield  {journal} {\bibinfo  {journal} {Bioinformation}\ }\textbf
  {\bibinfo {volume} {13}},\ \bibinfo {pages} {164} (\bibinfo {year}
  {2017})}\BibitemShut {NoStop}%
\bibitem [{\citenamefont {Gattinoni}\ and\ \citenamefont
  {Michaelides}(2015)}]{corrosion_vdW}%
  \BibitemOpen
  \bibfield  {author} {\bibinfo {author} {\bibfnamefont {C.}~\bibnamefont
  {Gattinoni}}\ and\ \bibinfo {author} {\bibfnamefont {A.}~\bibnamefont
  {Michaelides}},\ }\href@noop {} {\bibfield  {journal} {\bibinfo  {journal}
  {Faraday discussions}\ }\textbf {\bibinfo {volume} {180}},\ \bibinfo {pages}
  {439} (\bibinfo {year} {2015})}\BibitemShut {NoStop}%
\bibitem [{\citenamefont {Peng}\ and\ \citenamefont {Perdew}(2017)}]{vdwUsyn}%
  \BibitemOpen
  \bibfield  {author} {\bibinfo {author} {\bibfnamefont {H.}~\bibnamefont
  {Peng}}\ and\ \bibinfo {author} {\bibfnamefont {J.~P.}\ \bibnamefont
  {Perdew}},\ }\href@noop {} {\bibfield  {journal} {\bibinfo  {journal} {Phys.
  Rev. B}\ }\textbf {\bibinfo {volume} {96}},\ \bibinfo {pages} {100101}
  (\bibinfo {year} {2017})}\BibitemShut {NoStop}%
\bibitem [{\citenamefont {Tao}\ \emph {et~al.}(2010)\citenamefont {Tao},
  \citenamefont {Perdew},\ and\ \citenamefont {Ruzsinszky}}]{vdw_alkali}%
  \BibitemOpen
  \bibfield  {author} {\bibinfo {author} {\bibfnamefont {J.}~\bibnamefont
  {Tao}}, \bibinfo {author} {\bibfnamefont {J.~P.}\ \bibnamefont {Perdew}}, \
  and\ \bibinfo {author} {\bibfnamefont {A.}~\bibnamefont {Ruzsinszky}},\
  }\href@noop {} {\bibfield  {journal} {\bibinfo  {journal} {Phys. Rev. B}\
  }\textbf {\bibinfo {volume} {81}},\ \bibinfo {pages} {233102} (\bibinfo
  {year} {2010})}\BibitemShut {NoStop}%
\bibitem [{\citenamefont {Klime{\v{s}}}\ \emph {et~al.}(2011)\citenamefont
  {Klime{\v{s}}}, \citenamefont {Bowler},\ and\ \citenamefont
  {Michaelides}}]{vdw_solids}%
  \BibitemOpen
  \bibfield  {author} {\bibinfo {author} {\bibfnamefont {J.}~\bibnamefont
  {Klime{\v{s}}}}, \bibinfo {author} {\bibfnamefont {D.~R.}\ \bibnamefont
  {Bowler}}, \ and\ \bibinfo {author} {\bibfnamefont {A.}~\bibnamefont
  {Michaelides}},\ }\href@noop {} {\bibfield  {journal} {\bibinfo  {journal}
  {Phys. Rev. B}\ }\textbf {\bibinfo {volume} {83}},\ \bibinfo {pages} {195131}
  (\bibinfo {year} {2011})}\BibitemShut {NoStop}%
\bibitem [{\citenamefont {Kohn}\ and\ \citenamefont
  {Sham}(1965)}]{KohnShamDFT}%
  \BibitemOpen
  \bibfield  {author} {\bibinfo {author} {\bibfnamefont {W.}~\bibnamefont
  {Kohn}}\ and\ \bibinfo {author} {\bibfnamefont {L.~J.}\ \bibnamefont
  {Sham}},\ }\href {\doibase 10.1103/PhysRev.140.A1133} {\bibfield  {journal}
  {\bibinfo  {journal} {Phys. Rev.}\ }\textbf {\bibinfo {volume} {140}},\
  \bibinfo {pages} {A1133} (\bibinfo {year} {1965})}\BibitemShut {NoStop}%
\bibitem [{\citenamefont {Foulkes}\ \emph {et~al.}(2001)\citenamefont
  {Foulkes}, \citenamefont {Mitas}, \citenamefont {Needs},\ and\ \citenamefont
  {Rajagopal}}]{QMC}%
  \BibitemOpen
  \bibfield  {author} {\bibinfo {author} {\bibfnamefont {W.}~\bibnamefont
  {Foulkes}}, \bibinfo {author} {\bibfnamefont {L.}~\bibnamefont {Mitas}},
  \bibinfo {author} {\bibfnamefont {R.}~\bibnamefont {Needs}}, \ and\ \bibinfo
  {author} {\bibfnamefont {G.}~\bibnamefont {Rajagopal}},\ }\href@noop {}
  {\bibfield  {journal} {\bibinfo  {journal} {Rev. Mod. Phys.}\ }\textbf
  {\bibinfo {volume} {73}},\ \bibinfo {pages} {33} (\bibinfo {year}
  {2001})}\BibitemShut {NoStop}%
\bibitem [{\citenamefont {Raghavachari}\ \emph {et~al.}(1989)\citenamefont
  {Raghavachari}, \citenamefont {Trucks}, \citenamefont {Pople},\ and\
  \citenamefont {Head-Gordon}}]{CCSDT}%
  \BibitemOpen
  \bibfield  {author} {\bibinfo {author} {\bibfnamefont {K.}~\bibnamefont
  {Raghavachari}}, \bibinfo {author} {\bibfnamefont {G.~W.}\ \bibnamefont
  {Trucks}}, \bibinfo {author} {\bibfnamefont {J.~A.}\ \bibnamefont {Pople}}, \
  and\ \bibinfo {author} {\bibfnamefont {M.}~\bibnamefont {Head-Gordon}},\
  }\href@noop {} {\bibfield  {journal} {\bibinfo  {journal} {Chem. Phys.
  Lett.}\ }\textbf {\bibinfo {volume} {157}},\ \bibinfo {pages} {479} (\bibinfo
  {year} {1989})}\BibitemShut {NoStop}%
\bibitem [{\citenamefont {Eshuis}\ \emph {et~al.}(2012)\citenamefont {Eshuis},
  \citenamefont {Bates},\ and\ \citenamefont {Furche}}]{RPA}%
  \BibitemOpen
  \bibfield  {author} {\bibinfo {author} {\bibfnamefont {H.}~\bibnamefont
  {Eshuis}}, \bibinfo {author} {\bibfnamefont {J.~E.}\ \bibnamefont {Bates}}, \
  and\ \bibinfo {author} {\bibfnamefont {F.}~\bibnamefont {Furche}},\
  }\href@noop {} {\bibfield  {journal} {\bibinfo  {journal} {Theoret. Chem.
  Acc.}\ }\textbf {\bibinfo {volume} {131}},\ \bibinfo {pages} {1} (\bibinfo
  {year} {2012})}\BibitemShut {NoStop}%
\bibitem [{\citenamefont {Grimme}(2006)}]{D2}%
  \BibitemOpen
  \bibfield  {author} {\bibinfo {author} {\bibfnamefont {S.}~\bibnamefont
  {Grimme}},\ }\href@noop {} {\bibfield  {journal} {\bibinfo  {journal} {J.
  Comput. Chem.}\ }\textbf {\bibinfo {volume} {27}},\ \bibinfo {pages} {1787}
  (\bibinfo {year} {2006})}\BibitemShut {NoStop}%
\bibitem [{\citenamefont {Grimme}\ \emph {et~al.}(2010)\citenamefont {Grimme},
  \citenamefont {Antony}, \citenamefont {Ehrlich},\ and\ \citenamefont
  {Krieg}}]{D3}%
  \BibitemOpen
  \bibfield  {author} {\bibinfo {author} {\bibfnamefont {S.}~\bibnamefont
  {Grimme}}, \bibinfo {author} {\bibfnamefont {J.}~\bibnamefont {Antony}},
  \bibinfo {author} {\bibfnamefont {S.}~\bibnamefont {Ehrlich}}, \ and\
  \bibinfo {author} {\bibfnamefont {H.}~\bibnamefont {Krieg}},\ }\href@noop {}
  {\bibfield  {journal} {\bibinfo  {journal} {J. Chem. Phys.}\ }\textbf
  {\bibinfo {volume} {132}},\ \bibinfo {pages} {154104} (\bibinfo {year}
  {2010})}\BibitemShut {NoStop}%
\bibitem [{\citenamefont {Caldeweyher}\ \emph {et~al.}(2019)\citenamefont
  {Caldeweyher}, \citenamefont {Ehlert}, \citenamefont {Hansen}, \citenamefont
  {Neugebauer}, \citenamefont {Spicher}, \citenamefont {Bannwarth},\ and\
  \citenamefont {Grimme}}]{D4}%
  \BibitemOpen
  \bibfield  {author} {\bibinfo {author} {\bibfnamefont {E.}~\bibnamefont
  {Caldeweyher}}, \bibinfo {author} {\bibfnamefont {S.}~\bibnamefont {Ehlert}},
  \bibinfo {author} {\bibfnamefont {A.}~\bibnamefont {Hansen}}, \bibinfo
  {author} {\bibfnamefont {H.}~\bibnamefont {Neugebauer}}, \bibinfo {author}
  {\bibfnamefont {S.}~\bibnamefont {Spicher}}, \bibinfo {author} {\bibfnamefont
  {C.}~\bibnamefont {Bannwarth}}, \ and\ \bibinfo {author} {\bibfnamefont
  {S.}~\bibnamefont {Grimme}},\ }\href@noop {} {\bibfield  {journal} {\bibinfo
  {journal} {J. Chem. Phys.}\ }\textbf {\bibinfo {volume} {150}},\ \bibinfo
  {pages} {154122} (\bibinfo {year} {2019})}\BibitemShut {NoStop}%
\bibitem [{\citenamefont {Ehlert}\ \emph {et~al.}(2021)\citenamefont {Ehlert},
  \citenamefont {Huniar}, \citenamefont {Ning}, \citenamefont {Furness},
  \citenamefont {Sun}, \citenamefont {Kaplan}, \citenamefont {Perdew},\ and\
  \citenamefont {Brandenburg}}]{r2SCAND4}%
  \BibitemOpen
  \bibfield  {author} {\bibinfo {author} {\bibfnamefont {S.}~\bibnamefont
  {Ehlert}}, \bibinfo {author} {\bibfnamefont {U.}~\bibnamefont {Huniar}},
  \bibinfo {author} {\bibfnamefont {J.}~\bibnamefont {Ning}}, \bibinfo {author}
  {\bibfnamefont {J.~W.}\ \bibnamefont {Furness}}, \bibinfo {author}
  {\bibfnamefont {J.}~\bibnamefont {Sun}}, \bibinfo {author} {\bibfnamefont
  {A.~D.}\ \bibnamefont {Kaplan}}, \bibinfo {author} {\bibfnamefont {J.~P.}\
  \bibnamefont {Perdew}}, \ and\ \bibinfo {author} {\bibfnamefont {J.~G.}\
  \bibnamefont {Brandenburg}},\ }\href@noop {} {\bibfield  {journal} {\bibinfo
  {journal} {J. Chem. Phys.}\ }\textbf {\bibinfo {volume} {154}},\ \bibinfo
  {pages} {061101} (\bibinfo {year} {2021})}\BibitemShut {NoStop}%
\bibitem [{\citenamefont {Tkatchenko}\ and\ \citenamefont
  {Scheffler}(2009)}]{TS1}%
  \BibitemOpen
  \bibfield  {author} {\bibinfo {author} {\bibfnamefont {A.}~\bibnamefont
  {Tkatchenko}}\ and\ \bibinfo {author} {\bibfnamefont {M.}~\bibnamefont
  {Scheffler}},\ }\href@noop {} {\bibfield  {journal} {\bibinfo  {journal}
  {Phys. Rev. Lett.}\ }\textbf {\bibinfo {volume} {102}},\ \bibinfo {pages}
  {073005} (\bibinfo {year} {2009})}\BibitemShut {NoStop}%
\bibitem [{\citenamefont {Tkatchenko}\ \emph {et~al.}(2012)\citenamefont
  {Tkatchenko}, \citenamefont {DiStasio~Jr.}, \citenamefont {Car},\ and\
  \citenamefont {Scheffler}}]{TS2}%
  \BibitemOpen
  \bibfield  {author} {\bibinfo {author} {\bibfnamefont {A.}~\bibnamefont
  {Tkatchenko}}, \bibinfo {author} {\bibfnamefont {R.~A.}\ \bibnamefont
  {DiStasio~Jr.}}, \bibinfo {author} {\bibfnamefont {R.}~\bibnamefont {Car}}, \
  and\ \bibinfo {author} {\bibfnamefont {M.}~\bibnamefont {Scheffler}},\
  }\href@noop {} {\bibfield  {journal} {\bibinfo  {journal} {Phys. Rev. Lett.}\
  }\textbf {\bibinfo {volume} {108}},\ \bibinfo {pages} {236402} (\bibinfo
  {year} {2012})}\BibitemShut {NoStop}%
\bibitem [{\citenamefont {Liu}\ \emph {et~al.}(2015)\citenamefont {Liu},
  \citenamefont {Maa{\ss}}, \citenamefont {Willenbockel}, \citenamefont
  {Bronner}, \citenamefont {Schulze}, \citenamefont {Soubatch}, \citenamefont
  {Tautz}, \citenamefont {Tegeder},\ and\ \citenamefont {Tkatchenko}}]{TS3}%
  \BibitemOpen
  \bibfield  {author} {\bibinfo {author} {\bibfnamefont {W.}~\bibnamefont
  {Liu}}, \bibinfo {author} {\bibfnamefont {F.}~\bibnamefont {Maa{\ss}}},
  \bibinfo {author} {\bibfnamefont {M.}~\bibnamefont {Willenbockel}}, \bibinfo
  {author} {\bibfnamefont {C.}~\bibnamefont {Bronner}}, \bibinfo {author}
  {\bibfnamefont {M.}~\bibnamefont {Schulze}}, \bibinfo {author} {\bibfnamefont
  {S.}~\bibnamefont {Soubatch}}, \bibinfo {author} {\bibfnamefont {F.~S.}\
  \bibnamefont {Tautz}}, \bibinfo {author} {\bibfnamefont {P.}~\bibnamefont
  {Tegeder}}, \ and\ \bibinfo {author} {\bibfnamefont {A.}~\bibnamefont
  {Tkatchenko}},\ }\href@noop {} {\bibfield  {journal} {\bibinfo  {journal}
  {Phys. Rev. Lett.}\ }\textbf {\bibinfo {volume} {115}},\ \bibinfo {pages}
  {036104} (\bibinfo {year} {2015})}\BibitemShut {NoStop}%
\bibitem [{\citenamefont {Berland}\ \emph {et~al.}(2015)\citenamefont
  {Berland}, \citenamefont {Cooper}, \citenamefont {Lee}, \citenamefont
  {Schr{\"o}der}, \citenamefont {Thonhauser}, \citenamefont {Hyldgaard},\ and\
  \citenamefont {Lundqvist}}]{vdw-df}%
  \BibitemOpen
  \bibfield  {author} {\bibinfo {author} {\bibfnamefont {K.}~\bibnamefont
  {Berland}}, \bibinfo {author} {\bibfnamefont {V.~R.}\ \bibnamefont {Cooper}},
  \bibinfo {author} {\bibfnamefont {K.}~\bibnamefont {Lee}}, \bibinfo {author}
  {\bibfnamefont {E.}~\bibnamefont {Schr{\"o}der}}, \bibinfo {author}
  {\bibfnamefont {T.}~\bibnamefont {Thonhauser}}, \bibinfo {author}
  {\bibfnamefont {P.}~\bibnamefont {Hyldgaard}}, \ and\ \bibinfo {author}
  {\bibfnamefont {B.~I.}\ \bibnamefont {Lundqvist}},\ }\href@noop {} {\bibfield
   {journal} {\bibinfo  {journal} {Rep. Prog. Phys.}\ }\textbf {\bibinfo
  {volume} {78}},\ \bibinfo {pages} {066501} (\bibinfo {year}
  {2015})}\BibitemShut {NoStop}%
\bibitem [{\citenamefont {Vydrov}\ and\ \citenamefont
  {Van~Voorhis}(2010)}]{vv10}%
  \BibitemOpen
  \bibfield  {author} {\bibinfo {author} {\bibfnamefont {O.~A.}\ \bibnamefont
  {Vydrov}}\ and\ \bibinfo {author} {\bibfnamefont {T.}~\bibnamefont
  {Van~Voorhis}},\ }\href@noop {} {\bibfield  {journal} {\bibinfo  {journal}
  {J. Chem. Phys.}\ }\textbf {\bibinfo {volume} {133}},\ \bibinfo {pages}
  {244103} (\bibinfo {year} {2010})}\BibitemShut {NoStop}%
\bibitem [{\citenamefont {Sabatini}\ \emph {et~al.}(2013)\citenamefont
  {Sabatini}, \citenamefont {Gorni},\ and\ \citenamefont
  {De~Gironcoli}}]{rvv10}%
  \BibitemOpen
  \bibfield  {author} {\bibinfo {author} {\bibfnamefont {R.}~\bibnamefont
  {Sabatini}}, \bibinfo {author} {\bibfnamefont {T.}~\bibnamefont {Gorni}}, \
  and\ \bibinfo {author} {\bibfnamefont {S.}~\bibnamefont {De~Gironcoli}},\
  }\href@noop {} {\bibfield  {journal} {\bibinfo  {journal} {Phys. Rev. B}\
  }\textbf {\bibinfo {volume} {87}},\ \bibinfo {pages} {041108} (\bibinfo
  {year} {2013})}\BibitemShut {NoStop}%
\bibitem [{\citenamefont {Becke}\ and\ \citenamefont
  {Johnson}(2005)}]{Becke2005}%
  \BibitemOpen
  \bibfield  {author} {\bibinfo {author} {\bibfnamefont {A.~D.}\ \bibnamefont
  {Becke}}\ and\ \bibinfo {author} {\bibfnamefont {E.~R.}\ \bibnamefont
  {Johnson}},\ }\href {\doibase 10.1063/1.1884601} {\bibfield  {journal}
  {\bibinfo  {journal} {J. Chem. Phys.}\ }\textbf {\bibinfo {volume} {122}},\
  \bibinfo {pages} {154104} (\bibinfo {year} {2005})}\BibitemShut {NoStop}%
\bibitem [{\citenamefont {Becke}\ and\ \citenamefont
  {Johnson}(2007)}]{Becke2007}%
  \BibitemOpen
  \bibfield  {author} {\bibinfo {author} {\bibfnamefont {A.~D.}\ \bibnamefont
  {Becke}}\ and\ \bibinfo {author} {\bibfnamefont {E.~R.}\ \bibnamefont
  {Johnson}},\ }\href {\doibase 10.1063/1.2768530} {\bibfield  {journal}
  {\bibinfo  {journal} {J. Chem. Phys.}\ }\textbf {\bibinfo {volume} {127}},\
  \bibinfo {pages} {124108} (\bibinfo {year} {2007})}\BibitemShut {NoStop}%
\bibitem [{\citenamefont {Tang}\ \emph {et~al.}(2020)\citenamefont {Tang},
  \citenamefont {Chowdhury}, \citenamefont {Tao},\ and\ \citenamefont
  {Perdew}}]{tang2020}%
  \BibitemOpen
  \bibfield  {author} {\bibinfo {author} {\bibfnamefont {H.}~\bibnamefont
  {Tang}}, \bibinfo {author} {\bibfnamefont {S.~T. u.~R.}\ \bibnamefont
  {Chowdhury}}, \bibinfo {author} {\bibfnamefont {J.}~\bibnamefont {Tao}}, \
  and\ \bibinfo {author} {\bibfnamefont {J.~P.}\ \bibnamefont {Perdew}},\
  }\href {\doibase 10.1103/PhysRevB.101.195426} {\bibfield  {journal} {\bibinfo
   {journal} {Phys. Rev. B}\ }\textbf {\bibinfo {volume} {101}},\ \bibinfo
  {pages} {195426} (\bibinfo {year} {2020})}\BibitemShut {NoStop}%
\bibitem [{\citenamefont {Chowdhury}\ \emph {et~al.}(2021)\citenamefont
  {Chowdhury}, \citenamefont {Tang},\ and\ \citenamefont
  {Perdew}}]{chowdhury2021}%
  \BibitemOpen
  \bibfield  {author} {\bibinfo {author} {\bibfnamefont {S.~T. u.~R.}\
  \bibnamefont {Chowdhury}}, \bibinfo {author} {\bibfnamefont {H.}~\bibnamefont
  {Tang}}, \ and\ \bibinfo {author} {\bibfnamefont {J.~P.}\ \bibnamefont
  {Perdew}},\ }\href {\doibase 10.1103/PhysRevB.103.195410} {\bibfield
  {journal} {\bibinfo  {journal} {Phys. Rev. B}\ }\textbf {\bibinfo {volume}
  {103}},\ \bibinfo {pages} {195410} (\bibinfo {year} {2021})}\BibitemShut
  {NoStop}%
\bibitem [{\citenamefont {Sun}\ \emph {et~al.}(2015)\citenamefont {Sun},
  \citenamefont {Ruzsinszky},\ and\ \citenamefont {Perdew}}]{SCAN}%
  \BibitemOpen
  \bibfield  {author} {\bibinfo {author} {\bibfnamefont {J.}~\bibnamefont
  {Sun}}, \bibinfo {author} {\bibfnamefont {A.}~\bibnamefont {Ruzsinszky}}, \
  and\ \bibinfo {author} {\bibfnamefont {J.~P.}\ \bibnamefont {Perdew}},\
  }\href@noop {} {\bibfield  {journal} {\bibinfo  {journal} {Phys. Rev. Lett.}\
  }\textbf {\bibinfo {volume} {115}},\ \bibinfo {pages} {036402} (\bibinfo
  {year} {2015})}\BibitemShut {NoStop}%
\bibitem [{\citenamefont {Sun}\ \emph {et~al.}(2016)\citenamefont {Sun},
  \citenamefont {Remsing}, \citenamefont {Zhang}, \citenamefont {Sun},
  \citenamefont {Ruzsinszky}, \citenamefont {Peng}, \citenamefont {Yang},
  \citenamefont {Paul}, \citenamefont {Waghmare}, \citenamefont {Wu} \emph
  {et~al.}}]{SCAN_NChem}%
  \BibitemOpen
  \bibfield  {author} {\bibinfo {author} {\bibfnamefont {J.}~\bibnamefont
  {Sun}}, \bibinfo {author} {\bibfnamefont {R.~C.}\ \bibnamefont {Remsing}},
  \bibinfo {author} {\bibfnamefont {Y.}~\bibnamefont {Zhang}}, \bibinfo
  {author} {\bibfnamefont {Z.}~\bibnamefont {Sun}}, \bibinfo {author}
  {\bibfnamefont {A.}~\bibnamefont {Ruzsinszky}}, \bibinfo {author}
  {\bibfnamefont {H.}~\bibnamefont {Peng}}, \bibinfo {author} {\bibfnamefont
  {Z.}~\bibnamefont {Yang}}, \bibinfo {author} {\bibfnamefont {A.}~\bibnamefont
  {Paul}}, \bibinfo {author} {\bibfnamefont {U.}~\bibnamefont {Waghmare}},
  \bibinfo {author} {\bibfnamefont {X.}~\bibnamefont {Wu}},  \emph {et~al.},\
  }\href@noop {} {\bibfield  {journal} {\bibinfo  {journal} {Nature Chem.}\
  }\textbf {\bibinfo {volume} {8}},\ \bibinfo {pages} {831} (\bibinfo {year}
  {2016})}\BibitemShut {NoStop}%
\bibitem [{\citenamefont {Peng}\ \emph {et~al.}(2016)\citenamefont {Peng},
  \citenamefont {Yang}, \citenamefont {Perdew},\ and\ \citenamefont
  {Sun}}]{SCAN+rVV10}%
  \BibitemOpen
  \bibfield  {author} {\bibinfo {author} {\bibfnamefont {H.}~\bibnamefont
  {Peng}}, \bibinfo {author} {\bibfnamefont {Z.-H.}\ \bibnamefont {Yang}},
  \bibinfo {author} {\bibfnamefont {J.~P.}\ \bibnamefont {Perdew}}, \ and\
  \bibinfo {author} {\bibfnamefont {J.}~\bibnamefont {Sun}},\ }\href@noop {}
  {\bibfield  {journal} {\bibinfo  {journal} {Phys. Rev. X}\ }\textbf {\bibinfo
  {volume} {6}},\ \bibinfo {pages} {041005} (\bibinfo {year}
  {2016})}\BibitemShut {NoStop}%
\bibitem [{\citenamefont {Furness}\ and\ \citenamefont
  {Sun}(2019)}]{beta_indicator}%
  \BibitemOpen
  \bibfield  {author} {\bibinfo {author} {\bibfnamefont {J.~W.}\ \bibnamefont
  {Furness}}\ and\ \bibinfo {author} {\bibfnamefont {J.}~\bibnamefont {Sun}},\
  }\href {\doibase 10.1103/PhysRevB.99.041119} {\bibfield  {journal} {\bibinfo
  {journal} {Phys. Rev. B}\ }\textbf {\bibinfo {volume} {99}},\ \bibinfo
  {pages} {041119} (\bibinfo {year} {2019})}\BibitemShut {NoStop}%
\bibitem [{\citenamefont {Bartók}\ and\ \citenamefont
  {Yates}(2019)}]{bartok2019}%
  \BibitemOpen
  \bibfield  {author} {\bibinfo {author} {\bibfnamefont {A.~P.}\ \bibnamefont
  {Bartók}}\ and\ \bibinfo {author} {\bibfnamefont {J.~R.}\ \bibnamefont
  {Yates}},\ }\href {\doibase 10.1063/1.5094646} {\bibfield  {journal}
  {\bibinfo  {journal} {J. Chem. Phys.}\ }\textbf {\bibinfo {volume} {150}},\
  \bibinfo {pages} {161101} (\bibinfo {year} {2019})}\BibitemShut {NoStop}%
\bibitem [{\citenamefont {Yu}\ \emph {et~al.}(2020)\citenamefont {Yu},
  \citenamefont {Fiorin}, \citenamefont {Peng}, \citenamefont {Klein},\ and\
  \citenamefont {Perdew}}]{Yu2020}%
  \BibitemOpen
  \bibfield  {author} {\bibinfo {author} {\bibfnamefont {J.}~\bibnamefont
  {Yu}}, \bibinfo {author} {\bibfnamefont {G.}~\bibnamefont {Fiorin}}, \bibinfo
  {author} {\bibfnamefont {H.}~\bibnamefont {Peng}}, \bibinfo {author}
  {\bibfnamefont {M.~L.}\ \bibnamefont {Klein}}, \ and\ \bibinfo {author}
  {\bibfnamefont {J.~P.}\ \bibnamefont {Perdew}},\ }\href {\doibase
  10.1103/PhysRevMaterials.4.055601} {\bibfield  {journal} {\bibinfo  {journal}
  {Phys. Rev. Materials}\ }\textbf {\bibinfo {volume} {4}},\ \bibinfo {pages}
  {055601} (\bibinfo {year} {2020})}\BibitemShut {NoStop}%
\bibitem [{\citenamefont {Brandenburg}\ \emph {et~al.}(2016)\citenamefont
  {Brandenburg}, \citenamefont {Bates}, \citenamefont {Sun},\ and\
  \citenamefont {Perdew}}]{scand3}%
  \BibitemOpen
  \bibfield  {author} {\bibinfo {author} {\bibfnamefont {J.~G.}\ \bibnamefont
  {Brandenburg}}, \bibinfo {author} {\bibfnamefont {J.~E.}\ \bibnamefont
  {Bates}}, \bibinfo {author} {\bibfnamefont {J.}~\bibnamefont {Sun}}, \ and\
  \bibinfo {author} {\bibfnamefont {J.~P.}\ \bibnamefont {Perdew}},\
  }\href@noop {} {\bibfield  {journal} {\bibinfo  {journal} {{Phys. Rev. B}}\
  }\textbf {\bibinfo {volume} {94}},\ \bibinfo {pages} {115144} (\bibinfo
  {year} {2016})}\BibitemShut {NoStop}%
\bibitem [{\citenamefont {Wiktor}\ \emph {et~al.}(2017)\citenamefont {Wiktor},
  \citenamefont {Ambrosio},\ and\ \citenamefont {Pasquarello}}]{Wiktor2017}%
  \BibitemOpen
  \bibfield  {author} {\bibinfo {author} {\bibfnamefont {J.}~\bibnamefont
  {Wiktor}}, \bibinfo {author} {\bibfnamefont {F.}~\bibnamefont {Ambrosio}}, \
  and\ \bibinfo {author} {\bibfnamefont {A.}~\bibnamefont {Pasquarello}},\
  }\href {\doibase 10.1063/1.5006146} {\bibfield  {journal} {\bibinfo
  {journal} {J. Chem. Phys.}\ }\textbf {\bibinfo {volume} {147}},\ \bibinfo
  {pages} {216101} (\bibinfo {year} {2017})}\BibitemShut {NoStop}%
\bibitem [{\citenamefont {Hermann}\ and\ \citenamefont
  {Tkatchenko}(2018)}]{Hermann2018}%
  \BibitemOpen
  \bibfield  {author} {\bibinfo {author} {\bibfnamefont {J.}~\bibnamefont
  {Hermann}}\ and\ \bibinfo {author} {\bibfnamefont {A.}~\bibnamefont
  {Tkatchenko}},\ }\href {\doibase 10.1021/acs.jctc.7b01172} {\bibfield
  {journal} {\bibinfo  {journal} {J. Chem. Theory Comput.}\ }\textbf {\bibinfo
  {volume} {14}},\ \bibinfo {pages} {1361} (\bibinfo {year}
  {2018})}\BibitemShut {NoStop}%
\bibitem [{\citenamefont {Dasgupta}\ \emph {et~al.}(2021)\citenamefont
  {Dasgupta}, \citenamefont {Lambros}, \citenamefont {Perdew},\ and\
  \citenamefont {Paesani}}]{Dasgupta2021}%
  \BibitemOpen
  \bibfield  {author} {\bibinfo {author} {\bibfnamefont {S.}~\bibnamefont
  {Dasgupta}}, \bibinfo {author} {\bibfnamefont {E.}~\bibnamefont {Lambros}},
  \bibinfo {author} {\bibfnamefont {J.}~\bibnamefont {Perdew}}, \ and\ \bibinfo
  {author} {\bibfnamefont {F.}~\bibnamefont {Paesani}},\ }\href {\doibase
  10.1038/s41467-021-26618-9} {\bibfield  {journal} {\bibinfo  {journal}
  {Nature Commun.}\ }\textbf {\bibinfo {volume} {12}},\ \bibinfo {pages} {6359}
  (\bibinfo {year} {2021})}\BibitemShut {NoStop}%
\bibitem [{\citenamefont {Mej{\'{i}}a-Rodr{\'{i}}guez}\ and\ \citenamefont
  {Trickey}(2019)}]{Mejia-Rodriguez2019}%
  \BibitemOpen
  \bibfield  {author} {\bibinfo {author} {\bibfnamefont {D.}~\bibnamefont
  {Mej{\'{i}}a-Rodr{\'{i}}guez}}\ and\ \bibinfo {author} {\bibfnamefont
  {S.~B.}\ \bibnamefont {Trickey}},\ }\href {\doibase 10.1063/1.5120408}
  {\bibfield  {journal} {\bibinfo  {journal} {J. Chem. Phys.}\ }\textbf
  {\bibinfo {volume} {151}},\ \bibinfo {pages} {207101} (\bibinfo {year}
  {2019})}\BibitemShut {NoStop}%
\bibitem [{\citenamefont {Bart{\'{o}}k}\ and\ \citenamefont
  {Yates}(2019)}]{Bartok2019a}%
  \BibitemOpen
  \bibfield  {author} {\bibinfo {author} {\bibfnamefont {A.~P.}\ \bibnamefont
  {Bart{\'{o}}k}}\ and\ \bibinfo {author} {\bibfnamefont {J.~R.}\ \bibnamefont
  {Yates}},\ }\href {\doibase 10.1063/1.5128484} {\bibfield  {journal}
  {\bibinfo  {journal} {J. Chem. Phys.}\ }\textbf {\bibinfo {volume} {151}},\
  \bibinfo {pages} {207102} (\bibinfo {year} {2019})}\BibitemShut {NoStop}%
\bibitem [{\citenamefont {Mej\'{\i}a-Rodr\'{\i}guez}\ and\ \citenamefont
  {Trickey}(2020)}]{mejia-rodriguez2020a}%
  \BibitemOpen
  \bibfield  {author} {\bibinfo {author} {\bibfnamefont {D.}~\bibnamefont
  {Mej\'{\i}a-Rodr\'{\i}guez}}\ and\ \bibinfo {author} {\bibfnamefont {S.~B.}\
  \bibnamefont {Trickey}},\ }\href {\doibase 10.1021/acs.jpca.0c08883}
  {\bibfield  {journal} {\bibinfo  {journal} {J. Phys. Chem. A}\ }\textbf
  {\bibinfo {volume} {124}},\ \bibinfo {pages} {9889} (\bibinfo {year}
  {2020})}\BibitemShut {NoStop}%
\bibitem [{\citenamefont {Levy}\ and\ \citenamefont
  {Perdew}(1985)}]{Levy1985a}%
  \BibitemOpen
  \bibfield  {author} {\bibinfo {author} {\bibfnamefont {M.}~\bibnamefont
  {Levy}}\ and\ \bibinfo {author} {\bibfnamefont {J.~P.}\ \bibnamefont
  {Perdew}},\ }\href {\doibase 10.1103/PhysRevA.32.2010} {\bibfield  {journal}
  {\bibinfo  {journal} {Phys. Rev. A}\ }\textbf {\bibinfo {volume} {32}},\
  \bibinfo {pages} {2010} (\bibinfo {year} {1985})}\BibitemShut {NoStop}%
\bibitem [{\citenamefont {G{\"{o}}rling}\ and\ \citenamefont
  {Levy}(1993)}]{Gorling1993a}%
  \BibitemOpen
  \bibfield  {author} {\bibinfo {author} {\bibfnamefont {A.}~\bibnamefont
  {G{\"{o}}rling}}\ and\ \bibinfo {author} {\bibfnamefont {M.}~\bibnamefont
  {Levy}},\ }\href {\doibase 10.1103/PhysRevB.47.13105} {\bibfield  {journal}
  {\bibinfo  {journal} {Phys. Rev. B}\ }\textbf {\bibinfo {volume} {47}},\
  \bibinfo {pages} {13105} (\bibinfo {year} {1993})}\BibitemShut {NoStop}%
\bibitem [{\citenamefont {Pollack}\ and\ \citenamefont
  {Perdew}(2000)}]{Pollack2000}%
  \BibitemOpen
  \bibfield  {author} {\bibinfo {author} {\bibfnamefont {L.}~\bibnamefont
  {Pollack}}\ and\ \bibinfo {author} {\bibfnamefont {J.~P.}\ \bibnamefont
  {Perdew}},\ }\href {\doibase 10.1088/0953-8984/12/7/308} {\bibfield
  {journal} {\bibinfo  {journal} {J. Phys. Condens. Matter}\ }\textbf {\bibinfo
  {volume} {12}},\ \bibinfo {pages} {1239} (\bibinfo {year}
  {2000})}\BibitemShut {NoStop}%
\bibitem [{\citenamefont {Svendsen}\ and\ \citenamefont {von
  Barth}(1996)}]{Svendsen1996}%
  \BibitemOpen
  \bibfield  {author} {\bibinfo {author} {\bibfnamefont {P.}~\bibnamefont
  {Svendsen}}\ and\ \bibinfo {author} {\bibfnamefont {U.}~\bibnamefont {von
  Barth}},\ }\href {\doibase 10.1103/PhysRevB.54.17402} {\bibfield  {journal}
  {\bibinfo  {journal} {Phys. Rev. B}\ }\textbf {\bibinfo {volume} {54}},\
  \bibinfo {pages} {17402} (\bibinfo {year} {1996})}\BibitemShut {NoStop}%
\bibitem [{\citenamefont {Perdew}\ and\ \citenamefont
  {Wang}(1992)}]{Perdew1992a}%
  \BibitemOpen
  \bibfield  {author} {\bibinfo {author} {\bibfnamefont {J.~P.}\ \bibnamefont
  {Perdew}}\ and\ \bibinfo {author} {\bibfnamefont {Y.}~\bibnamefont {Wang}},\
  }\href {\doibase 10.1103/PhysRevB.45.13244} {\bibfield  {journal} {\bibinfo
  {journal} {Phys. Rev. B}\ }\textbf {\bibinfo {volume} {45}},\ \bibinfo
  {pages} {13244} (\bibinfo {year} {1992})}\BibitemShut {NoStop}%
\bibitem [{\citenamefont {Furness}\ \emph {et~al.}(2020)\citenamefont
  {Furness}, \citenamefont {Kaplan}, \citenamefont {Ning}, \citenamefont
  {Perdew},\ and\ \citenamefont {Sun}}]{r2SCAN}%
  \BibitemOpen
  \bibfield  {author} {\bibinfo {author} {\bibfnamefont {J.~W.}\ \bibnamefont
  {Furness}}, \bibinfo {author} {\bibfnamefont {A.~D.}\ \bibnamefont {Kaplan}},
  \bibinfo {author} {\bibfnamefont {J.}~\bibnamefont {Ning}}, \bibinfo {author}
  {\bibfnamefont {J.~P.}\ \bibnamefont {Perdew}}, \ and\ \bibinfo {author}
  {\bibfnamefont {J.}~\bibnamefont {Sun}},\ }\href
  {https://doi.org/10.1021/acs.jpclett.0c02405} {\bibfield  {journal} {\bibinfo
   {journal} {J. Phys. Chem. Lett.}\ }\textbf {\bibinfo {volume} {11}},\
  \bibinfo {pages} {8208} (\bibinfo {year} {2020})},\ \bibinfo {note}
  {\textit{ibid.} \textbf{11}, 9248 (2020).}\BibitemShut {Stop}%
\bibitem [{\citenamefont {Furness}\ \emph {et~al.}(2022)\citenamefont
  {Furness}, \citenamefont {Kaplan}, \citenamefont {Ning}, \citenamefont
  {Perdew},\ and\ \citenamefont {Sun}}]{furness2021}%
  \BibitemOpen
  \bibfield  {author} {\bibinfo {author} {\bibfnamefont {J.~W.}\ \bibnamefont
  {Furness}}, \bibinfo {author} {\bibfnamefont {A.~D.}\ \bibnamefont {Kaplan}},
  \bibinfo {author} {\bibfnamefont {J.}~\bibnamefont {Ning}}, \bibinfo {author}
  {\bibfnamefont {J.~P.}\ \bibnamefont {Perdew}}, \ and\ \bibinfo {author}
  {\bibfnamefont {J.}~\bibnamefont {Sun}},\ }\href {\doibase 10.1063/5.0073623}
  {\bibfield  {journal} {\bibinfo  {journal} {J. Chem. Phys.}\ }\textbf
  {\bibinfo {volume} {156}},\ \bibinfo {pages} {034109} (\bibinfo {year}
  {2022})}\BibitemShut {NoStop}%
\bibitem [{\citenamefont {Grimme}\ \emph {et~al.}(2021)\citenamefont {Grimme},
  \citenamefont {Hansen}, \citenamefont {Ehlert},\ and\ \citenamefont
  {Mewes}}]{Grimme2021}%
  \BibitemOpen
  \bibfield  {author} {\bibinfo {author} {\bibfnamefont {S.}~\bibnamefont
  {Grimme}}, \bibinfo {author} {\bibfnamefont {A.}~\bibnamefont {Hansen}},
  \bibinfo {author} {\bibfnamefont {S.}~\bibnamefont {Ehlert}}, \ and\ \bibinfo
  {author} {\bibfnamefont {J.-M.}\ \bibnamefont {Mewes}},\ }\href {\doibase
  10.1063/5.0040021} {\bibfield  {journal} {\bibinfo  {journal} {J. Chem.
  Phys.}\ }\textbf {\bibinfo {volume} {154}},\ \bibinfo {pages} {064103}
  (\bibinfo {year} {2021})}\BibitemShut {NoStop}%
\bibitem [{\citenamefont {Holzwarth}\ \emph {et~al.}(2022)\citenamefont
  {Holzwarth}, \citenamefont {Torrent}, \citenamefont {Charraud},\ and\
  \citenamefont {C\^ot\'e}}]{holzwarth2022}%
  \BibitemOpen
  \bibfield  {author} {\bibinfo {author} {\bibfnamefont {N.~A.~W.}\
  \bibnamefont {Holzwarth}}, \bibinfo {author} {\bibfnamefont {M.}~\bibnamefont
  {Torrent}}, \bibinfo {author} {\bibfnamefont {J.-B.}\ \bibnamefont
  {Charraud}}, \ and\ \bibinfo {author} {\bibfnamefont {M.}~\bibnamefont
  {C\^ot\'e}},\ }\href {\doibase 10.1103/PhysRevB.105.125144} {\bibfield
  {journal} {\bibinfo  {journal} {Phys. Rev. B}\ }\textbf {\bibinfo {volume}
  {105}},\ \bibinfo {pages} {125144} (\bibinfo {year} {2022})}\BibitemShut
  {NoStop}%
\bibitem [{\citenamefont {Mardirossian}\ and\ \citenamefont
  {Head-Gordon}(2017)}]{mardirossian2017a}%
  \BibitemOpen
  \bibfield  {author} {\bibinfo {author} {\bibfnamefont {N.}~\bibnamefont
  {Mardirossian}}\ and\ \bibinfo {author} {\bibfnamefont {M.}~\bibnamefont
  {Head-Gordon}},\ }\href {\doibase 10.1080/00268976.2017.1333644} {\bibfield
  {journal} {\bibinfo  {journal} {Mol. Phys.}\ }\textbf {\bibinfo {volume}
  {115}},\ \bibinfo {pages} {2315} (\bibinfo {year} {2017})}\BibitemShut
  {NoStop}%
\bibitem [{\citenamefont {Mardirossian}\ \emph {et~al.}(2017)\citenamefont
  {Mardirossian}, \citenamefont {Ruiz~Pestana}, \citenamefont {Womack},
  \citenamefont {Skylaris}, \citenamefont {Head-Gordon},\ and\ \citenamefont
  {Head-Gordon}}]{mardirossian2017}%
  \BibitemOpen
  \bibfield  {author} {\bibinfo {author} {\bibfnamefont {N.}~\bibnamefont
  {Mardirossian}}, \bibinfo {author} {\bibfnamefont {L.}~\bibnamefont
  {Ruiz~Pestana}}, \bibinfo {author} {\bibfnamefont {J.~C.}\ \bibnamefont
  {Womack}}, \bibinfo {author} {\bibfnamefont {C.-K.}\ \bibnamefont
  {Skylaris}}, \bibinfo {author} {\bibfnamefont {T.}~\bibnamefont
  {Head-Gordon}}, \ and\ \bibinfo {author} {\bibfnamefont {M.}~\bibnamefont
  {Head-Gordon}},\ }\href {\doibase 10.1021/acs.jpclett.6b02527} {\bibfield
  {journal} {\bibinfo  {journal} {J. Phys. Chem. Lett.}\ }\textbf {\bibinfo
  {volume} {8}},\ \bibinfo {pages} {35} (\bibinfo {year} {2017})}\BibitemShut
  {NoStop}%
\bibitem [{\citenamefont {Terentjev}\ \emph {et~al.}(2019)\citenamefont
  {Terentjev}, \citenamefont {Constantin}, \citenamefont {Artacho},\ and\
  \citenamefont {Pitarke}}]{terentjev2019}%
  \BibitemOpen
  \bibfield  {author} {\bibinfo {author} {\bibfnamefont {A.~V.}\ \bibnamefont
  {Terentjev}}, \bibinfo {author} {\bibfnamefont {L.~A.}\ \bibnamefont
  {Constantin}}, \bibinfo {author} {\bibfnamefont {E.}~\bibnamefont {Artacho}},
  \ and\ \bibinfo {author} {\bibfnamefont {J.~M.}\ \bibnamefont {Pitarke}},\
  }\href {\doibase 10.1103/PhysRevB.100.235439} {\bibfield  {journal} {\bibinfo
   {journal} {Phys. Rev. B}\ }\textbf {\bibinfo {volume} {100}},\ \bibinfo
  {pages} {235439} (\bibinfo {year} {2019})}\BibitemShut {NoStop}%
\bibitem [{\citenamefont {Axilrod}\ and\ \citenamefont
  {Teller}(1943)}]{axilrod1943}%
  \BibitemOpen
  \bibfield  {author} {\bibinfo {author} {\bibfnamefont {B.~M.}\ \bibnamefont
  {Axilrod}}\ and\ \bibinfo {author} {\bibfnamefont {E.}~\bibnamefont
  {Teller}},\ }\href {\doibase 10.1063/1.1723844} {\bibfield  {journal}
  {\bibinfo  {journal} {J. Chem. Phys.}\ }\textbf {\bibinfo {volume} {11}},\
  \bibinfo {pages} {299} (\bibinfo {year} {1943})}\BibitemShut {NoStop}%
\bibitem [{\citenamefont {Novoselov}\ \emph {et~al.}(2004)\citenamefont
  {Novoselov}, \citenamefont {Geim}, \citenamefont {Morozov}, \citenamefont
  {Jiang}, \citenamefont {Zhang}, \citenamefont {Dubonos}, \citenamefont
  {Grigorieva},\ and\ \citenamefont {Firsov}}]{Graphene}%
  \BibitemOpen
  \bibfield  {author} {\bibinfo {author} {\bibfnamefont {K.~S.}\ \bibnamefont
  {Novoselov}}, \bibinfo {author} {\bibfnamefont {A.~K.}\ \bibnamefont {Geim}},
  \bibinfo {author} {\bibfnamefont {S.~V.}\ \bibnamefont {Morozov}}, \bibinfo
  {author} {\bibfnamefont {D.}~\bibnamefont {Jiang}}, \bibinfo {author}
  {\bibfnamefont {Y.}~\bibnamefont {Zhang}}, \bibinfo {author} {\bibfnamefont
  {S.~V.}\ \bibnamefont {Dubonos}}, \bibinfo {author} {\bibfnamefont {I.~V.}\
  \bibnamefont {Grigorieva}}, \ and\ \bibinfo {author} {\bibfnamefont {A.~A.}\
  \bibnamefont {Firsov}},\ }\href@noop {} {\bibfield  {journal} {\bibinfo
  {journal} {Science}\ }\textbf {\bibinfo {volume} {306}},\ \bibinfo {pages}
  {666} (\bibinfo {year} {2004})}\BibitemShut {NoStop}%
\bibitem [{\citenamefont {Novoselov}\ \emph {et~al.}(2016)\citenamefont
  {Novoselov}, \citenamefont {Mishchenko}, \citenamefont {Carvalho},\ and\
  \citenamefont {Castro~Neto}}]{2D}%
  \BibitemOpen
  \bibfield  {author} {\bibinfo {author} {\bibfnamefont {K.~S.}\ \bibnamefont
  {Novoselov}}, \bibinfo {author} {\bibfnamefont {A.}~\bibnamefont
  {Mishchenko}}, \bibinfo {author} {\bibfnamefont {A.}~\bibnamefont
  {Carvalho}}, \ and\ \bibinfo {author} {\bibfnamefont {A.~H.}\ \bibnamefont
  {Castro~Neto}},\ }\href@noop {} {\bibfield  {journal} {\bibinfo  {journal}
  {Science}\ }\textbf {\bibinfo {volume} {353}} (\bibinfo {year}
  {2016})}\BibitemShut {NoStop}%
\bibitem [{\citenamefont {Perdew}\ \emph {et~al.}(1996)\citenamefont {Perdew},
  \citenamefont {Burke},\ and\ \citenamefont {Ernzerhof}}]{PBE}%
  \BibitemOpen
  \bibfield  {author} {\bibinfo {author} {\bibfnamefont {J.~P.}\ \bibnamefont
  {Perdew}}, \bibinfo {author} {\bibfnamefont {K.}~\bibnamefont {Burke}}, \
  and\ \bibinfo {author} {\bibfnamefont {M.}~\bibnamefont {Ernzerhof}},\
  }\href@noop {} {\bibfield  {journal} {\bibinfo  {journal} {Phys. Rev. Lett.}\
  }\textbf {\bibinfo {volume} {77}},\ \bibinfo {pages} {3865} (\bibinfo {year}
  {1996})}\BibitemShut {NoStop}%
\bibitem [{\citenamefont {Murray}\ \emph {et~al.}(2009)\citenamefont {Murray},
  \citenamefont {Lee},\ and\ \citenamefont {Langreth}}]{xforvdw}%
  \BibitemOpen
  \bibfield  {author} {\bibinfo {author} {\bibfnamefont {{\'E}.~D.}\
  \bibnamefont {Murray}}, \bibinfo {author} {\bibfnamefont {K.}~\bibnamefont
  {Lee}}, \ and\ \bibinfo {author} {\bibfnamefont {D.~C.}\ \bibnamefont
  {Langreth}},\ }\href@noop {} {\bibfield  {journal} {\bibinfo  {journal} {J.
  Chem. Theory Comput.}\ }\textbf {\bibinfo {volume} {5}},\ \bibinfo {pages}
  {2754} (\bibinfo {year} {2009})}\BibitemShut {NoStop}%
\bibitem [{\citenamefont {Podeszwa}\ \emph {et~al.}(2010)\citenamefont
  {Podeszwa}, \citenamefont {Patkowski},\ and\ \citenamefont
  {Szalewicz}}]{S22}%
  \BibitemOpen
  \bibfield  {author} {\bibinfo {author} {\bibfnamefont {R.}~\bibnamefont
  {Podeszwa}}, \bibinfo {author} {\bibfnamefont {K.}~\bibnamefont {Patkowski}},
  \ and\ \bibinfo {author} {\bibfnamefont {K.}~\bibnamefont {Szalewicz}},\
  }\href@noop {} {\bibfield  {journal} {\bibinfo  {journal} {Phys. Chem. Chem.
  Phys.}\ }\textbf {\bibinfo {volume} {12}},\ \bibinfo {pages} {5974} (\bibinfo
  {year} {2010})}\BibitemShut {NoStop}%
\bibitem [{\citenamefont {Takatani}\ \emph {et~al.}(2010)\citenamefont
  {Takatani}, \citenamefont {Hohenstein}, \citenamefont {Malagoli},
  \citenamefont {Marshall},\ and\ \citenamefont {Sherrill}}]{S22A}%
  \BibitemOpen
  \bibfield  {author} {\bibinfo {author} {\bibfnamefont {T.}~\bibnamefont
  {Takatani}}, \bibinfo {author} {\bibfnamefont {E.~G.}\ \bibnamefont
  {Hohenstein}}, \bibinfo {author} {\bibfnamefont {M.}~\bibnamefont
  {Malagoli}}, \bibinfo {author} {\bibfnamefont {M.~S.}\ \bibnamefont
  {Marshall}}, \ and\ \bibinfo {author} {\bibfnamefont {C.~D.}\ \bibnamefont
  {Sherrill}},\ }\href@noop {} {\bibfield  {journal} {\bibinfo  {journal} {J.
  Chem. Phys.}\ }\textbf {\bibinfo {volume} {132}},\ \bibinfo {pages} {144104}
  (\bibinfo {year} {2010})}\BibitemShut {NoStop}%
\bibitem [{\citenamefont {Hankins}\ \emph {et~al.}(1970)\citenamefont
  {Hankins}, \citenamefont {Moskowitz},\ and\ \citenamefont
  {Stillinger}}]{hankins1979}%
  \BibitemOpen
  \bibfield  {author} {\bibinfo {author} {\bibfnamefont {D.}~\bibnamefont
  {Hankins}}, \bibinfo {author} {\bibfnamefont {J.~W.}\ \bibnamefont
  {Moskowitz}}, \ and\ \bibinfo {author} {\bibfnamefont {F.~H.}\ \bibnamefont
  {Stillinger}},\ }\href {\doibase 10.1063/1.1673986} {\bibfield  {journal}
  {\bibinfo  {journal} {J. Chem. Phys.}\ }\textbf {\bibinfo {volume} {53}},\
  \bibinfo {pages} {4544} (\bibinfo {year} {1970})}\BibitemShut {NoStop}%
\bibitem [{\citenamefont {Dobson}(2014)}]{dobson2014}%
  \BibitemOpen
  \bibfield  {author} {\bibinfo {author} {\bibfnamefont {J.~F.}\ \bibnamefont
  {Dobson}},\ }\href {\doibase https://doi.org/10.1002/qua.24635} {\bibfield
  {journal} {\bibinfo  {journal} {Int. J. Quantum Chem.}\ }\textbf {\bibinfo
  {volume} {114}},\ \bibinfo {pages} {1157} (\bibinfo {year}
  {2014})}\BibitemShut {NoStop}%
\bibitem [{\citenamefont {Patkowski}\ \emph {et~al.}(2005)\citenamefont
  {Patkowski}, \citenamefont {Murdachaew}, \citenamefont {Fou},\ and\
  \citenamefont {Szalewicz*}}]{Ar2CCSD_1}%
  \BibitemOpen
  \bibfield  {author} {\bibinfo {author} {\bibfnamefont {K.}~\bibnamefont
  {Patkowski}}, \bibinfo {author} {\bibfnamefont {G.}~\bibnamefont
  {Murdachaew}}, \bibinfo {author} {\bibfnamefont {C.-M.}\ \bibnamefont {Fou}},
  \ and\ \bibinfo {author} {\bibfnamefont {K.}~\bibnamefont {Szalewicz*}},\
  }\href@noop {} {\bibfield  {journal} {\bibinfo  {journal} {Mol. Phys.}\
  }\textbf {\bibinfo {volume} {103}},\ \bibinfo {pages} {2031} (\bibinfo {year}
  {2005})}\BibitemShut {NoStop}%
\bibitem [{\citenamefont {Slav{\'i}́{\v{c}}ek}\ \emph
  {et~al.}(2003)\citenamefont {Slav{\'i}́{\v{c}}ek}, \citenamefont {Kalus},
  \citenamefont {Pa{\v{s}}ka}, \citenamefont {Odv{\'{a}}rkov{\'{a}}},
  \citenamefont {Hobza},\ and\ \citenamefont {Malijevský}}]{Ar2CCSD_2}%
  \BibitemOpen
  \bibfield  {author} {\bibinfo {author} {\bibfnamefont {P.}~\bibnamefont
  {Slav{\'i}́{\v{c}}ek}}, \bibinfo {author} {\bibfnamefont {R.}~\bibnamefont
  {Kalus}}, \bibinfo {author} {\bibfnamefont {P.}~\bibnamefont {Pa{\v{s}}ka}},
  \bibinfo {author} {\bibfnamefont {I.}~\bibnamefont {Odv{\'{a}}rkov{\'{a}}}},
  \bibinfo {author} {\bibfnamefont {P.}~\bibnamefont {Hobza}}, \ and\ \bibinfo
  {author} {\bibfnamefont {A.}~\bibnamefont {Malijevský}},\ }\href@noop {}
  {\bibfield  {journal} {\bibinfo  {journal} {J. Chem. Phys.}\ }\textbf
  {\bibinfo {volume} {119}},\ \bibinfo {pages} {2102} (\bibinfo {year}
  {2003})}\BibitemShut {NoStop}%
\bibitem [{r2s(2022)}]{r2scan_d4_repo}%
  \BibitemOpen
  \href@noop {} {} (\bibinfo {year} {2022}),\ \bibinfo {note} {see Table S22x5
  of the r$^2$SCAN-D4 code repository:
  \url{https://github.com/awvwgk/r2scan-d4-paper}}\BibitemShut {NoStop}%
\bibitem [{\citenamefont {Kresse}\ and\ \citenamefont
  {Furthm\"uller}(1996)}]{VASP}%
  \BibitemOpen
  \bibfield  {author} {\bibinfo {author} {\bibfnamefont {G.}~\bibnamefont
  {Kresse}}\ and\ \bibinfo {author} {\bibfnamefont {J.}~\bibnamefont
  {Furthm\"uller}},\ }\href {\doibase 10.1103/PhysRevB.54.11169} {\bibfield
  {journal} {\bibinfo  {journal} {Phys. Rev. B}\ }\textbf {\bibinfo {volume}
  {54}},\ \bibinfo {pages} {11169} (\bibinfo {year} {1996})}\BibitemShut
  {NoStop}%
\bibitem [{vas(2022)}]{vasp_stress_bug}%
  \BibitemOpen
  \href@noop {} {} (\bibinfo {year} {2022}),\ \bibinfo {note} {the rVV10
  correlation stress tensor elements in VASP versions 6.2.0 and below were
  incorrectly coded, see
  \url{https://www.vasp.at/wiki/index.php/Nonlocal_vdW-DF_functionals}}\BibitemShut
  {NoStop}%
\bibitem [{\citenamefont {Marshall}\ \emph {et~al.}(2011)\citenamefont
  {Marshall}, \citenamefont {Burns},\ and\ \citenamefont {Sherrill}}]{S22B}%
  \BibitemOpen
  \bibfield  {author} {\bibinfo {author} {\bibfnamefont {M.~S.}\ \bibnamefont
  {Marshall}}, \bibinfo {author} {\bibfnamefont {L.~A.}\ \bibnamefont {Burns}},
  \ and\ \bibinfo {author} {\bibfnamefont {C.~D.}\ \bibnamefont {Sherrill}},\
  }\href@noop {} {\bibfield  {journal} {\bibinfo  {journal} {J. Chem. Phys.}\
  }\textbf {\bibinfo {volume} {135}},\ \bibinfo {pages} {194102} (\bibinfo
  {year} {2011})}\BibitemShut {NoStop}%
\bibitem [{\citenamefont {Bj{\"o}rkman}(2014)}]{L28_jcp}%
  \BibitemOpen
  \bibfield  {author} {\bibinfo {author} {\bibfnamefont {T.}~\bibnamefont
  {Bj{\"o}rkman}},\ }\href@noop {} {\bibfield  {journal} {\bibinfo  {journal}
  {J. Chem. Phys.}\ }\textbf {\bibinfo {volume} {141}},\ \bibinfo {pages}
  {074708} (\bibinfo {year} {2014})}\BibitemShut {NoStop}%
\bibitem [{\citenamefont {Ning}\ \emph
  {et~al.}(2022{\natexlab{a}})\citenamefont {Ning}, \citenamefont {Kothakonda},
  \citenamefont {Furness}, \citenamefont {Kaplan}, \citenamefont {Ehlert},
  \citenamefont {Brandenburg}, \citenamefont {Perdew},\ and\ \citenamefont
  {Sun}}]{ning2022}%
  \BibitemOpen
  \bibfield  {author} {\bibinfo {author} {\bibfnamefont {J.}~\bibnamefont
  {Ning}}, \bibinfo {author} {\bibfnamefont {M.}~\bibnamefont {Kothakonda}},
  \bibinfo {author} {\bibfnamefont {J.~W.}\ \bibnamefont {Furness}}, \bibinfo
  {author} {\bibfnamefont {A.~D.}\ \bibnamefont {Kaplan}}, \bibinfo {author}
  {\bibfnamefont {S.}~\bibnamefont {Ehlert}}, \bibinfo {author} {\bibfnamefont
  {J.~G.}\ \bibnamefont {Brandenburg}}, \bibinfo {author} {\bibfnamefont
  {J.~P.}\ \bibnamefont {Perdew}}, \ and\ \bibinfo {author} {\bibfnamefont
  {J.}~\bibnamefont {Sun}},\ }\href {\doibase 10.5281/zenodo.6871949} {\enquote
  {\bibinfo {title} {{Data for "Workhorse minimally-empirical
  dispersion-corrected density functional, with tests for weakly-bound systems:
  r2 SCAN + rVV10"}},}\ } (\bibinfo {year} {2022}{\natexlab{a}}),\ \bibinfo
  {note} {{DOI: 10.5281/zenodo.6871949}}\BibitemShut {NoStop}%
\bibitem [{\citenamefont {Bj\"orkman}\ \emph {et~al.}(2012)\citenamefont
  {Bj\"orkman}, \citenamefont {Gulans}, \citenamefont {Krasheninnikov},\ and\
  \citenamefont {Nieminen}}]{Eb_RPA}%
  \BibitemOpen
  \bibfield  {author} {\bibinfo {author} {\bibfnamefont {T.}~\bibnamefont
  {Bj\"orkman}}, \bibinfo {author} {\bibfnamefont {A.}~\bibnamefont {Gulans}},
  \bibinfo {author} {\bibfnamefont {A.~V.}\ \bibnamefont {Krasheninnikov}}, \
  and\ \bibinfo {author} {\bibfnamefont {R.~M.}\ \bibnamefont {Nieminen}},\
  }\href {\doibase 10.1103/PhysRevLett.108.235502} {\bibfield  {journal}
  {\bibinfo  {journal} {Phys. Rev. Lett.}\ }\textbf {\bibinfo {volume} {108}},\
  \bibinfo {pages} {235502} (\bibinfo {year} {2012})}\BibitemShut {NoStop}%
\bibitem [{\citenamefont {Togo}\ and\ \citenamefont {Tanaka}(2015)}]{phonopy}%
  \BibitemOpen
  \bibfield  {author} {\bibinfo {author} {\bibfnamefont {A.}~\bibnamefont
  {Togo}}\ and\ \bibinfo {author} {\bibfnamefont {I.}~\bibnamefont {Tanaka}},\
  }\href@noop {} {\bibfield  {journal} {\bibinfo  {journal} {Scripta
  Materialia}\ }\textbf {\bibinfo {volume} {108}},\ \bibinfo {pages} {1}
  (\bibinfo {year} {2015})}\BibitemShut {NoStop}%
\bibitem [{\citenamefont {Lee}\ \emph {et~al.}(2010)\citenamefont {Lee},
  \citenamefont {Murray}, \citenamefont {Kong}, \citenamefont {Lundqvist},\
  and\ \citenamefont {Langreth}}]{vdW_DF2}%
  \BibitemOpen
  \bibfield  {author} {\bibinfo {author} {\bibfnamefont {K.}~\bibnamefont
  {Lee}}, \bibinfo {author} {\bibfnamefont {E.~D.}\ \bibnamefont {Murray}},
  \bibinfo {author} {\bibfnamefont {L.}~\bibnamefont {Kong}}, \bibinfo {author}
  {\bibfnamefont {B.~I.}\ \bibnamefont {Lundqvist}}, \ and\ \bibinfo {author}
  {\bibfnamefont {D.~C.}\ \bibnamefont {Langreth}},\ }\href {\doibase
  10.1103/PhysRevB.82.081101} {\bibfield  {journal} {\bibinfo  {journal} {Phys.
  Rev. B}\ }\textbf {\bibinfo {volume} {82}},\ \bibinfo {pages} {081101}
  (\bibinfo {year} {2010})}\BibitemShut {NoStop}%
\bibitem [{\citenamefont {Gould}\ \emph {et~al.}(2018)\citenamefont {Gould},
  \citenamefont {Johnson},\ and\ \citenamefont {Tawfik}}]{gould2018}%
  \BibitemOpen
  \bibfield  {author} {\bibinfo {author} {\bibfnamefont {T.}~\bibnamefont
  {Gould}}, \bibinfo {author} {\bibfnamefont {E.~R.}\ \bibnamefont {Johnson}},
  \ and\ \bibinfo {author} {\bibfnamefont {S.~A.}\ \bibnamefont {Tawfik}},\
  }\href {\doibase 10.3762/bjoc.14.99} {\bibfield  {journal} {\bibinfo
  {journal} {Beilstein J. Org. Chem.}\ }\textbf {\bibinfo {volume} {14}},\
  \bibinfo {pages} {1181–1191} (\bibinfo {year} {2018})}\BibitemShut
  {NoStop}%
\bibitem [{\citenamefont {Dasgupta}\ \emph {et~al.}(2022)\citenamefont
  {Dasgupta}, \citenamefont {Shahi}, \citenamefont {Bhetwal}, \citenamefont
  {Perdew},\ and\ \citenamefont {Paesani}}]{dasgupta2022}%
  \BibitemOpen
  \bibfield  {author} {\bibinfo {author} {\bibfnamefont {S.}~\bibnamefont
  {Dasgupta}}, \bibinfo {author} {\bibfnamefont {S.}~\bibnamefont {Shahi}},
  \bibinfo {author} {\bibfnamefont {P.}~\bibnamefont {Bhetwal}}, \bibinfo
  {author} {\bibfnamefont {J.}~\bibnamefont {Perdew}}, \ and\ \bibinfo {author}
  {\bibfnamefont {F.}~\bibnamefont {Paesani}},\ }\href {\doibase
  10.26434/chemrxiv-2022-8r5v9} {\enquote {\bibinfo {title} {How good is the
  density-corrected scan functional for neutral and ionic aqueous systems, and
  what is so right about the hartree-fock density?}}\ } (\bibinfo {year}
  {2022}),\ \bibinfo {note} {submitted to J. Chem. Theory Comput.}\BibitemShut
  {Stop}%
\bibitem [{\citenamefont {Ning}\ \emph
  {et~al.}(2022{\natexlab{b}})\citenamefont {Ning}, \citenamefont {Furness},\
  and\ \citenamefont {Sun}}]{r2SCAN_phonon}%
  \BibitemOpen
  \bibfield  {author} {\bibinfo {author} {\bibfnamefont {J.}~\bibnamefont
  {Ning}}, \bibinfo {author} {\bibfnamefont {J.~W.}\ \bibnamefont {Furness}}, \
  and\ \bibinfo {author} {\bibfnamefont {J.}~\bibnamefont {Sun}},\ }\href
  {\doibase 10.1021/acs.chemmater.1c03222} {\bibfield  {journal} {\bibinfo
  {journal} {Chem. Mater.}\ }\textbf {\bibinfo {volume} {34}},\ \bibinfo
  {pages} {2562} (\bibinfo {year} {2022}{\natexlab{b}})}\BibitemShut {NoStop}%
\bibitem [{\citenamefont {Sinnokrot}\ and\ \citenamefont
  {Sherrill}(2004)}]{Tben_CCSDT}%
  \BibitemOpen
  \bibfield  {author} {\bibinfo {author} {\bibfnamefont {M.~O.}\ \bibnamefont
  {Sinnokrot}}\ and\ \bibinfo {author} {\bibfnamefont {C.~D.}\ \bibnamefont
  {Sherrill}},\ }\href@noop {} {\bibfield  {journal} {\bibinfo  {journal} {The
  Journal of Physical Chemistry A}\ }\textbf {\bibinfo {volume} {108}},\
  \bibinfo {pages} {10200} (\bibinfo {year} {2004})}\BibitemShut {NoStop}%
\bibitem [{\citenamefont {Maultzsch}\ \emph {et~al.}(2004)\citenamefont
  {Maultzsch}, \citenamefont {Reich}, \citenamefont {Thomsen}, \citenamefont
  {Requardt},\ and\ \citenamefont {Ordej{\'o}n}}]{G_phonon_PRL}%
  \BibitemOpen
  \bibfield  {author} {\bibinfo {author} {\bibfnamefont {J.}~\bibnamefont
  {Maultzsch}}, \bibinfo {author} {\bibfnamefont {S.}~\bibnamefont {Reich}},
  \bibinfo {author} {\bibfnamefont {C.}~\bibnamefont {Thomsen}}, \bibinfo
  {author} {\bibfnamefont {H.}~\bibnamefont {Requardt}}, \ and\ \bibinfo
  {author} {\bibfnamefont {P.}~\bibnamefont {Ordej{\'o}n}},\ }\href@noop {}
  {\bibfield  {journal} {\bibinfo  {journal} {Phys. Rev. Lett.}\ }\textbf
  {\bibinfo {volume} {92}},\ \bibinfo {pages} {075501} (\bibinfo {year}
  {2004})}\BibitemShut {NoStop}%
\bibitem [{\citenamefont {Mohr}\ \emph {et~al.}(2007)\citenamefont {Mohr},
  \citenamefont {Maultzsch}, \citenamefont {Dobard{\v{z}}i{\'c}}, \citenamefont
  {Reich}, \citenamefont {Milo{\v{s}}evi{\'c}}, \citenamefont
  {Damnjanovi{\'c}}, \citenamefont {Bosak}, \citenamefont {Krisch},\ and\
  \citenamefont {Thomsen}}]{G_phonon_PRB}%
  \BibitemOpen
  \bibfield  {author} {\bibinfo {author} {\bibfnamefont {M.}~\bibnamefont
  {Mohr}}, \bibinfo {author} {\bibfnamefont {J.}~\bibnamefont {Maultzsch}},
  \bibinfo {author} {\bibfnamefont {E.}~\bibnamefont {Dobard{\v{z}}i{\'c}}},
  \bibinfo {author} {\bibfnamefont {S.}~\bibnamefont {Reich}}, \bibinfo
  {author} {\bibfnamefont {I.}~\bibnamefont {Milo{\v{s}}evi{\'c}}}, \bibinfo
  {author} {\bibfnamefont {M.}~\bibnamefont {Damnjanovi{\'c}}}, \bibinfo
  {author} {\bibfnamefont {A.}~\bibnamefont {Bosak}}, \bibinfo {author}
  {\bibfnamefont {M.}~\bibnamefont {Krisch}}, \ and\ \bibinfo {author}
  {\bibfnamefont {C.}~\bibnamefont {Thomsen}},\ }\href@noop {} {\bibfield
  {journal} {\bibinfo  {journal} {Phys. Rev. B}\ }\textbf {\bibinfo {volume}
  {76}},\ \bibinfo {pages} {035439} (\bibinfo {year} {2007})}\BibitemShut
  {NoStop}%
\bibitem [{\citenamefont {Nicklow}\ \emph {et~al.}(1972)\citenamefont
  {Nicklow}, \citenamefont {Wakabayashi},\ and\ \citenamefont
  {Smith}}]{G_phonon_1972PRB}%
  \BibitemOpen
  \bibfield  {author} {\bibinfo {author} {\bibfnamefont {R.}~\bibnamefont
  {Nicklow}}, \bibinfo {author} {\bibfnamefont {N.}~\bibnamefont
  {Wakabayashi}}, \ and\ \bibinfo {author} {\bibfnamefont {H.}~\bibnamefont
  {Smith}},\ }\href@noop {} {\bibfield  {journal} {\bibinfo  {journal} {Phys.
  Rev. B}\ }\textbf {\bibinfo {volume} {5}},\ \bibinfo {pages} {4951} (\bibinfo
  {year} {1972})}\BibitemShut {NoStop}%
\bibitem [{\citenamefont {Tornatzky}\ \emph {et~al.}(2019)\citenamefont
  {Tornatzky}, \citenamefont {Gillen}, \citenamefont {Uchiyama},\ and\
  \citenamefont {Maultzsch}}]{MoS2_2019phonon}%
  \BibitemOpen
  \bibfield  {author} {\bibinfo {author} {\bibfnamefont {H.}~\bibnamefont
  {Tornatzky}}, \bibinfo {author} {\bibfnamefont {R.}~\bibnamefont {Gillen}},
  \bibinfo {author} {\bibfnamefont {H.}~\bibnamefont {Uchiyama}}, \ and\
  \bibinfo {author} {\bibfnamefont {J.}~\bibnamefont {Maultzsch}},\ }\href@noop
  {} {\bibfield  {journal} {\bibinfo  {journal} {Phys. Rev. B}\ }\textbf
  {\bibinfo {volume} {99}},\ \bibinfo {pages} {144309} (\bibinfo {year}
  {2019})}\BibitemShut {NoStop}%
\bibitem [{\citenamefont {Scheuschner}\ \emph {et~al.}(2015)\citenamefont
  {Scheuschner}, \citenamefont {Gillen}, \citenamefont {Staiger},\ and\
  \citenamefont {Maultzsch}}]{MoS2_R2015phonon}%
  \BibitemOpen
  \bibfield  {author} {\bibinfo {author} {\bibfnamefont {N.}~\bibnamefont
  {Scheuschner}}, \bibinfo {author} {\bibfnamefont {R.}~\bibnamefont {Gillen}},
  \bibinfo {author} {\bibfnamefont {M.}~\bibnamefont {Staiger}}, \ and\
  \bibinfo {author} {\bibfnamefont {J.}~\bibnamefont {Maultzsch}},\ }\href
  {\doibase 10.1103/PhysRevB.91.235409} {\bibfield  {journal} {\bibinfo
  {journal} {Phys. Rev. B}\ }\textbf {\bibinfo {volume} {91}},\ \bibinfo
  {pages} {235409} (\bibinfo {year} {2015})}\BibitemShut {NoStop}%
\bibitem [{\citenamefont {Wieting}\ and\ \citenamefont
  {Verble}(1971)}]{MoS2_IR1971phonon}%
  \BibitemOpen
  \bibfield  {author} {\bibinfo {author} {\bibfnamefont {T.~J.}\ \bibnamefont
  {Wieting}}\ and\ \bibinfo {author} {\bibfnamefont {J.~L.}\ \bibnamefont
  {Verble}},\ }\href {\doibase 10.1103/PhysRevB.3.4286} {\bibfield  {journal}
  {\bibinfo  {journal} {Phys. Rev. B}\ }\textbf {\bibinfo {volume} {3}},\
  \bibinfo {pages} {4286} (\bibinfo {year} {1971})}\BibitemShut {NoStop}%
\bibitem [{\citenamefont {Chen}\ and\ \citenamefont
  {Wang}(1974)}]{MoS2_R1974phonon}%
  \BibitemOpen
  \bibfield  {author} {\bibinfo {author} {\bibfnamefont {J.}~\bibnamefont
  {Chen}}\ and\ \bibinfo {author} {\bibfnamefont {C.}~\bibnamefont {Wang}},\
  }\href@noop {} {\bibfield  {journal} {\bibinfo  {journal} {Solid State
  Commun.}\ }\textbf {\bibinfo {volume} {14}},\ \bibinfo {pages} {857}
  (\bibinfo {year} {1974})}\BibitemShut {NoStop}%
\bibitem [{\citenamefont {Sch\"{u}tz}\ \emph {et~al.}(2017)\citenamefont
  {Sch\"{u}tz}, \citenamefont {Maschio}, \citenamefont {Karttunen},\ and\
  \citenamefont {Usvyat}}]{BlackP_LdrCCD}%
  \BibitemOpen
  \bibfield  {author} {\bibinfo {author} {\bibfnamefont {M.}~\bibnamefont
  {Sch\"{u}tz}}, \bibinfo {author} {\bibfnamefont {L.}~\bibnamefont {Maschio}},
  \bibinfo {author} {\bibfnamefont {A.~J.}\ \bibnamefont {Karttunen}}, \ and\
  \bibinfo {author} {\bibfnamefont {D.}~\bibnamefont {Usvyat}},\ }\href@noop {}
  {\bibfield  {journal} {\bibinfo  {journal} {J. Phys. Chem. Lett.}\ }\textbf
  {\bibinfo {volume} {8}},\ \bibinfo {pages} {1290} (\bibinfo {year}
  {2017})}\BibitemShut {NoStop}%
\bibitem [{\citenamefont {Hamada}(2014)}]{hamada2014}%
  \BibitemOpen
  \bibfield  {author} {\bibinfo {author} {\bibfnamefont {I.}~\bibnamefont
  {Hamada}},\ }\href {\doibase 10.1103/PhysRevB.89.121103} {\bibfield
  {journal} {\bibinfo  {journal} {Phys. Rev. B}\ }\textbf {\bibinfo {volume}
  {89}},\ \bibinfo {pages} {121103} (\bibinfo {year} {2014})}\BibitemShut
  {NoStop}%
\bibitem [{\citenamefont {Zacharia}\ \emph {et~al.}(2004)\citenamefont
  {Zacharia}, \citenamefont {Ulbricht},\ and\ \citenamefont {Hertel}}]{G_expt}%
  \BibitemOpen
  \bibfield  {author} {\bibinfo {author} {\bibfnamefont {R.}~\bibnamefont
  {Zacharia}}, \bibinfo {author} {\bibfnamefont {H.}~\bibnamefont {Ulbricht}},
  \ and\ \bibinfo {author} {\bibfnamefont {T.}~\bibnamefont {Hertel}},\ }\href
  {\doibase 10.1103/PhysRevB.69.155406} {\bibfield  {journal} {\bibinfo
  {journal} {Phys. Rev. B}\ }\textbf {\bibinfo {volume} {69}},\ \bibinfo
  {pages} {155406} (\bibinfo {year} {2004})}\BibitemShut {NoStop}%
\bibitem [{\citenamefont {Spanu}\ \emph {et~al.}(2009)\citenamefont {Spanu},
  \citenamefont {Sorella},\ and\ \citenamefont {Galli}}]{G_QMC}%
  \BibitemOpen
  \bibfield  {author} {\bibinfo {author} {\bibfnamefont {L.}~\bibnamefont
  {Spanu}}, \bibinfo {author} {\bibfnamefont {S.}~\bibnamefont {Sorella}}, \
  and\ \bibinfo {author} {\bibfnamefont {G.}~\bibnamefont {Galli}},\ }\href
  {\doibase 10.1103/PhysRevLett.103.196401} {\bibfield  {journal} {\bibinfo
  {journal} {Phys. Rev. Lett.}\ }\textbf {\bibinfo {volume} {103}},\ \bibinfo
  {pages} {196401} (\bibinfo {year} {2009})}\BibitemShut {NoStop}%
\bibitem [{\citenamefont {Leb\`egue}\ \emph {et~al.}(2010)\citenamefont
  {Leb\`egue}, \citenamefont {Harl}, \citenamefont {Gould}, \citenamefont
  {\'Angy\'an}, \citenamefont {Kresse},\ and\ \citenamefont
  {Dobson}}]{lebegue2010}%
  \BibitemOpen
  \bibfield  {author} {\bibinfo {author} {\bibfnamefont {S.}~\bibnamefont
  {Leb\`egue}}, \bibinfo {author} {\bibfnamefont {J.}~\bibnamefont {Harl}},
  \bibinfo {author} {\bibfnamefont {T.}~\bibnamefont {Gould}}, \bibinfo
  {author} {\bibfnamefont {J.~G.}\ \bibnamefont {\'Angy\'an}}, \bibinfo
  {author} {\bibfnamefont {G.}~\bibnamefont {Kresse}}, \ and\ \bibinfo {author}
  {\bibfnamefont {J.~F.}\ \bibnamefont {Dobson}},\ }\href {\doibase
  10.1103/PhysRevLett.105.196401} {\bibfield  {journal} {\bibinfo  {journal}
  {Phys. Rev. Lett.}\ }\textbf {\bibinfo {volume} {105}},\ \bibinfo {pages}
  {196401} (\bibinfo {year} {2010})}\BibitemShut {NoStop}%
\bibitem [{\citenamefont {Mostaani}\ \emph {et~al.}(2015)\citenamefont
  {Mostaani}, \citenamefont {Drummond},\ and\ \citenamefont
  {Fal'ko}}]{G2L_QMC}%
  \BibitemOpen
  \bibfield  {author} {\bibinfo {author} {\bibfnamefont {E.}~\bibnamefont
  {Mostaani}}, \bibinfo {author} {\bibfnamefont {N.~D.}\ \bibnamefont
  {Drummond}}, \ and\ \bibinfo {author} {\bibfnamefont {V.~I.}\ \bibnamefont
  {Fal'ko}},\ }\href {\doibase 10.1103/PhysRevLett.115.115501} {\bibfield
  {journal} {\bibinfo  {journal} {Phys. Rev. Lett.}\ }\textbf {\bibinfo
  {volume} {115}},\ \bibinfo {pages} {115501} (\bibinfo {year}
  {2015})}\BibitemShut {NoStop}%
\bibitem [{\citenamefont {Hsing}\ \emph {et~al.}(2014)\citenamefont {Hsing},
  \citenamefont {Cheng}, \citenamefont {Chou}, \citenamefont {Chang},\ and\
  \citenamefont {Wei}}]{hBN2L_QMC}%
  \BibitemOpen
  \bibfield  {author} {\bibinfo {author} {\bibfnamefont {C.-R.}\ \bibnamefont
  {Hsing}}, \bibinfo {author} {\bibfnamefont {C.}~\bibnamefont {Cheng}},
  \bibinfo {author} {\bibfnamefont {J.-P.}\ \bibnamefont {Chou}}, \bibinfo
  {author} {\bibfnamefont {C.-M.}\ \bibnamefont {Chang}}, \ and\ \bibinfo
  {author} {\bibfnamefont {C.-M.}\ \bibnamefont {Wei}},\ }\href@noop {}
  {\bibfield  {journal} {\bibinfo  {journal} {New J. Phys.}\ }\textbf {\bibinfo
  {volume} {16}},\ \bibinfo {pages} {113015} (\bibinfo {year}
  {2014})}\BibitemShut {NoStop}%
\bibitem [{\citenamefont {Fang}\ \emph {et~al.}(2020)\citenamefont {Fang},
  \citenamefont {Li}, \citenamefont {Shi}, \citenamefont {Li}, \citenamefont
  {Guo}, \citenamefont {Chen}, \citenamefont {Peng},\ and\ \citenamefont
  {Wei}}]{MoS2_expt}%
  \BibitemOpen
  \bibfield  {author} {\bibinfo {author} {\bibfnamefont {Z.}~\bibnamefont
  {Fang}}, \bibinfo {author} {\bibfnamefont {X.}~\bibnamefont {Li}}, \bibinfo
  {author} {\bibfnamefont {W.}~\bibnamefont {Shi}}, \bibinfo {author}
  {\bibfnamefont {Z.}~\bibnamefont {Li}}, \bibinfo {author} {\bibfnamefont
  {Y.}~\bibnamefont {Guo}}, \bibinfo {author} {\bibfnamefont {Q.}~\bibnamefont
  {Chen}}, \bibinfo {author} {\bibfnamefont {L.}~\bibnamefont {Peng}}, \ and\
  \bibinfo {author} {\bibfnamefont {X.}~\bibnamefont {Wei}},\ }\href@noop {}
  {\bibfield  {journal} {\bibinfo  {journal} {J. Phys. Chem. C}\ }\textbf
  {\bibinfo {volume} {124}},\ \bibinfo {pages} {23419} (\bibinfo {year}
  {2020})}\BibitemShut {NoStop}%
\bibitem [{\citenamefont {Krogel}\ \emph {et~al.}(2020)\citenamefont {Krogel},
  \citenamefont {Yuk}, \citenamefont {Kent},\ and\ \citenamefont
  {Cooper}}]{TiS2_QMC}%
  \BibitemOpen
  \bibfield  {author} {\bibinfo {author} {\bibfnamefont {J.~T.}\ \bibnamefont
  {Krogel}}, \bibinfo {author} {\bibfnamefont {S.~F.}\ \bibnamefont {Yuk}},
  \bibinfo {author} {\bibfnamefont {P.~R.}\ \bibnamefont {Kent}}, \ and\
  \bibinfo {author} {\bibfnamefont {V.~R.}\ \bibnamefont {Cooper}},\
  }\href@noop {} {\bibfield  {journal} {\bibinfo  {journal} {J. Phys. Chem. A}\
  }\textbf {\bibinfo {volume} {124}},\ \bibinfo {pages} {9867} (\bibinfo {year}
  {2020})}\BibitemShut {NoStop}%
\bibitem [{\citenamefont {Shulenburger}\ \emph {et~al.}(2015)\citenamefont
  {Shulenburger}, \citenamefont {Baczewski}, \citenamefont {Zhu}, \citenamefont
  {Guan},\ and\ \citenamefont {Tomanek}}]{BlackP_QMC}%
  \BibitemOpen
  \bibfield  {author} {\bibinfo {author} {\bibfnamefont {L.}~\bibnamefont
  {Shulenburger}}, \bibinfo {author} {\bibfnamefont {A.~D.}\ \bibnamefont
  {Baczewski}}, \bibinfo {author} {\bibfnamefont {Z.}~\bibnamefont {Zhu}},
  \bibinfo {author} {\bibfnamefont {J.}~\bibnamefont {Guan}}, \ and\ \bibinfo
  {author} {\bibfnamefont {D.}~\bibnamefont {Tomanek}},\ }\href@noop {}
  {\bibfield  {journal} {\bibinfo  {journal} {Nano Lett.}\ }\textbf {\bibinfo
  {volume} {15}},\ \bibinfo {pages} {8170} (\bibinfo {year}
  {2015})}\BibitemShut {NoStop}%
\bibitem [{\citenamefont {Guillam{\'o}n}\ \emph {et~al.}(2008)\citenamefont
  {Guillam{\'o}n}, \citenamefont {Suderow}, \citenamefont {Vieira},
  \citenamefont {Cario}, \citenamefont {Diener},\ and\ \citenamefont
  {Rodiere}}]{NbS2_sup}%
  \BibitemOpen
  \bibfield  {author} {\bibinfo {author} {\bibfnamefont {I.}~\bibnamefont
  {Guillam{\'o}n}}, \bibinfo {author} {\bibfnamefont {H.}~\bibnamefont
  {Suderow}}, \bibinfo {author} {\bibfnamefont {S.}~\bibnamefont {Vieira}},
  \bibinfo {author} {\bibfnamefont {L.}~\bibnamefont {Cario}}, \bibinfo
  {author} {\bibfnamefont {P.}~\bibnamefont {Diener}}, \ and\ \bibinfo {author}
  {\bibfnamefont {P.}~\bibnamefont {Rodiere}},\ }\href@noop {} {\bibfield
  {journal} {\bibinfo  {journal} {Phys. Rev. Lett.}\ }\textbf {\bibinfo
  {volume} {101}},\ \bibinfo {pages} {166407} (\bibinfo {year}
  {2008})}\BibitemShut {NoStop}%
\bibitem [{\citenamefont {Johannes}\ \emph {et~al.}(2006)\citenamefont
  {Johannes}, \citenamefont {Mazin},\ and\ \citenamefont
  {Howells}}]{NbSe2_CDW}%
  \BibitemOpen
  \bibfield  {author} {\bibinfo {author} {\bibfnamefont {M.}~\bibnamefont
  {Johannes}}, \bibinfo {author} {\bibfnamefont {I.}~\bibnamefont {Mazin}}, \
  and\ \bibinfo {author} {\bibfnamefont {C.}~\bibnamefont {Howells}},\
  }\href@noop {} {\bibfield  {journal} {\bibinfo  {journal} {Phys. Rev. B}\
  }\textbf {\bibinfo {volume} {73}},\ \bibinfo {pages} {205102} (\bibinfo
  {year} {2006})}\BibitemShut {NoStop}%
\bibitem [{\citenamefont {Battaglia}\ \emph {et~al.}(2005)\citenamefont
  {Battaglia}, \citenamefont {Cercellier}, \citenamefont {Clerc}, \citenamefont
  {Despont}, \citenamefont {Garnier}, \citenamefont {Koitzsch}, \citenamefont
  {Aebi}, \citenamefont {Berger}, \citenamefont {Forr{\'o}},\ and\
  \citenamefont {Ambrosch-Draxl}}]{NbTe2_CDW}%
  \BibitemOpen
  \bibfield  {author} {\bibinfo {author} {\bibfnamefont {C.}~\bibnamefont
  {Battaglia}}, \bibinfo {author} {\bibfnamefont {H.}~\bibnamefont
  {Cercellier}}, \bibinfo {author} {\bibfnamefont {F.}~\bibnamefont {Clerc}},
  \bibinfo {author} {\bibfnamefont {L.}~\bibnamefont {Despont}}, \bibinfo
  {author} {\bibfnamefont {M.~G.}\ \bibnamefont {Garnier}}, \bibinfo {author}
  {\bibfnamefont {C.}~\bibnamefont {Koitzsch}}, \bibinfo {author}
  {\bibfnamefont {P.}~\bibnamefont {Aebi}}, \bibinfo {author} {\bibfnamefont
  {H.}~\bibnamefont {Berger}}, \bibinfo {author} {\bibfnamefont
  {L.}~\bibnamefont {Forr{\'o}}}, \ and\ \bibinfo {author} {\bibfnamefont
  {C.}~\bibnamefont {Ambrosch-Draxl}},\ }\href@noop {} {\bibfield  {journal}
  {\bibinfo  {journal} {Phys. Rev. B}\ }\textbf {\bibinfo {volume} {72}},\
  \bibinfo {pages} {195114} (\bibinfo {year} {2005})}\BibitemShut {NoStop}%
\bibitem [{\citenamefont {Dobson}\ and\ \citenamefont
  {Wang}(1999)}]{dobson1999}%
  \BibitemOpen
  \bibfield  {author} {\bibinfo {author} {\bibfnamefont {J.~F.}\ \bibnamefont
  {Dobson}}\ and\ \bibinfo {author} {\bibfnamefont {J.}~\bibnamefont {Wang}},\
  }\href {\doibase 10.1103/PhysRevLett.82.2123} {\bibfield  {journal} {\bibinfo
   {journal} {Phys. Rev. Lett.}\ }\textbf {\bibinfo {volume} {82}},\ \bibinfo
  {pages} {2123} (\bibinfo {year} {1999})}\BibitemShut {NoStop}%
\bibitem [{\citenamefont {Kurth}\ and\ \citenamefont
  {Perdew}(1999)}]{kurth1999}%
  \BibitemOpen
  \bibfield  {author} {\bibinfo {author} {\bibfnamefont {S.}~\bibnamefont
  {Kurth}}\ and\ \bibinfo {author} {\bibfnamefont {J.~P.}\ \bibnamefont
  {Perdew}},\ }\href {\doibase 10.1103/PhysRevB.59.10461} {\bibfield  {journal}
  {\bibinfo  {journal} {Phys. Rev. B}\ }\textbf {\bibinfo {volume} {59}},\
  \bibinfo {pages} {10461} (\bibinfo {year} {1999})}\BibitemShut {NoStop}%
\bibitem [{\citenamefont {Gould}(2012)}]{gould2012}%
  \BibitemOpen
  \bibfield  {author} {\bibinfo {author} {\bibfnamefont {T.}~\bibnamefont
  {Gould}},\ }\href {\doibase 10.1063/1.4755286} {\bibfield  {journal}
  {\bibinfo  {journal} {J. Chem. Phys.}\ }\textbf {\bibinfo {volume} {137}},\
  \bibinfo {pages} {111101} (\bibinfo {year} {2012})}\BibitemShut {NoStop}%
\end{thebibliography}%
